\documentclass[letterpaper,twocolumn,10pt]{article}
\usepackage{usenix}

\usepackage{subcaption}
\usepackage{xspace}
\usepackage{array}
\usepackage{tabularx}
\usepackage{enumitem}
\usepackage{float}
\usepackage{pgfplots}
\usepackage{tikz}
\usepackage{amsmath}
\usepackage{cleveref}
\usepackage{tablefootnote}
\usepackage{titlesec}
\usepackage{booktabs}
\usepackage{makecell}
\usepackage[fixed]{fontawesome5}
\usetikzlibrary{tikzmark}
\usetikzlibrary{shapes.geometric}
\usetikzlibrary {arrows.meta}
\usepackage[english]{babel}
\usepackage{blindtext}

\pgfplotsset{
    width=\linewidth,
    height=27mm,
    compat=1.18,
    tick label style={font=\footnotesize},
    label style={font=\footnotesize},
    legend style={font=\footnotesize},
    title style={font=\footnotesize,yshift=-3mm}}

\definecolor{myblue}{HTML}{082a54}
\definecolor{myred}{HTML}{e02b35}
\definecolor{myyellow}{HTML}{f0c571}
\definecolor{mygreen}{HTML}{59a89c}
\definecolor{myviolet}{HTML}{7e4794}
\definecolor{mylightgreen}{HTML}{36b700}
\definecolor{mylightblue}{HTML}{0b81a2}
\definecolor{mydarkred}{HTML}{9d2c00}

\crefname{section}{\S}{\S}
\Crefname{section}{§}{§}
\crefname{appendix}{§}{§}
\Crefname{appendix}{Appendix}{Appendix}
\Crefname{figure}{Fig.}{Fig.}
\crefname{figure}{Figure}{Fig.}
\Crefname{equation}{Eq.}{Eq.}
\crefname{equation}{Eq.}{Eq.}
\crefformat{footnote}{#2\footnotemark[#1]#3} % Referencing footnotes
\usepackage{cleveref}

\newcommand{\eg}{\textsl{e.g.},\xspace}

\newcommand{\etc}{etc.\xspace}

\titleformat{\subsubsection}[runin]{\bfseries}{}{0pt}{}[.]
\titlespacing{\subsubsection}{0pt}{*1}{*1}
\titleformat{\paragraph}[runin]{\itshape}{}{0pt}{}[.]
\titlespacing{\paragraph}{0pt}{*0.5}{*0.5}

\def\Lightpass/{\textit{Fastroute}}
\def\Ideal/{\textit{Ideal}}
\def\Baseline/{\textit{State-of-the-Art}}
\def\Static/{\textit{Oversubscribed}}
\def\LightpassPlot/{Fastroute}
\def\IdealPlot/{Ideal}
\def\BaselinePlot/{State-of-the-Art}
\def\StaticPlot/{Oversubscribed}
\def\ASIC/{chiplet}

\definecolor{uchu-blue-1}{HTML}{ccdefc}
\definecolor{uchu-blue-2}{HTML}{9bc0f9}
\definecolor{uchu-blue-3}{HTML}{6aa2f5}
\definecolor{uchu-blue-4}{HTML}{3984f2}
\definecolor{uchu-blue-5}{HTML}{0965ef}
\definecolor{uchu-blue-6}{HTML}{085cd8}
\definecolor{uchu-blue-7}{HTML}{0853c1}
\definecolor{uchu-blue-8}{HTML}{0949ac}
\definecolor{uchu-blue-9}{HTML}{084095}
\definecolor{uchu-general-yang}{HTML}{fdfdfd}
\definecolor{uchu-general-yin}{HTML}{080a0d}
\definecolor{uchu-gray-1}{HTML}{f0f0f2}
\definecolor{uchu-gray-2}{HTML}{e3e5e5}
\definecolor{uchu-gray-3}{HTML}{d8d8da}
\definecolor{uchu-gray-4}{HTML}{cbcdcd}
\definecolor{uchu-gray-5}{HTML}{bfc0c1}
\definecolor{uchu-gray-6}{HTML}{adaeaf}
\definecolor{uchu-gray-7}{HTML}{9b9b9d}
\definecolor{uchu-gray-8}{HTML}{878a8b}
\definecolor{uchu-gray-9}{HTML}{757779}
\definecolor{uchu-green-1}{HTML}{d5f5d9}
\definecolor{uchu-green-2}{HTML}{afecb6}
\definecolor{uchu-green-3}{HTML}{8ae293}
\definecolor{uchu-green-4}{HTML}{64d970}
\definecolor{uchu-green-5}{HTML}{3fcf4e}
\definecolor{uchu-green-6}{HTML}{39bc47}
\definecolor{uchu-green-7}{HTML}{34a741}
\definecolor{uchu-green-8}{HTML}{2e943a}
\definecolor{uchu-green-9}{HTML}{297f34}
\definecolor{uchu-pink-1}{HTML}{ffebf2}
\definecolor{uchu-pink-2}{HTML}{ffd9e8}
\definecolor{uchu-pink-3}{HTML}{ffc9dd}
\definecolor{uchu-pink-4}{HTML}{ffb7d3}
\definecolor{uchu-pink-5}{HTML}{ffa6c8}
\definecolor{uchu-pink-6}{HTML}{e697b5}
\definecolor{uchu-pink-7}{HTML}{cd87a2}
\definecolor{uchu-pink-8}{HTML}{b57790}
\definecolor{uchu-pink-9}{HTML}{9c677d}
\definecolor{uchu-purple-1}{HTML}{e2d4f4}
\definecolor{uchu-purple-2}{HTML}{c7abe9}
\definecolor{uchu-purple-3}{HTML}{ac83de}
\definecolor{uchu-purple-4}{HTML}{915ad3}
\definecolor{uchu-purple-5}{HTML}{7532c8}
\definecolor{uchu-purple-6}{HTML}{6a2eb5}
\definecolor{uchu-purple-7}{HTML}{5f2aa2}
\definecolor{uchu-purple-8}{HTML}{542690}
\definecolor{uchu-purple-9}{HTML}{49227d}
\definecolor{uchu-orange-1}{HTML}{ffe5d3}
\definecolor{uchu-orange-2}{HTML}{ffcdab}
\definecolor{uchu-orange-3}{HTML}{ffb783}
\definecolor{uchu-orange-4}{HTML}{ff9f5b}
\definecolor{uchu-orange-5}{HTML}{ff8834}
\definecolor{uchu-orange-6}{HTML}{e67c2f}
\definecolor{uchu-orange-7}{HTML}{cd6f2c}
\definecolor{uchu-orange-8}{HTML}{b56227}
\definecolor{uchu-orange-9}{HTML}{9c5524}
\definecolor{uchu-red-1}{HTML}{facdd7}
\definecolor{uchu-red-2}{HTML}{f59cb1}
\definecolor{uchu-red-3}{HTML}{ef6d8b}
\definecolor{uchu-red-4}{HTML}{ea3c65}
\definecolor{uchu-red-5}{HTML}{e50e3f}
\definecolor{uchu-red-6}{HTML}{cf0c3a}
\definecolor{uchu-red-7}{HTML}{b80c35}
\definecolor{uchu-red-8}{HTML}{a30d30}
\definecolor{uchu-red-9}{HTML}{8c0c2b}
\definecolor{uchu-yellow-1}{HTML}{fff5d8}
\definecolor{uchu-yellow-2}{HTML}{ffeeb9}
\definecolor{uchu-yellow-3}{HTML}{fee69a}
\definecolor{uchu-yellow-4}{HTML}{fedf7b}
\definecolor{uchu-yellow-5}{HTML}{fed75c}
\definecolor{uchu-yellow-6}{HTML}{e5c255}
\definecolor{uchu-yellow-7}{HTML}{ccae4b}
\definecolor{uchu-yellow-8}{HTML}{b59944}
\definecolor{uchu-yellow-9}{HTML}{9c853c}
\definecolor{uchu-yin-1}{HTML}{e3e4e6}
\definecolor{uchu-yin-2}{HTML}{cccccf}
\definecolor{uchu-yin-3}{HTML}{b2b4b6}
\definecolor{uchu-yin-4}{HTML}{9a9c9e}
\definecolor{uchu-yin-5}{HTML}{828386}
\definecolor{uchu-yin-6}{HTML}{6a6b6e}
\definecolor{uchu-yin-7}{HTML}{515255}
\definecolor{uchu-yin-8}{HTML}{383b3d}
\definecolor{uchu-yin-9}{HTML}{202225}

\definecolor{slate50}{HTML}{f8fafc}
\definecolor{slate100}{HTML}{f1f5f9}
\definecolor{slate200}{HTML}{e2e8f0}
\definecolor{slate300}{HTML}{cbd5e1}
\definecolor{slate400}{HTML}{94a3b8}
\definecolor{slate500}{HTML}{64748b}
\definecolor{slate600}{HTML}{475569}
\definecolor{slate700}{HTML}{334155}
\definecolor{slate800}{HTML}{1e293b}
\definecolor{slate900}{HTML}{0f172a}
\definecolor{slate950}{HTML}{020617}

\definecolor{gray50}{HTML}{f9fafb}
\definecolor{gray100}{HTML}{f3f4f6}
\definecolor{gray200}{HTML}{e5e7eb}
\definecolor{gray300}{HTML}{d1d5db}
\definecolor{gray400}{HTML}{9ca3af}
\definecolor{gray500}{HTML}{6b7280}
\definecolor{gray600}{HTML}{4b5563}
\definecolor{gray700}{HTML}{374151}
\definecolor{gray800}{HTML}{1f2937}
\definecolor{gray900}{HTML}{111827}
\definecolor{gray950}{HTML}{030712}

\definecolor{zinc50}{HTML}{fafafa}
\definecolor{zinc100}{HTML}{f4f4f5}
\definecolor{zinc200}{HTML}{e4e4e7}
\definecolor{zinc300}{HTML}{d4d4d8}
\definecolor{zinc400}{HTML}{a1a1aa}
\definecolor{zinc500}{HTML}{71717a}
\definecolor{zinc600}{HTML}{52525b}
\definecolor{zinc700}{HTML}{3f3f46}
\definecolor{zinc800}{HTML}{27272a}
\definecolor{zinc900}{HTML}{18181b}
\definecolor{zinc950}{HTML}{09090b}

\definecolor{neutral50}{HTML}{fafafa}
\definecolor{neutral100}{HTML}{f5f5f5}
\definecolor{neutral200}{HTML}{e5e5e5}
\definecolor{neutral300}{HTML}{d4d4d4}
\definecolor{neutral400}{HTML}{a3a3a3}
\definecolor{neutral500}{HTML}{737373}
\definecolor{neutral600}{HTML}{525252}
\definecolor{neutral700}{HTML}{404040}
\definecolor{neutral800}{HTML}{262626}
\definecolor{neutral900}{HTML}{171717}
\definecolor{neutral950}{HTML}{0a0a0a}

\definecolor{stone50}{HTML}{fafaf9}
\definecolor{stone100}{HTML}{f5f5f4}
\definecolor{stone200}{HTML}{e7e5e4}
\definecolor{stone300}{HTML}{d6d3d1}
\definecolor{stone400}{HTML}{a8a29e}
\definecolor{stone500}{HTML}{78716c}
\definecolor{stone600}{HTML}{57534e}
\definecolor{stone700}{HTML}{44403c}
\definecolor{stone800}{HTML}{292524}
\definecolor{stone900}{HTML}{1c1917}
\definecolor{stone950}{HTML}{0c0a09}

\definecolor{red50}{HTML}{fef2f2}
\definecolor{red100}{HTML}{fee2e2}
\definecolor{red200}{HTML}{fecaca}
\definecolor{red300}{HTML}{fca5a5}
\definecolor{red400}{HTML}{f87171}
\definecolor{red500}{HTML}{ef4444}
\definecolor{red600}{HTML}{dc2626}
\definecolor{red700}{HTML}{b91c1c}
\definecolor{red800}{HTML}{991b1b}
\definecolor{red900}{HTML}{7f1d1d}
\definecolor{red950}{HTML}{450a0a}

\definecolor{orange50}{HTML}{fff7ed}
\definecolor{orange100}{HTML}{ffedd5}
\definecolor{orange200}{HTML}{fed7aa}
\definecolor{orange300}{HTML}{fdba74}
\definecolor{orange400}{HTML}{fb923c}
\definecolor{orange500}{HTML}{f97316}
\definecolor{orange600}{HTML}{ea580c}
\definecolor{orange700}{HTML}{c2410c}
\definecolor{orange800}{HTML}{9a3412}
\definecolor{orange900}{HTML}{7c2d12}
\definecolor{orange950}{HTML}{431407}

\definecolor{amber50}{HTML}{fffbeb}
\definecolor{amber100}{HTML}{fef3c7}
\definecolor{amber200}{HTML}{fde68a}
\definecolor{amber300}{HTML}{fcd34d}
\definecolor{amber400}{HTML}{fbbf24}
\definecolor{amber500}{HTML}{f59e0b}
\definecolor{amber600}{HTML}{d97706}
\definecolor{amber700}{HTML}{b45309}
\definecolor{amber800}{HTML}{92400e}
\definecolor{amber900}{HTML}{78350f}
\definecolor{amber950}{HTML}{451a03}

\definecolor{yellow50}{HTML}{fefce8}
\definecolor{yellow100}{HTML}{fef9c3}
\definecolor{yellow200}{HTML}{fef08a}
\definecolor{yellow300}{HTML}{fde047}
\definecolor{yellow400}{HTML}{facc15}
\definecolor{yellow500}{HTML}{eab308}
\definecolor{yellow600}{HTML}{ca8a04}
\definecolor{yellow700}{HTML}{a16207}
\definecolor{yellow800}{HTML}{854d0e}
\definecolor{yellow900}{HTML}{713f12}
\definecolor{yellow950}{HTML}{422006}

\definecolor{lime50}{HTML}{f7fee7}
\definecolor{lime100}{HTML}{ecfccb}
\definecolor{lime200}{HTML}{d9f99d}
\definecolor{lime300}{HTML}{bef264}
\definecolor{lime400}{HTML}{a3e635}
\definecolor{lime500}{HTML}{84cc16}
\definecolor{lime600}{HTML}{65a30d}
\definecolor{lime700}{HTML}{4d7c0f}
\definecolor{lime800}{HTML}{3f6212}
\definecolor{lime900}{HTML}{365314}
\definecolor{lime950}{HTML}{1a2e05}

\definecolor{green50}{HTML}{f0fdf4}
\definecolor{green100}{HTML}{dcfce7}
\definecolor{green200}{HTML}{bbf7d0}
\definecolor{green300}{HTML}{86efac}
\definecolor{green400}{HTML}{4ade80}
\definecolor{green500}{HTML}{22c55e}
\definecolor{green600}{HTML}{16a34a}
\definecolor{green700}{HTML}{15803d}
\definecolor{green800}{HTML}{166534}
\definecolor{green900}{HTML}{14532d}
\definecolor{green950}{HTML}{052e16}

\definecolor{emerald50}{HTML}{ecfdf5}
\definecolor{emerald100}{HTML}{d1fae5}
\definecolor{emerald200}{HTML}{a7f3d0}
\definecolor{emerald300}{HTML}{6ee7b7}
\definecolor{emerald400}{HTML}{34d399}
\definecolor{emerald500}{HTML}{10b981}
\definecolor{emerald600}{HTML}{059669}
\definecolor{emerald700}{HTML}{047857}
\definecolor{emerald800}{HTML}{065f46}
\definecolor{emerald900}{HTML}{064e3b}
\definecolor{emerald950}{HTML}{022c22}

\definecolor{teal50}{HTML}{f0fdfa}
\definecolor{teal100}{HTML}{ccfbf1}
\definecolor{teal200}{HTML}{99f6e4}
\definecolor{teal300}{HTML}{5eead4}
\definecolor{teal400}{HTML}{2dd4bf}
\definecolor{teal500}{HTML}{14b8a6}
\definecolor{teal600}{HTML}{0d9488}
\definecolor{teal700}{HTML}{0f766e}
\definecolor{teal800}{HTML}{115e59}
\definecolor{teal900}{HTML}{134e4a}
\definecolor{teal950}{HTML}{042f2e}

\definecolor{cyan50}{HTML}{ecfeff}
\definecolor{cyan100}{HTML}{cffafe}
\definecolor{cyan200}{HTML}{a5f3fc}
\definecolor{cyan300}{HTML}{67e8f9}
\definecolor{cyan400}{HTML}{22d3ee}
\definecolor{cyan500}{HTML}{06b6d4}
\definecolor{cyan600}{HTML}{0891b2}
\definecolor{cyan700}{HTML}{0e7490}
\definecolor{cyan800}{HTML}{155e75}
\definecolor{cyan900}{HTML}{164e63}
\definecolor{cyan950}{HTML}{083344}

\definecolor{sky50}{HTML}{f0f9ff}
\definecolor{sky100}{HTML}{e0f2fe}
\definecolor{sky200}{HTML}{bae6fd}
\definecolor{sky300}{HTML}{7dd3fc}
\definecolor{sky400}{HTML}{38bdf8}
\definecolor{sky500}{HTML}{0ea5e9}
\definecolor{sky600}{HTML}{0284c7}
\definecolor{sky700}{HTML}{0369a1}
\definecolor{sky800}{HTML}{075985}
\definecolor{sky900}{HTML}{0c4a6e}
\definecolor{sky950}{HTML}{082f49}

\definecolor{blue50}{HTML}{eff6ff}
\definecolor{blue100}{HTML}{dbeafe}
\definecolor{blue200}{HTML}{bfdbfe}
\definecolor{blue300}{HTML}{93c5fd}
\definecolor{blue400}{HTML}{60a5fa}
\definecolor{blue500}{HTML}{3b82f6}
\definecolor{blue600}{HTML}{2563eb}
\definecolor{blue700}{HTML}{1d4ed8}
\definecolor{blue800}{HTML}{1e40af}
\definecolor{blue900}{HTML}{1e3a8a}
\definecolor{blue950}{HTML}{172554}

\definecolor{indigo50}{HTML}{eef2ff}
\definecolor{indigo100}{HTML}{e0e7ff}
\definecolor{indigo200}{HTML}{c7d2fe}
\definecolor{indigo300}{HTML}{a5b4fc}
\definecolor{indigo400}{HTML}{818cf8}
\definecolor{indigo500}{HTML}{6366f1}
\definecolor{indigo600}{HTML}{4f46e5}
\definecolor{indigo700}{HTML}{4338ca}
\definecolor{indigo800}{HTML}{3730a3}
\definecolor{indigo900}{HTML}{312e81}
\definecolor{indigo950}{HTML}{1e1b4b}

\definecolor{violet50}{HTML}{f5f3ff}
\definecolor{violet100}{HTML}{ede9fe}
\definecolor{violet200}{HTML}{ddd6fe}
\definecolor{violet300}{HTML}{c4b5fd}
\definecolor{violet400}{HTML}{a78bfa}
\definecolor{violet500}{HTML}{8b5cf6}
\definecolor{violet600}{HTML}{7c3aed}
\definecolor{violet700}{HTML}{6d28d9}
\definecolor{violet800}{HTML}{5b21b6}
\definecolor{violet900}{HTML}{4c1d95}
\definecolor{violet950}{HTML}{2e1065}

\definecolor{purple50}{HTML}{faf5ff}
\definecolor{purple100}{HTML}{f3e8ff}
\definecolor{purple200}{HTML}{e9d5ff}
\definecolor{purple300}{HTML}{d8b4fe}
\definecolor{purple400}{HTML}{c084fc}
\definecolor{purple500}{HTML}{a855f7}
\definecolor{purple600}{HTML}{9333ea}
\definecolor{purple700}{HTML}{7e22ce}
\definecolor{purple800}{HTML}{6b21a8}
\definecolor{purple900}{HTML}{581c87}
\definecolor{purple950}{HTML}{3b0764}

\definecolor{fuchsia50}{HTML}{fdf4ff}
\definecolor{fuchsia100}{HTML}{fae8ff}
\definecolor{fuchsia200}{HTML}{f5d0fe}
\definecolor{fuchsia300}{HTML}{f0abfc}
\definecolor{fuchsia400}{HTML}{e879f9}
\definecolor{fuchsia500}{HTML}{d946ef}
\definecolor{fuchsia600}{HTML}{c026d3}
\definecolor{fuchsia700}{HTML}{a21caf}
\definecolor{fuchsia800}{HTML}{86198f}
\definecolor{fuchsia900}{HTML}{701a75}
\definecolor{fuchsia950}{HTML}{4a044e}

\definecolor{pink50}{HTML}{fdf2f8}
\definecolor{pink100}{HTML}{fce7f3}
\definecolor{pink200}{HTML}{fbcfe8}
\definecolor{pink300}{HTML}{f9a8d4}
\definecolor{pink400}{HTML}{f472b6}
\definecolor{pink500}{HTML}{ec4899}
\definecolor{pink600}{HTML}{db2777}
\definecolor{pink700}{HTML}{be185d}
\definecolor{pink800}{HTML}{9d174d}
\definecolor{pink900}{HTML}{831843}
\definecolor{pink950}{HTML}{500724}

\definecolor{rose50}{HTML}{fff1f2}
\definecolor{rose100}{HTML}{ffe4e6}
\definecolor{rose200}{HTML}{fecdd3}
\definecolor{rose300}{HTML}{fda4af}
\definecolor{rose400}{HTML}{fb7185}
\definecolor{rose500}{HTML}{f43f5e}
\definecolor{rose600}{HTML}{e11d48}
\definecolor{rose700}{HTML}{be123c}
\definecolor{rose800}{HTML}{9f1239}
\definecolor{rose900}{HTML}{881337}
\definecolor{rose950}{HTML}{4c0519}

\newcommand{\circuitswitch}[1][1]{%
  \begin{tikzpicture}[scale=#1, baseline=-0.5ex]
    \draw[line width=0.2pt*#1, fill=white, rounded corners=1pt*#1] (-1,-1) rectangle (1,1);
    
    \tikzset{
      switch arrow/.style={
        line width=0.3pt*#1,
        {Stealth[length=1pt*#1, width=1pt*#1,inset=0pt]}-{Stealth[length=1pt*#1, width=1pt*#1,inset=0pt]},
      }
    }
    
    \draw[switch arrow] (-0.85, 0.5) -- (-0.35, 0.5) -- (0.35, -0.5) -- (0.85, -0.5);
    
    \draw[switch arrow] (-0.85, -0.5) -- (-0.35, -0.5) -- (0.35, 0.5) -- (0.85, 0.5);
  \end{tikzpicture}%
}

\newcommand{\packetswitch}[1][1]{%
    \begin{tikzpicture}[scale=#1, line width=0.4pt*#1, line cap=round, line join=round]
            \draw[rounded corners=0.5pt] (-1,-1) rectangle (1,1);
            
            \fill[rounded corners=0.25pt] (-0.45,-0.45) rectangle (0.45,0.45);
            
            \fill (-0.75,0.75) -- (-0.35,0.75) -- (-0.75,0.35) -- cycle;
        
            \draw (-0.5,1)  -- (-0.5,1.3)  (0,1)  -- (0,1.3)  (0.5,1)  -- (0.5,1.3);
            \draw (-0.5,-1) -- (-0.5,-1.3) (0,-1) -- (0,-1.3) (0.5,-1) -- (0.5,-1.3);
            \draw (-1,0.5)  -- (-1.3,0.5)  (-1,0) -- (-1.3,0) (-1,-0.5) -- (-1.3,-0.5);
            \draw (1,0.5)   -- (1.3,0.5)   (1,0)  -- (1.3,0)  (1,-0.5)  -- (1.3,-0.5);
    \end{tikzpicture}
}

\def\cornerrad{1mm}
\def\archcol{sky300}
\def\archline{sky900}
\def\asiccol{blue400}
\def\asicline{blue900}
\def\control{green300}
\def\controlline{green900}
\def\controlaccent{green900}

\newcommand{\susdas}[1]{{\textcolor{red}{\textbf{[Sushovan: #1]}}}} % comments Sushovan
\newcommand{\lukas}[1]{{\textcolor{blue}{\textbf{[Lukas: #1]}}}} % comments Lukas
\newcommand{\gf}[1]{\textcolor{orange}{\textbf{G:} #1}}

\renewcommand{\susdas}[1]{} % disable comments Sushovan
\renewcommand{\lukas}[1]{} % disable comments Lukas
\renewcommand{\gf}[1]{} % disable comments Georgia

\crefname{lstlisting}{listing}{listings}
\Crefname{lstlisting}{Listing}{Listings}
\usepackage{listings}
\begin{document}
%-------------------------------------------------------------------------------

%don't want date printed
\date{}

%\title[\Lightpass/]{Enabling Next-Generation Multi-ASIC Switches with Flexible Indirection}
%\title[\Lightpass/]{Flexible Indirection: A Scalable Architecture for Next-Generation Multi-ASIC Switches.}
%\title[]{Taking Network Switches to the Beyond Moore Era}
%\title[]{Lightpass: Scalability Beyond the Silicon Boundary}
%\title[]{Gone with the Overhead: A New Era for multi-ASIC Switches}
%\title[]{Beyond the Die:\\ Scaling Network Silicon with Multi-ASIC Integration}
%\title[]{Thinking Outside the Chip:\\ A New Paradigm for Multi-ASIC Network Silicon}
%\title{\Large \bf Enabling Next-Generation Switching with Chiplets: The Power of Indirection}
%\title{\Large \bf Lightpass: Packet or Circuit switching, why not both?}
%\title[]{Flexible Indirection: Scaling Network Switches Beyond Silicon Limits}
%\title[]{Unlocking the Potential of Multi-ASIC Switching: Scaling Beyond Silicon Limits}

%\title[]{The Power of Indirection:\\ Scaling Switches Beyond Silicon Boundaries}
%\title[]{The Case for Port Indirection:\\ Scaling Switches Beyond Silicon Boundaries}
\title{\Large \bf The Power of Indirection: Scaling Switches Beyond Silicon Boundaries}

%\title{\Large \bf Indirection Is All You Need: Scaling Switches Beyond Silicon Boundaries}
%\title{\Large \bf Beyond the Die: Scaling Switches Beyond Silicon Boundaries}
%\title{\Large Unlocking the Potential of Multi-ASIC Switches: Scaling Beyond Silicon Limits}
%\title{\Large Forwarding Beyond the Silicon Boundary:\\ Efficient Multi-ASIC Switches with Dynamic Port Indirection}

\iftrue
\author{
{\rm Lukas Röllin}\\
ETH Zürich\\
\and
{\rm Sushovan Das}\\
ETH Zürich
\and
{\rm Paolo Costa}\\
Microsoft Research
\and
{\rm Laurent Vanbever}\\
ETH Zürich
% copy the following lines to add more authors
% \and
% {\rm Name}\\
%Name Institution
} % end author
\else
\author{\rm Paper \# 340, 12 pages}
%\titlenote{Produces the permission block, and copyright information}
%\subtitle{Extended Abstract}
%\author{Paper \#, 12 pages}
% \author{Firstname Lastname}
% \authornote{Note}
% \affiliation{%
%   \institution{Affiliation}
%   \streetaddress{Address}
%   \city{City} 
%   \state{State} 
%   \postcode{Zipcode}
% }
% \email{email@domain.com}
\fi

\maketitle

%-------------------------------------------------------------------------------
\begin{abstract}

% Context
The slowdown of Moore’s law and the area limit of monolithic integration have made chiplet-based designs inevitable across many domains, including network ASICs.
% Problem
However, combining multiple network ASICs together poses a fundamental challenge: maintaining sufficient inter-ASIC bandwidth to match the performance of an idealistic single-ASIC design. 
Providing full bandwidth is prohibitively expensive as it requires valuable forwarding capacity, while reducing inter-ASIC bandwidth creates severe performance bottlenecks.

% Task
We propose a novel multi-ASIC switch architecture that introduces a circuit-switched indirection layer in front of the ASICs. 
This layer flexibly remaps ingress ports across ASICs, localizing traffic and minimizing inter-ASIC communication based on observed patterns.
% Findings
Our system, \Lightpass/, combines packet and circuit switching to deliver performance comparable to a single-ASIC switch while reducing inter-ASIC bandwidth requirements.
This frees up capacity for external network interfaces, allowing \Lightpass/ to outperform traditional non-oversubscribed multi-ASIC designs.
Our hardware prototype demonstrates the system's functional feasibility by evaluating it on an LLM training workload.

% Outlook
By reducing bandwidth and power overhead, \Lightpass/ bridges the gap between silicon fabrication limits and soaring application demands. 
It provides an efficient transition to multi-ASIC switches, enabling bandwidth and radix demand to be met without waiting for the next ASIC generation.
\end{abstract}

\section{Introduction}
\label{introduction}

Network ASICs, like other high-performance computing chips, have been increasingly constrained by the slowdown of Moore’s law, approaching the limits of monolithic integration~\cite{michael2025Tomahawk, moore2019Another}. 
While historically higher switch throughput was achieved by moving to a smaller technology node and scaling die size, both avenues are slowly (but surely) coming to an end. In particular, the end of Dennard scaling has caused transistor performance scaling to saturate~\cite{bohr2011evolutiona, markov2014Limits}, and pushed the die size to reach the ``reticle limit''---the maximum size of a single chip that can be fabricated~\cite{bailey2020Designs, esmaeilzadeh2011Dark}. 
Because of this, the pace at which switch throughput doubles has been slowing down. 
As an illustration, Broadcom released a new Tomahawk series every two years up to Tomahawk 4~\cite{throughput2019Broadcom}, while the following two generations each took three years~\cite{broadcom2026News}.

And yet, hyperscale data centers and emerging workloads, such as AI training and inference or high-performance computing (HPC), are driving unprecedented demand.
Not just for higher network bandwidth but also pushing the limits of scale, connecting more servers than ever~\cite{alizadeh2010Data, abadi2016TensorFlow, narayanan2021Efficient}.

\begin{figure}
    \centering
    \begin{tikzpicture}[x=0.1cm,y=0.1cm,scale=1]
    \tikzset{thick/.style = {line width=0.3mm}}
    \def\asiclen{1.2}
    \def\serdeslen{0.2}
    \def\serdesoffset{8.5}
    \def\xone{27}
    \def\xtwo{56}

    \node [draw=\archline, rounded corners=\cornerrad, fill=\archcol, rectangle, minimum width=\asiclen*1.5 cm, minimum height=\asiclen*1.5 cm] at (0,0) {};
    \node [rounded corners=\cornerrad, rectangle, fill=\asiccol,line width=0.3mm, minimum width=\asiclen cm, minimum height=\asiclen cm] at (0,0) {\footnotesize ASIC};
    \node [rectangle, inner sep=0cm, minimum height=0.4 cm] at (0,-7.5) {\footnotesize SerDes};
    \node [rectangle, inner sep=0cm, minimum height=0.4 cm] at (0,7.5) {\footnotesize SerDes};
    
    \node [] at (0,-14) {\footnotesize Single-ASIC};
    
    \node [draw=\archline, rounded corners=\cornerrad, inner sep=0.8mm, fill=\archcol, rectangle, minimum width=\asiclen cm, minimum height=\serdeslen cm] at (\xone,\serdesoffset) {\footnotesize SerDes};
    \node [draw=\archline, fill=\archcol, rounded corners=\cornerrad, inner sep=0.8mm, rectangle, minimum width=\asiclen cm, minimum height=\serdeslen cm] at (\xone,-\serdesoffset) {\footnotesize SerDes};
    \node [draw=\archline, fill=\archcol,rounded corners=\cornerrad, inner sep=0.8mm, rectangle, rotate=90,  minimum width=\asiclen cm, minimum height=\serdeslen cm] at (\xone-\serdesoffset,0) {\footnotesize SerDes};
    \node [draw=\archline, fill=\archcol, rounded corners=\cornerrad, inner sep=0.8mm, rectangle, rotate=90, minimum width=\asiclen cm, minimum height=\serdeslen cm] at (\xone+\serdesoffset,0) {\footnotesize SerDes};
    
    \node [draw=\asicline, fill=\asiccol, rounded corners=\cornerrad, rectangle, minimum width=\asiclen cm, minimum height=\asiclen cm] at (\xone,0) {\footnotesize ASIC};
    
    \node [] at (\xone,-14) {\footnotesize Tomahawk 6};
   
    \node [draw=\archline, fill=\archcol, rounded corners=\cornerrad, inner sep=0.8mm, rectangle, minimum width=\asiclen cm, minimum height=\serdeslen cm] (SerDes1) at (\xtwo,\serdesoffset) {\footnotesize SerDes};
    \node [draw=\archline, fill=\archcol, rounded corners=\cornerrad, inner sep=0.8mm, rectangle, minimum width=\asiclen cm, minimum height=\serdeslen cm] (SerDes2) at (\xtwo,-\serdesoffset) {\footnotesize SerDes};
    \node [draw=\archline, fill=\archcol, rounded corners=\cornerrad, inner sep=0.8mm, rectangle, rotate=90,  minimum width=\asiclen cm, minimum height=\serdeslen cm] (SerDes3) at (\xtwo-\serdesoffset,0) {\footnotesize SerDes};
    \node [draw=\archline, fill=\archcol,rounded corners=\cornerrad, inner sep=0.8mm, rectangle, rotate=90, minimum width=\asiclen cm, minimum height=\serdeslen cm] (SerDes4) at (\xtwo+\serdesoffset,0) {\footnotesize SerDes};
    
    \node [draw=\asicline, fill=\asiccol,rounded corners=\cornerrad, rectangle, minimum width=0.55 cm, minimum height=0.55cm] (ASIC1) at (52.75,3.125) {\footnotesize A1};
    \node [draw=\asicline, fill=\asiccol,rounded corners=\cornerrad, rectangle, minimum width=0.55 cm, minimum height=0.55cm] (ASIC2) at (59.25,3.125) {\footnotesize A2};
    \node [draw=\asicline, fill=\asiccol,rounded corners=\cornerrad, rectangle, minimum width=0.55 cm, minimum height=0.55cm] (ASIC1) at (52.75,-3.125) {\footnotesize A3};
    \node [draw=\asicline, fill=\asiccol,rounded corners=\cornerrad, rectangle, minimum width=0.55 cm, minimum height=0.55cm] (ASIC1) at (59.25,-3.125) {\footnotesize A4};

    \node [] at (\xtwo,-14) {\footnotesize Multi-ASIC Design};
    
    %\foreach \x in {-6,-5,-4,-3,-2,-1,0,1,2,3,4,5,6}{
        %\draw ([xshift=\x mm] SerDes1.south) -- ([xshift=\x mm] ASIC1.north);
        %\draw ([xshift=\x mm] SerDes2.north) -- ([xshift=\x mm] ASIC2.south);
        %\draw ([xshift=\x mm] ASIC1.south) -- ([xshift=\x mm] ASIC2.north);
    %}

    \path (current bounding box.south west) +(0,-1) (current bounding box.north east) +(0,0);

\end{tikzpicture}%
    \caption{To keep up with bandwidth demands and to avoid silicon limitations, switches need to be split up.}
    \label{fig:asictrend}
\end{figure}
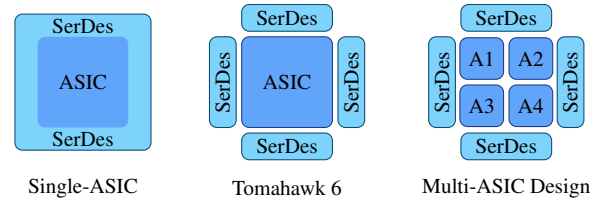

\begin{figure}
    \centering
    \pgfdeclarelayer{background}
\pgfsetlayers{background,main}

\tikzset{
  sota/.style={rectangle, fill=myviolet, inner sep=0.6mm},
  lightpass/.style={rectangle, fill=myred, inner sep=0.6mm},
  ic chip/.pic={
    % 'pic actions' ensures passing options like [mylightblue] colors the chip elements dynamically
    \begin{scope}[pic actions, line width=0.4pt, line cap=round, line join=round, x=0.5mm, y=0.5mm]
        % Main rounded body
        \draw[rounded corners=0.5pt] (-2,-2) rectangle (2,2);
        
        % Center solid square
        \fill[rounded corners=0.25pt] (-0.9,-0.9) rectangle (0.9,0.9);
        
        % Orientation Marker
        \fill (-1.5,1.5) -- (-0.7,1.5) -- (-1.5,0.7) -- cycle;
    
        % Pins
        \draw (-1,2)  -- (-1,2.6)  (0,2)  -- (0,2.6)  (1,2)  -- (1,2.6);
        \draw (-1,-2) -- (-1,-2.6) (0,-2) -- (0,-2.6) (1,-2) -- (1,-2.6);
        \draw (-2,1)  -- (-2.6,1)  (-2,0) -- (-2.6,0) (-2,-1) -- (-2.6,-1);
        \draw (2,1)   -- (2.6,1)   (2,0)  -- (2.6,0)  (2,-1)  -- (2.6,-1);
    \end{scope}
  }
}

\begin{tikzpicture}[scale=1,yscale=0.75]
    \newcommand{\yoffset}{-0.25}
    \newcommand{\xoffset}{0.25}
    % --- Legend ---
    \node[lightpass, minimum width=2mm] at (4.5, 1.2) {};
    \node[lightpass, minimum width=3mm] at (4.55, 1.4) {};
    \node[anchor=west] at (4.8, 1.3) {\footnotesize \Lightpass/ \strut};
    
    \node[sota, minimum width=2mm] at (4.5, 0.6) {};
    \node[sota, minimum width=3mm] at (4.55, 0.8) {};
    \node[anchor=west] at (4.8, 0.7) {\footnotesize \Baseline/ \strut};

    %\node[draw, minimum height=10mm, minimum width=25mm] at (5.5,2) {};

    % --- X-Axis (BW Increase) ---
    \draw[->, thick] (0,0) -- (6.8,0) node[right] {};
    \node[] at (3.4,-0.8) {\footnotesize Bandwidth increase over Generation 1};
    \foreach \x/\label in {0/+0\%, 2/+100\%, 4/+200\%, 6/+300\%} {
        \draw (\x+\xoffset, 2pt) -- (\x+\xoffset, -2pt) node[below=-0.3mm, font=\scriptsize] {\label};
    }

    % --- Y-Axis (Generations with Axis Break) ---
    \draw[thick, mylightblue] (0,0) -- (0,1.95+\yoffset);
    \draw[thick, mylightgreen] (0,2.05+\yoffset) -- (0,2.5+\yoffset);
    \draw[thick, mylightgreen, ->] (0,2.5+\yoffset) -- (0,4+\yoffset);
    
    % Timeline Break Symbol
    \draw[thick] (-0.15, 2+\yoffset) -- (0.15, 2.1+\yoffset);
    \draw[thick] (-0.15, 1.9+\yoffset) -- (0.15, 2+\yoffset);
    \node[right, font=\scriptsize] at (0.1, 2+\yoffset) {\textasciitilde 3 years};

    % --- Generation 1 Ticks & Labels ---
    \foreach \y/\mult in {0.5/1\times, 1/2\times, 1.5/3\times} {
        \draw (-2pt, \y+\yoffset) -- (2pt, \y+\yoffset);
        \node at (-0.6, \y+\yoffset) {\scriptsize $\mult$};
        \pic[mylightblue] at (-0.25, \y+\yoffset) {ic chip};
    }
    \node[text=mylightblue, align=center, font=\scriptsize, rotate=90] at (-1.25, 1+\yoffset) {Generation 1 \\ \tiny (\eg Tomahawk 6)};
    %\node[text=mylightblue, align=center, font=\scriptsize] at (5.8, 1+\yoffset) {(\eg Tomahawk 6)};

    % --- Generation 2 Ticks & Labels ---
    \foreach \y/\mult in {2.5/1\times, 3/2\times, 3.5/3\times} {
        \draw (-2pt, \y+\yoffset) -- (2pt, \y+\yoffset);
        \node at (-0.6, \y+\yoffset) {\scriptsize $\mult$};
        \pic[mylightgreen] at (-0.25, \y+\yoffset) {ic chip};
    }
    \node[text=mylightgreen, align=center, font=\scriptsize, rotate=90] at (-1.25, 3+\yoffset) {Generation 2 \\ \tiny (\eg Tomahawk 7)}; %\node[text=mylightgreen, align=center, font=\scriptsize] at (5.8, 3+\yoffset) {(\eg Tomahawk 7)};

    \begin{pgfonlayer}{background}
    % Gen 1 - 1x
    %\node[sota] at (0*2+\xoffset, 0.5+\yoffset-0.08) {};
    \draw[myviolet, line width=1.4mm] (0, 0.5+\yoffset-0.08) -- (0*2+\xoffset, 0.5+\yoffset-0.08);

    % Gen 1 - 2x
    %\node[sota] at (0*2+\xoffset, 1+\yoffset-0.08) {};
    \draw[myviolet, line width=1.4mm] (0, 1+\yoffset-0.08) -- (0*2+\xoffset, 1+\yoffset-0.08);
    
    %\node[lightpass] at (0.33*2+\xoffset, 1+\yoffset+0.08) {}; 
    \draw[myred, line width=1.4mm] (0, 1+\yoffset+0.08) -- (0.33*2+\xoffset, 1+\yoffset+0.08);
    \node[right, font=\scriptsize, text=myred] at (0.33*2+\xoffset, 1+\yoffset) {+33\%};

    % Gen 1 - 3x
    %\node[sota] at (0.5*2+\xoffset, 1.5+\yoffset-0.08) {};
    \draw[rectangle, myviolet, line width=1.4mm] (0, 1.5+\yoffset-0.08) -- (0.5*2+\xoffset, 1.5+\yoffset-0.08);
    
    %\node[lightpass] at (0.8*2+\xoffset, 1.5+\yoffset+0.08) {};
    \draw[myred, line width=1.4mm] (0, 1.5+\yoffset+0.08) -- (0.8*2+\xoffset, 1.5+\yoffset+0.08);
    \node[right, font=\scriptsize, text=myred] at (0.8*2+\xoffset, 1.5+\yoffset) {+30\%};

    % Gen 2 - 1x
    %\node[sota] at (1*2+\xoffset, 3+\yoffset-0.08) {};
    \draw[myviolet, line width=1.4mm] (0, 2.5+\yoffset-0.08) -- (1*2+\xoffset, 2.5+\yoffset-0.08);

    % Gen 2 - 2x
    %\node[sota] at (1*2+\xoffset, 3.5+\yoffset-0.08) {}; 
    \draw[myviolet, line width=1.4mm] (0, 3+\yoffset-0.08) -- (1*2+\xoffset, 3+\yoffset-0.08);
    
    %\node[lightpass] at (1.66*2+\xoffset, 3.5+\yoffset+0.08) {};
    \draw[myred, line width=1.4mm] (0, 3+\yoffset+0.08) -- (1.66*2+\xoffset, 3+\yoffset+0.08);
    \node[right, font=\scriptsize, text=myred] at (1.66*2+\xoffset, 3+\yoffset) {+66\%};

    % Gen 2 - 3x
    %\node[sota] at (2*2+\xoffset, 4+\yoffset-0.08) {}; 
    \draw[myviolet, line width=1.4mm] (0, 3.5+\yoffset-0.08) -- (2*2+\xoffset, 3.5+\yoffset-0.08);
    
    %\node[lightpass] at (2.6*2+\xoffset, 4+\yoffset+0.08) {};
    \draw[myred, line width=1.4mm] (0, 3.5+\yoffset+0.08) -- (2.6*2+\xoffset, 3.5+\yoffset+0.08);
    \node[right, font=\scriptsize, text=myred] at (2.6*2+\xoffset, 3.5+\yoffset) {+60\%};
    \end{pgfonlayer}

    \path (current bounding box.south west) +(0,-0.1) (current bounding box.north east) +(0,0);
\end{tikzpicture}%
    \caption{\Lightpass/ increases achievable bandwidth between generations by enabling more efficient multi-ASIC designs.}
    \label{fig:generations}
\end{figure}
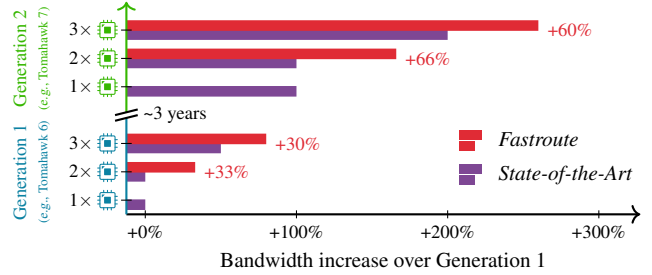

To keep up with the demand, device manufacturers have been increasingly moving towards ``chiplet-based'' architectures~\cite{borkar2007Thousand}. Such architectures spread the switch ingress ports across several ASICs that are then connected through an internal fabric. 
By combining multiple ASICs this way, manufacturers can build switches whose performance vastly exceeds that of single ASIC switches---both in capacity and number of ports (radix)---without waiting for the next generation (\Cref{fig:generations}).
Broadcom’s Tomahawk~6 (\Cref{fig:asictrend}) is an example of such an architecture in which the Serializers/Deserializers~(SerDes) have been turned into IO chiplets that sit next to the forwarding ASIC, freeing up valuable silicon area~\cite{michael2025Tomahawk}. 
Juniper’s Express~5~\cite{yeluri2023Chiplets} is another example of an architecture that goes one step further by splitting the forwarding ASIC itself (in addition to the SerDes), as shown in~\Cref{fig:asictrend}.

One of the key challenges in building multi-ASIC switches that offer higher throughput than single-ASIC designs is correspondingly scaling the inter-ASIC bandwidth.
Architects face a clear trade-off between non-blocking behavior and usable bandwidth.
%\lukas{add trade off line}
Providing non-blocking inter-ASIC bandwidth in a Clos-like internal topology (\Baseline/) requires scaling up fabric links, thereby diverting scarce ASIC forwarding capacity toward internal interconnects.
The upside is non-blocking behavior, but it comes at the cost of reduced usable bandwidth, especially at low ASIC counts ($2\mbox{ -- }5$), where creating proper Clos topologies is not always possible.
This is highlighted in \Cref{fig:generations}, which shows the total usable bandwidth achieved by combining multiple ASICs in the \Baseline/ way.
Oversubscribing inter-ASIC bandwidth can partly mitigate this, but at the cost of worst-case performance, as certain traffic patterns cause congestion at the inter-ASIC links. 
Our simulations show that even reasonable oversubscriptions can cause workloads to experience throughput reductions of up to $4\times$ and $9\times$ increases in tail flow completion time (FCT).

We propose a new approach that enables multi-ASIC switches with performance that closely matches that of an ideal single-ASIC \textit{even} when oversubscribing the inter-ASIC bandwidth.
Our key insight is to introduce a circuit-switched indirection layer that dynamically remaps switch ingress ports across ASICs based on observed traffic, thereby minimizing inter-ASIC traffic. 
It outperforms traditional non-blocking multi-ASIC designs~(\Baseline/) by using the available ASIC forwarding capacity more efficiently, extracting around $30\,\%$ more bandwidth and radix from the same ASICs (\Cref{fig:generations}).
In addition to improving performance, reducing the inter-ASIC bandwidth also improves energy efficiency.
%\lukas{change to make more explicit about the so what}

We present \Lightpass/, a complete system that realizes a circuit-switched indirection layer inside a multi-ASIC switch.
Our analysis shows that this indirection layer reduces inter-ASIC traffic while requiring minimal ASIC changes (\Cref{subsec:dataplane}).
To keep up with dynamic traffic patterns, the indirection layer is rapidly reconfigured.
Operating within a single device enables fast, integrated electrical/optical circuit switches with nanosecond/microsecond reconfiguration downtime, enabling rapid reconfigurations without disrupting traffic.
This fast reconfiguration layer requires algorithms that run in the order of microseconds while still providing good traffic localization.
We achieve this by mapping only the ingress port between ASICs and \textit{leaving the egress port fixed}. 
This enables a novel heuristic that achieves near-optimal traffic localization (within $0.5\,\%$) while meeting the stricter runtime targets.

Our evaluation demonstrates that \Lightpass/ delivers performance within $1\,\%$ of an ideal, non-manufacturable~(due to size constraints) single-ASIC switch, while cutting inter-ASIC bandwidth by up to half.
This lower bandwidth requirement allows our design to outperform traditional non-blocking multi-ASIC designs while achieving hundreds of watts of power savings per switch~(\Cref{subsec:powerandcost}), thanks to the indirection layer itself incurring negligible overhead ($<0.5\,\%$) in both chip area and power.
We show that \Lightpass/ can handle diverse application scenarios (from structured ML patterns to bursty DCN workloads), thanks to its fast control plane. 
Our hardware prototype running a PyTorch~\cite{paszke2019PyTorcha} LLM workload achieves $3\times$ higher token throughput than traditional designs.

The largest benefits of \Lightpass/ arise in the $2\mbox{ -- }4$ ASIC regime, where it delivers near single-ASIC performance while substantially reducing interconnect bandwidth and power overhead.
In this regime, the higher possible radix of \Lightpass/ increases the maximum number of hosts in a 2-layer Fat-Tree network by $30\mbox{ -- }78\,\%$ and in a 3-layer Fat-Tree network by $49\mbox{ -- }138\,\%$ compared to the \Baseline/, enabling considerably larger deployments.
This represents \textit{the relevant ASIC counts} for extending switch bandwidth before packaging complexity and cost make larger designs increasingly unattractive.
As demonstrated by NVIDIA shifting from a planned 4-ASIC chip to a simpler 2-ASIC design~\cite{2026NVIDIAs}.
\lukas{rewrite this and add reference here instead}

Importantly, \Lightpass/ provides not only a one-time benefit but smooths the transition between any successive ASIC generations, creating more efficient options.
As demand grows, a switch designer can first migrate from a single ASIC to an efficient multi-ASIC design using \Lightpass/, and later adopt the next ASIC generation when it becomes available. 
The same approach can then be repeated for the new ASIC~(\Cref{fig:generations}).

Overall, our contributions are as follows:
\begin{itemize}[topsep=2pt]
    \itemsep 0em 
    \item Enabling high-performance multi-ASIC switch designs with a limited inter-ASIC bandwidth requirement.
    \item Designing a feasible data plane and lightweight control plane to enable dynamic port-to-ASIC mapping~(\Cref{Lightpass}).  
    \item Developing an optimal algorithm to minimize inter-ASIC traffic, along with an efficient heuristic~(\Cref{design}).
    \item Showing near single-ASIC performance under diverse DCN and HPC/ML workloads with $2$-ASIC and $n$-ASIC switches~(\Cref{implementation} and \Cref{evaluation}).
    \item Validating our approach by benchmarking a PyTorch LLM workload on our hardware prototype~(\Cref{implementation} and \Cref{evaluation}).
\end{itemize}

%\textit{This work does not raise any ethical issues.}

\section{Motivation}\label{motivation}
The increased bandwidth demands from emerging applications, such as AI training, have put pressure on network switches to increase both bandwidth and radix (number of ports) to accommodate ever-larger-scale GPU server deployments~\cite{si2026Collective,discuss2026Microsoft}.
This is due to different factors: 
First, high bandwidth is necessary due to higher demand per host, with GPUs easily saturating 800G links and beyond.
Second, due to the scale of the deployments, involving hundreds of thousands of GPUs, the switch radix has become increasingly important. 
Low-radix switches complicate network architecture by requiring more layers to interconnect all hosts.
Multi-plane network topologies, such as Rail-only~\cite{wang2024Railonly} and P-FatTree~\cite{mellette2016PFatTree}, aim to alleviate this problem but incur additional complexity due to host-level load balancing.
But access to high-radix switches even simplifies multi-plane network designs by reducing the number of layers or increasing the scale.

\subsubsection{Transition to Multiple ASICs}
\begin{figure}
    \centering
    \begin{tikzpicture}[x=0.1cm,y=0.1cm,scale=1]

    \tikzset{thick/.style = {line width=0.3mm}}

    \node [draw=\asicline, rounded corners=\cornerrad, rectangle, fill=\asiccol, inner sep=0cm, minimum width=34mm, minimum height=4mm] (con1) at (0, 10) {\footnotesize Chiplet Interconnect};

    \node [draw=\asicline, rounded corners=\cornerrad, rectangle, fill=\asiccol, inner sep=0cm, minimum width=34mm, minimum height=4mm] (int1) at (0, -12) {\footnotesize Network Interfaces};

    \node [draw=\asicline, rounded corners=\cornerrad, rectangle, fill=\asiccol, inner sep=0cm, minimum width=10mm, minimum height=8mm] (ing1) at (-12, -4) {\footnotesize Ingress};

    \node [draw=\asicline, rounded corners=\cornerrad, rectangle, fill=\asiccol, inner sep=0cm, minimum width=10mm, minimum height=8mm] (egr1) at (12, -4) {\footnotesize Egress};

    \node [draw=\asicline, rounded corners=\cornerrad, rectangle, fill=\asiccol, inner sep=0cm, minimum width=10mm, minimum height=8mm, text width=10mm, align=center] (lt1) at (0, -4) {\footnotesize Lookup Tables};

    \node [draw=\asicline, rounded corners=\cornerrad, rectangle, fill=\asiccol, inner sep=0cm, minimum width=34mm, minimum height=4mm, align=center] (buf1) at (0, 4) {\footnotesize Buffer};

    \node [draw, thick,rounded corners=\cornerrad, rectangle,inner sep=0cm, minimum width=36mm, minimum height=26mm] (ASIC1) at (0, -1) {};

    \draw[thick, ->] ([xshift=-6 mm] buf1.north) -- ([xshift=-6 mm] con1.south);
    \draw[thick, <-] ([xshift=6 mm] buf1.north) -- ([xshift=6 mm] con1.south);

    \draw[thick, ->] ([xshift=-12 mm] int1.north) -- (ing1.south);
    \draw[thick, <-] ([xshift=12 mm] int1.north) -- (egr1.south);

    \draw[thick, ->] ([xshift=12 mm] buf1.south) -- (egr1.north);
    \draw[thick, <-] ([xshift=-12 mm] buf1.south) -- (ing1.north);

    \draw[->,thick] ([xshift=-6 mm] ASIC1.north) -- +(0,2);
    \draw[<-,thick] ([xshift=6 mm] ASIC1.north) -- +(0,2);

    \node [rectangle,inner sep=0cm, minimum width=38mm] () at (0, -17) {\footnotesize Packet Switch ASIC};
    \node [rectangle,inner sep=0cm, minimum width=38mm] () at (-43, -17) {\footnotesize $n$-ASIC Switch};

    \node [draw=\archline, rounded corners=\cornerrad, rectangle, fill=\archcol, inner sep=0cm, minimum width=8mm, minimum height=6mm] (asic4) at (-59, -3){\footnotesize $A_1$};
    \node [draw=\archline, rounded corners=\cornerrad, rectangle, fill=\archcol, inner sep=0cm, minimum width=8mm, minimum height=6mm] (asic3) at (-50, -3){\footnotesize $A_2$};
    \node [draw=\archline, rounded corners=\cornerrad, rectangle, fill=\archcol, inner sep=0cm, minimum width=8mm, minimum height=6mm] (asic2) at (-36, -3){\footnotesize $A_{n-1}$};
    \node [draw=\archline, rounded corners=\cornerrad, rectangle, fill=\archcol, inner sep=0cm, minimum width=8mm, minimum height=6mm] (asic1) at (-27, -3){\footnotesize $A_n$};

    \node [draw=\archline, rounded corners=\cornerrad, rectangle, fill=\archcol, inner sep=0cm, minimum width=40mm, minimum height=4mm] (port) at (-43, -12) {\footnotesize Physical Ports};
    \node [draw=\archline, rounded corners=\cornerrad, rectangle, fill=\archcol, inner sep=0cm, minimum width=8mm, minimum height=6mm] (fab1) at (-33, 10){\footnotesize Fabric};
    \node [draw=\archline,rounded corners=\cornerrad, rectangle, fill=\archcol, inner sep=0cm, minimum width=8mm, minimum height=6mm] (fab2) at (-53, 10){\footnotesize Fabric};

    \node [rectangle, inner sep=0cm, minimum width=5mm] () at (-43,10){\dots};
    \node [rectangle, inner sep=0cm, minimum width=5mm] () at (-43,-3){\dots};

    \draw [] (asic1.north) -- (fab1.south);
    \draw [] (asic2.north) -- (fab1.south);
    \draw [] (asic3.north) -- (fab1.south);
    \draw [] (asic4.north) -- (fab1.south);

    \draw [] (asic1.north) -- (fab2.south);
    \draw [] (asic2.north) -- (fab2.south);
    \draw [] (asic3.north) -- (fab2.south);
    \draw [] (asic4.north) -- (fab2.south);

    \draw [thick] (asic1.south) -- (asic1.south |- port.north);
    \draw [thick] (asic2.south) -- (asic2.south |- port.north);
    \draw [thick] (asic3.south) -- (asic3.south |- port.north);
    \draw [thick] (asic4.south) -- (asic4.south |- port.north);
    
    \draw[] (ASIC1.north west)++(-0.5,-1) -- (asic1.north east);
    \draw[] (ASIC1.south west)++(-0.5,1) -- (asic1.south east);

    \path (current bounding box.south west) +(0,-2) (current bounding box.north east) +(0,0);

\end{tikzpicture}%
    \caption{Chiplet-based $n$-ASIC packet switch setup.}
    \label{fig:asic}
\end{figure}
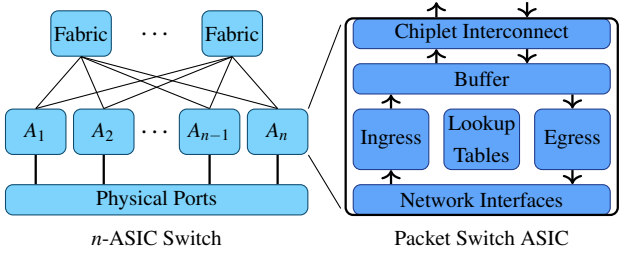

Achieving both higher bandwidth per port and higher radix simultaneously is challenging in a single ASIC due to limited I/O bandwidth and area constraints.
To keep up with today's demand, a shift towards multi-ASIC designs is required; an example of a multi-ASIC network switch is shown in~\Cref{fig:asic}; it 
%The multi-ASIC switch 
consists of $n$ packet-switch ASICs ($A_1 \cdots A_n$), each similar to a single-ASIC packet switch, with the addition of a chiplet interconnect.
With it, the packet-switch ASICs are connected to the fabric ASICs, which route traffic between them, typically via cell switching~\cite{zilberman2019Stardust}. 
For 2 ASICs, the chiplet interconnects can be directly connected, with no fabric required.

\begin{figure*}
    \centering
    \input{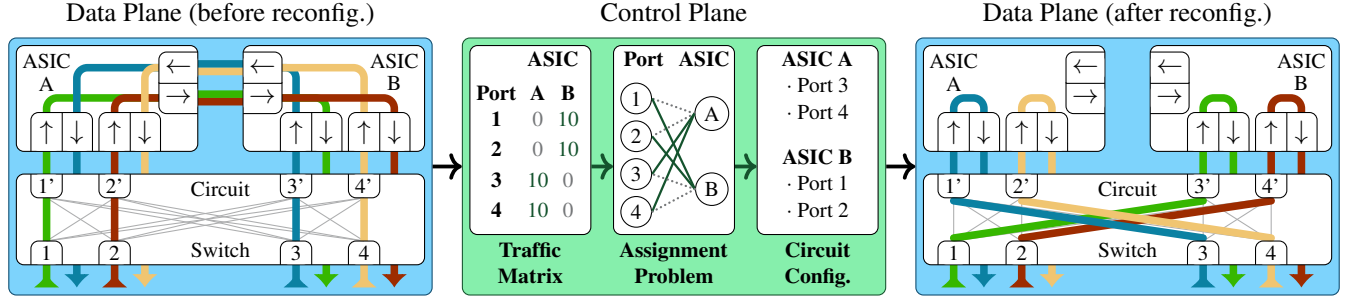}%
    \caption{The \Lightpass/ data plane adds a circuit-switched indirection layer enabling ingress port-to-ASIC mapping. The control plane computes circuit configurations based on historical traffic, minimizing inter-ASIC bandwidth. 
    }
    \label{fig:system}
\end{figure*}

\subsubsection{The Need for Inter-ASIC Bandwidth}
Multi-ASIC designs come with a set of challenges. 
Primarily, providing sufficient bandwidth between network chiplets is crucial to maintain performance comparable to that of a single ASIC.
The reason is simple: a packet that arrives at one ASIC might be destined to a port connected to another ASIC and therefore needs to cross the chiplet interconnect.
Now, if there is not enough bandwidth between the ASICs to accommodate all such packets, then the packets might need to be buffered or even dropped~(leading to a reduction in throughput of up to $4\times$; \Cref{subsec:performance}).
In the worst case, each ASIC must provide as much bandwidth to other ASICs as it provides to the outside network interfaces to maintain full bisection bandwidth, since every packet could need to be sent to a different ASIC.

\subsubsection{Bandwidth Creates Overhead}
Providing non-blocking behavior increases the needed chiplet interconnect and fabric capacity. 
This increases not only power~(by hundreds of watts per switch;\Cref{subsec:powerandcost}) but also uses up more of the ASIC's forwarding capacity, reducing the available bandwidth for external network interfaces.
In addition, it reduces the available space at the chip's edge that could be used for I/O to other components, such as off-chip High Bandwidth Memory~(HBM) packet buffers.
Meanwhile, the internal buffers must handle additional traffic from both the network interfaces and the chiplet interconnect, since packets that traverse the chiplet interconnect are written to a packet buffer, not once but twice, unless complex buffering logic is added to mitigate this issue.
%Switch latency has become an important aspect in HPC and ML workloads, with multi-ASIC designs introducing additional latency due to cross-chiplet communication.
%Localizing traffic within a single ASIC reduces the additional latency introduced by cross-ASIC traffic.

\subsubsection{In Pursuit of the Root Cause}
The root cause for performance degradation in oversubscribed multi-ASIC switches is that packets arrive at the \emph{wrong} ASIC and must cross the inter-ASIC boundary, overwhelming inter-ASIC links.
Therefore, closing the performance gap requires taking action as early as possible: traffic must be localized within a single ASIC whenever feasible. 
Achieving such localization requires an indirection layer between the physical switch ports and the packet-switch ASICs, which can dynamically remap ingress ports to ASICs based on observed traffic patterns.
A naïve realization of this indirection layer would be to move the fabric layer (shown in
\Cref{fig:asic}) between the ports and the ASICs.
However, this approach merely shifts the problem rather than solving it.
To make correct forwarding decisions, the indirection layer performs per-packet processing such as longest-prefix lookups, reintroducing the same bandwidth, power and design complexity that we aim to avoid.

We propose implementing the indirection layer using a circuit switch instead, which is much simpler and, unlike a packet-switching fabric, is agnostic to individual packets while still enabling dynamic remapping.
This poses new challenges, such as obtaining the traffic information essential to computing a new configuration.
This new configuration should be accurate enough to localize most traffic under the same ASIC, while being applied before the traffic information becomes outdated.
Additional logic is needed to handle the transient traffic during the downtime associated with the circuit-switch reconfiguration.
Our design~(\Cref{Lightpass} and \Cref{design}) addresses these challenges, and the evaluation~(\Cref{evaluation}) demonstrates close to single-ASIC performance with low overhead.

\section{\LightpassPlot/}\label{Lightpass}

We discuss the \Lightpass/ architecture with its data plane (\Cref{subsec:dataplane}) and control plane (\Cref{subsec:controlplane}). 
%We also discuss the practical viability of \Lightpass/ (\Cref{subsec:powerandcost}). 
The primary component of the data plane is the circuit-switched indirection layer; with the control plane being responsible for dynamically updating the circuit reconfiguration based on the traffic. 
\Cref{fig:system} shows the system overview, with the abstraction of circuit-switched indirection layer and a simplified reconfiguration example.

\subsection{Data Plane}
\label{subsec:dataplane}

The \Lightpass/ data plane consists of a circuit-switched indirection layer between switch ingress ports and ASICs, which enables flexible ingress port-to-ASIC mapping.
\Cref{fig:system} shows the data plane for a 4-port two-ASIC packet switch (ASICs A and B) with a circuit switch in between.
The two packet-switch ASICs (A and B) have an internal layout as shown in~\Cref{fig:asic}, which is inspired by an existing Cisco ASIC~\cite{cisco2026Cisco}. 
Each ASIC contains both an ingress and an egress packet processing pipeline with shared lookup tables. 
Hence, \Lightpass/ leaves the core packet-switch design largely unchanged,
requiring minor adjustments to add the indirection layer.
The egress scheduler, packet buffer, forwarding tables and similar components do not require any modifications beyond those required for traditional multi-ASIC designs.

\subsubsection{Flexible Port Assignment}
\label{subsec:flexportassignment}
Let's look at the red and dark blue flows. 
They arrive at ports 1 and 2; are destined to ports 3 and 4, respectively.
For the initial circuit-switch configuration (left side in \Cref{fig:system}), ingress ports 1 and 2 are mapped to internal ports 1' and 2'.
As a result, both the flows will enter ASIC A, and then need to traverse the inter-ASIC link to ASIC B. 
For the final circuit-switch configuration (right side in \Cref{fig:system}), the ingress ports 1 and 2 are now mapped to internal ports 3' and 4' respectively. 
By doing so, the flows do not need to traverse the inter-ASIC link and can be localized within ASIC B.
Such flexible ingress port-to-ASIC mapping can significantly reduce the inter-ASIC traffic volume.  
In \Cref{fig:system}, we consider one big circuit switch spanning all packet-switch ports for simplicity.  
But in practice, it might need to be split into multiple smaller ones, which  
\Lightpass/ can easily accommodate with little performance degradation~(\Cref{subsec:n-asic-case}).

%can accommodate multiple small circuit switches with little performance degradation~(\Cref{subsec:multiple-circuits}).

\subsubsection{Keeping the Egress Fixed}
A key insight of \Lightpass/ is only having a flexible ingress switch port-to-ASIC mapping, while keeping the egress ports fixed. Such a design choice provides many benefits:
\textbf{1)} If the egress were flexible, packets would need to follow the egress to wherever it is connected at the moment, creating the issue of packets moving between ASICs multiple times, in addition to the complex logic required to keep track of the egress.
Fixing the egress simplifies the hardware: packets always leave through the same egress without additional bookkeeping, just as in traditional forwarding.
\textbf{2)} It also avoids state migration: e.g., per-flow counters or scheduling state can be maintained at the fixed egress, rather than being shuffled across ASICs whenever an ingress is reassigned.
\textbf{3)} Finally, fixing egress ports shrinks the search space for the control plane algorithm, making the problem tractable~(\Cref{design}) and allowing for a heuristic running in the order of microseconds~(\Cref{subsec:sensitivityanalysis}).
Note that we can keep the egress fixed 
as \Lightpass/ operates inside a single device.
%in our single-device context, 
Whereas, previous optical edge works~\cite{wang2022RDC,das2024Rearchitecting} operate at network scale, forcing ingress and egress to be coupled.
%, as they operate at network scale. 
Separating the two there
is complicated: transceivers usually require bidirectional connectivity to establish links
(more differences in \Cref{subsec:discussion}).

\subsubsection{Realizing the Indirection Layer}
\label{subsec:realizationof}
The indirection layer is not explicitly tied to any specific circuit switch technology. 
%In the following, 
We discuss two potential realizations with %respective 
%the 
design trade-offs.

\begin{figure}
    \centering
    \begin{subfigure}{0.49\columnwidth}
    \centering
    \begin{tikzpicture}[x=0.1cm,y=0.1cm,scale=1]

\def\asiclen{2.2}
\def\serdeslen{1.2}
\def\elementwidth{0.8}
\def\serdesoffset{10}

\node[draw=\asicline, fill=\asiccol,rectangle,rounded corners=\cornerrad, minimum width=\asiclen cm, minimum height=\elementwidth cm] at (0,17) (ASIC) {};
\node[text width=1 cm, align=center] at (-17,17) {\footnotesize ASIC\strut};

\node[draw=\archline, fill=\archcol, rectangle,rounded corners=\cornerrad, minimum width=\asiclen cm, minimum height=\serdeslen cm] at (0,6) (SerDes) {};
\node[text width=1 cm, align=center] at (-17,6) {\footnotesize SerDes Chiplet\strut};

\node[text width=1 cm, align=center] at (-17,-2) {\footnotesize Ports\strut};

\node[rectangle, rounded corners=\cornerrad, minimum width= \asiclen cm, minimum height= 0.4 cm,inner sep=0mm] at (0,19) (pipe) {\footnotesize Forw. Pipelines};

\node[rectangle, rounded corners=\cornerrad, minimum width= \asiclen cm, minimum height= 0.4 cm,inner sep=0mm] at (0,15) {\footnotesize Interconnect};

\node[rounded corners=\cornerrad, rectangle, minimum width= \asiclen cm,minimum height=0.4 cm,inner sep=0mm] at (0,10) {\footnotesize Interconnect};

\node[rectangle, rounded corners=\cornerrad, minimum width= \asiclen cm, minimum height=0.4 cm,inner sep=0mm] at (0,6) (ECS) {\hspace*{-4mm}\footnotesize ECS};
\node[rectangle, minimum width= \asiclen cm, minimum height=0.4 cm,inner sep=0mm] at (3,6) {\circuitswitch[1.5]};

\node[rounded corners=\cornerrad, rectangle, minimum width= \asiclen cm, minimum height=0.4 cm,inner sep=0mm] at (0,2) {\footnotesize LR SerDes};

\draw[\archline] (pipe.south west) -- (pipe.south east);
\draw[\archline] (ECS.north west) -- (ECS.north east);
\draw[\archline] (ECS.south west) -- (ECS.south east);

\foreach \x in {-10,-8,-6,-4,-2,0,2,4,6,8,10}{

    \draw ([xshift=\x mm] ASIC.south) -- ([xshift=\x mm] SerDes.north);
    \draw ([xshift=\x mm] SerDes.south) -- ([xshift=\x mm] (0,-1););
    \node[draw,rectangle, rounded corners=0.3mm, minimum width=1mm, inner sep=0mm, minimum height=0.2 cm] at (\x,-2) {};
}

\end{tikzpicture}%
    \vspace{-0mm}
    \caption{Electrical Indirection}
    \label{fig:ecs}
    \end{subfigure}
    \begin{subfigure}{0.49\columnwidth}
    \centering
    \begin{tikzpicture}[x=0.1cm,y=0.1cm,scale=1]

\def\asiclen{2.2}
\def\serdeslen{1.6}
\def\elementwidth{0.8}

\node[draw=\asicline,fill=\asiccol, rectangle,rounded corners=\cornerrad, minimum width=\asiclen cm, minimum height=\elementwidth cm] at (0,17) (ASIC) {};
\node[text width=1 cm, align=center] at (-17,17) {\footnotesize ASIC\strut};

\node[draw=\archline, fill=\archcol, rectangle,rounded corners=\cornerrad, minimum width=\asiclen cm, minimum height=1.2 cm] at (0,6) (OE) {};
\node[text width=1 cm, align=center] at (-17,6) {\footnotesize Optical Engine\strut};

\node[rectangle,rounded corners=\cornerrad, minimum width=\asiclen cm, minimum height=0.4 cm] at (0,1) (OCS) {};

\node[text width=1 cm, align=center] at (-17,-2) {\footnotesize Ports\strut};

\node[rectangle,rounded corners=\cornerrad, minimum width= \asiclen cm, minimum height= 0.4cm,inner sep=0mm] at (0,19) (pipe) {\footnotesize Forw. Pipelines};

\node[rectangle,rounded corners=\cornerrad, minimum width= \asiclen cm, minimum height=0.4 cm,inner sep=0mm] at (0,15) {\footnotesize Interconnect};

\node[rectangle,rounded corners=\cornerrad, minimum width= \asiclen cm,minimum height=0.4 cm,inner sep=0mm] at (0,10) {\footnotesize Interconnect};

\node[rectangle, minimum width= \asiclen cm, minimum height=0.4 cm,inner sep=0mm] at (0,2) {\hspace*{-4mm}\footnotesize OCS};
\node[rectangle, minimum width= \asiclen cm, minimum height=0.4 cm,inner sep=0mm] at (3,2) {\circuitswitch[1.5]};

\node[rectangle,rounded corners=\cornerrad, minimum width= \asiclen cm, minimum height=0.4 cm,inner sep=0mm] at (0,6) (optical) {\footnotesize Optical TX/RX \strut};

\draw [\archline] (pipe.south west) -- (pipe.south east);
\draw [\archline] (optical.north west) -- (optical.north east);
\draw [\archline] (optical.south west) -- (optical.south east);

\foreach \x in {-10,-8,-6,-4,-2,0,2,4,6,8,10}{

    \draw ([xshift=\x mm] ASIC.south) -- ([xshift=\x mm] OE.north);
    \draw ([xshift=\x mm] OE.south) -- ([xshift=\x mm] (0,-1););
    \node[draw,rectangle, rounded corners=0.3mm, minimum width=1mm, inner sep=0mm, minimum height= 0.2 cm] at (\x,-2) {};
}

\end{tikzpicture}%
    \vspace{-0mm}
    \caption{Optical Indirection}
    \label{fig:ocs}
    \end{subfigure}
    \caption{\Lightpass/ can leverage different circuit switching technologies to realize the indirection layer.}
    \label{fig:circuitswitch}
\end{figure}
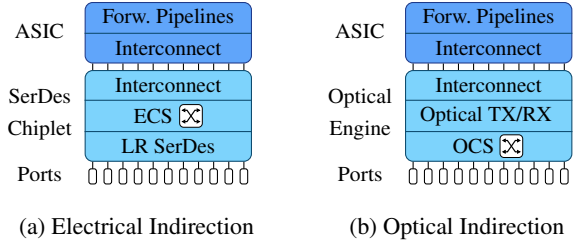

\paragraph{Electrical Indirection Layer}
Electrical circuit switches (ECS) can be added to the SerDes chiplets as shown in~\Cref{fig:ecs}.
The reconfiguration downtime is of the order of nanoseconds \cite{chatzieleftheriou2018Larrya, shrivastav2019Shoal}, while none of the SerDes connections need to be re-established, which is a primary advantage.
The ECS is after the long-reach SerDes~(connecting to the outside) and so the signal is already decoded when it reaches the circuit switch.
It is also before the interconnect~(connecting SerDes chiplet to the ASIC), where the bits are encoded again.

\paragraph{Optical Indirection Layer}
The indirection layer could be implemented using a silicon photonics optical circuit switch (OCS) \cite{seok2016Largescale,han2015Largescale,wu2016Largescale,ikeda2020Largescale}, as shown in~\Cref{fig:ocs}.
This will be a perfect fit for co-packaged optics-based packet switches. 
The primary benefit here is lower power consumption and high bandwidth that an OCS can provide, but the reconfiguration delay will be of the order of a few microseconds.
Lightmatter's emerging package technology~\cite{lightmatter2026Passage} already includes the necessary OCS, showing the feasibility of the design.

\subsection{Control Plane}
\label{subsec:controlplane}
% trying to make the traditional meaning bit more clear 

Unlike the traditional control plane in network switches, which governs network-wide
routing and forwarding policies, the \Lightpass/ control plane is a device-local controller
associated with the circuit-switched indirection layer. 
When multiple circuit switches are used,
each controller operates independently and is responsible for managing its
own circuit switch reconfiguration. 
We first explain the step-by-step workflow for computing the new circuit configuration, with more details about the algorithm in~\Cref{design}.
Next, we discuss how traffic should be handled during reconfiguration events.

\subsubsection{Workflow}
The workflow of the \Lightpass/ control plane has three steps. The controller $1$) first records the statistics about ingress port-to-ASIC traffic volumes, $2$) computes suitable ingress port-to-ASIC mapping, and $3$) finally updates the circuit switch configuration. 
%For reference, 
See~\Cref{fig:system} (middle one).

\paragraph{1. Collect Statistics}
\label{subsec:collectstatistics} 
To adapt the configuration, the control plane collects the port-level traffic statistics and records the traffic volume between ingress ports to different ASICs.
This scales with $\Theta(pn)$, where $p$ is the number of switch ports and $n$ is the number of ASICs; thus manageable to perform directly on the hardware.
It is comparable to having port counters inside the network device, which is a ubiquitous feature.
One way to implement this functionality is a small SRAM module that is shared by all ASICs, where the values are directly written to or updated frequently.
The circuit switch controller can then read this data from the memory to compute the mapping.
Even for a big switch with 512 ports, 8 ASICs and 32-bit integer entries~(4 bytes), the memory consumption would be only $512 \times 8 \times 4\,bytes = 16\,KB$ at $<1\,GB/s$ bandwidth ($\approx1000\times$ below on-chip SRAM).

\paragraph{2. Compute Ingress Port-to-ASIC Mapping}
\label{subsec:findassignment}
After recording the traffic, the controller computes a suitable ingress port-to-ASIC mapping that reduces the inter-ASIC traffic.
A detailed discussion of the algorithm can be found in~\Cref{design}.

\paragraph{3. Reconfiguring the Circuit Switch}
\label{subsec:reconfiguringthe}
After computing the suitable mapping, it is applied to the circuit switch.
Note that, only the ports with a new ASIC assignment need to be reconfigured.  
The rest do not incur any downtime.

\subsubsection{Handling the Transient}
During the circuit-switch reconfiguration, packets continue to arrive from outside; additional measures are required to handle them. 
We discuss those based on the circuit-switching technology used.

\paragraph{Electrical Indirection Layer}
While electrical circuit switches are very fast at reconfiguring, there is still a short downtime that might cause packet loss.
The benefit of electrical circuit switches is that reconfiguration downtime can be masked by adding packet buffers, whereas in optical circuit switches it cannot due to the lack of ``optical memory''.
So, in the electrical circuit-switch case, the controller can just reconfigure without dropping any packets if the reconfiguration is faster than the time it takes for those buffers to fill up.
Due to the fast reconfiguration, packet reordering might occur. 
This can be easily avoided by marking the last packet before reconfiguration. 
The egress ASIC then just needs to prioritize packets from the old ingress until the marked packet arrives.

\paragraph{Optical Indirection Layer}
Even though silicon photonics OCS are much faster than 3D-MEMS based OCS, they still have a reconfiguration downtime of the order of microseconds~\cite{seok2016Largescale,han2015Largescale,wu2016Largescale,ikeda2020Largescale,wang2022TopoOpta}.
To ensure lossless transient reconfiguration, the controller first sends a PAUSE frame to the neighboring switch. 
During reconfiguration, the other device requires enough space to buffer the packets. 
For example, at $800$ Gbps link speed, buffering packets for 1 microsecond requires only $100\,KB$ of buffer space,
while recent switch ASICs have $50-165\,MB$ of packet buffer~\cite{miao2017SilkRoad, intel2026Intel, fs2026N960064OD}.
After the PAUSE frame, the OCS is reconfigured, followed by a CONTINUE frame to the neighboring device to resume packet flow.
This mechanism is already present in Ethernet flow control~\cite{desanti20118021Qbb}.
The downtime during reconfiguration creates barriers, allowing in-flight packets to be drained before new ones arrive, effectively eliminating packet reordering.

%Unlike in lossless fabrics, where PFC deadlocks can occur, in our case, they are fixed between two devices and do not spread in the network. 
%They are active only during reconfiguration, placing an upper bound on their duration.
\lukas{get rid of PFC naming to avoid confusion}

\section{Control Algorithm Design} \label{design}

The primary objective of the control algorithm is to find the optimal ingress port-to-ASIC mapping that minimizes the inter-ASIC traffic volume.
Hence, the algorithm seeks to localize traffic within each ASIC as much as possible. 
To this end, we rely on the port-to-ASIC traffic matrix ($TM$), which records the traffic volume between ingress ports and ASICs. 

\subsection{Problem Formulation}
Given a multi-ASIC switch with $p$ ports and $n$ ASICs, and the port-to-ASIC traffic matrix ($TM$), we define a linear cost function $\Lambda$ representing the total inter-ASIC traffic as a function of the ingress port-to-ASIC mapping $\Phi$ (\Cref{eqn:cost}).
\setlength\abovedisplayskip{6pt}
\setlength\belowdisplayskip{6pt}
\begin{equation}
\begin{split}
    \Lambda(\Phi;\ TM) & = \sum_{i=1}^{p} \sum_{j=1}^{n} \Phi_{ij} \cdot TM_{ij} = \text{inter-ASIC traffic},\\   
    \text{with: } \quad \Phi_{ij} & =
    \begin{cases}
     0, & \text{if port $i$ is assigned to ASIC $j$},\\
     1, & \text{otherwise}.
    \end{cases} 
\end{split}
\label{eqn:cost}
\end{equation}

The optimization objective is to find the port-to-ASIC mapping $\Phi^*$ that minimizes the cost function subject to the constraint that each ASIC accommodates $\frac{p}{n}$ ports~(\Cref{eqn:constraint}). 
\begin{equation}
\begin{aligned}
    \min_{\Phi} \ \Lambda(\Phi;\ TM) \quad 
    \textrm{s.t.} \ \sum_{i=1}^{p} \Phi_{ij} = \frac{p}{n}, 
    \quad \forall~j \in \{1,\dots,n\}
\end{aligned}
\label{eqn:constraint}
\end{equation}

Note that, expressing costs in terms of port-to-ASIC traffic matrix, we pre-aggregate the underlying port-to-port traffic. 
Next, we discuss the optimal algorithm and heuristic in detail.

\subsection{Optimal Port-to-ASIC Mapping}
\label{subsec:matching}

At first glance, the port-to-ASIC mapping resembles a balanced graph partitioning problem: 
ports are modeled as vertices, edges represent traffic between ports, and the goal is to partition the graph into $n$ balanced groups (ASICs) while minimizing the weight of inter-group edges. 
However, this is known to be NP-hard, since the cost of assigning any given port depends on the placement of all other ports.

As discussed in~\Cref{subsec:dataplane}, \Lightpass/ indirection layer only allows ingress port-to-ASIC mapping, while the egress ports remain static.
Such a design choice substantially reduces the combinatorial complexity of the problem and enables a transformation into a linear assignment problem~\cite{kuhn2010Hungarian}: ports correspond to jobs, ASICs correspond to workers, and each edge weight represents the induced inter-ASIC traffic for that assignment.
The optimization goal is therefore to find a minimum-cost assignment that minimizes the total inter-ASIC traffic.
Since each ASIC must accommodate exactly $p/n$ ingress ports (where $p$ is the total number of switch ports), we reduce the problem to a one-to-one bipartite matching by replicating each ASIC node $p/n$ times.
The Hungarian algorithm then yields an optimal solution in polynomial time, with a worst-case complexity of $\Theta(p^3)$.

\subsection{Efficient Heuristic Design}
\label{subsec:efficient_heuristic_design}

\subsubsection{Greedy Heuristic}
To speed up the algorithm's runtime, we design a greedy heuristic~(\Cref{lst:heuristic}).
It assigns ingress ports to ASICs by prioritizing mappings that maximize \textit{net gain}, which represents the ``benefit'' of assigning port $i$ to ASIC $j$ i.e., the difference between localized and non-localized traffic.
It computes the net gain for every ingress port-ASIC pair~(line 3-6), sorts all entries in descending order~(line 8), and iterates once over them, assigning ports if not already been assigned and the ASIC still has capacity~(line 8-11).

\subsubsection{Accuracy}
For two ASICs, the greedy heuristic is provably optimal. 
Since each port can only be mapped to one of the two ASICs, ordering assignments by net gain always yields the minimum inter-ASIC traffic.
For more than 2 ASICs, the heuristic is not optimal, since the above argument no longer holds.
However, our evaluation of $1$k random traffic matrices shows that the heuristic is within $0.5\,\%$ of the optimal solution, even in $n$-ASIC scenarios.

\subsubsection{Time Complexity}
Given the switch consists of $p$ ports and $n$ ASICs, the heuristic runs in $\Theta(pn \log (pn))$ time: computing the net gain for all $p\times n$ pairs has a time complexity of $\Theta(pn)$, sorting the list dominates with $\Theta(pn \log (pn))$, and the final assignment pass takes $\Theta(pn)$ steps. 
This is a significant improvement over the $\Theta(p^3)$ complexity of the optimal algorithm, 
%especially since 
as the number of ASICs tends to be much smaller than the number of ports.
Runtime analysis for both optimal algorithm and greedy heuristic 
%is provided 
(\Cref{subsec:sensitivityanalysis}) shows that the heuristic scales almost linearly with number of switch ports.

\begin{lstlisting}[float=t,language=python, caption={Pseudocode of the greedy heuristic.}, label=lst:heuristic]
    def greedy_asic_mapping(TM):
        # compute net gain for each port-to-ASIC mapping
        for i in ports:
            row_sum = sum(TM[i,:])
            for j in asic:
                net_gain[i,j] = TM[i,j] - (row_sum - TM[i,j])
                
        # sort (port, ASIC) pairs by net gain, largest first
        for port, asic in argsort(net_gain):
            # assign if port is unassigned
            if not_mapped(port):
                # and if the ASIC has unassigned ports
                if free_slots(asic):
                    mapping.assign(port, asic)
                
        return mapping
\end{lstlisting}

\subsection{Difference from Optical Edge}
\label{subsec:discussion}

While similar in spirit, \Lightpass/ fundamentally differs from reconfigurable optical edge proposals, such as RDC~\cite{wang2022RDC} and OSSV~\cite{das2024Rearchitecting}, in three key aspects.
%as follows.

\subsubsection{Context}
RDC and OSSV operate at datacenter scale, placing optical circuit switches between servers and top-of-rack (ToR) switches to improve traffic locality across the network.
In contrast, \Lightpass/ operates entirely within a single multi-ASIC packet switch, localizing traffic across ports and ASICs within a single device.
This change in scope fundamentally alters both the control problem and the space of
feasible solutions.
This new context, for example, enables separating a port's ingress and egress, assigning them independently of each other.
This opens up new algorithmic options beyond the traditional optical edge work, where the two are coupled.

\subsubsection{Algorithmic Design}
Prior optical-edge approaches solve a variant of the Balanced Graph Partitioning problem that is NP-hard, requiring heavyweight heuristics over large inputs~(hosts and ToRs).
In contrast, our problem formulation of \Lightpass/ deliberately keeps the egress static and allows flexibility only at the ingress.
%Not that this is only possible due to our specific context within a single switch package, and would not be possible at the network level, where breaking up ingress and egress.
Such a design choice makes the problem tractable with cubic worst-case complexity. 
Finally, our efficient heuristic further reduces the complexity. %i.e., $\Theta(pn \log (pn))$, where $p$ and $n$ denote the number of ports and ASICs, respectively.
\lukas{add argument that splitting ingress and egress is only possible in this context}

\subsubsection{Performance and Practicality}
To our knowledge, \Lightpass/ is the first to propose fast integrated-circuit-switches inside a single switch package, enabling microsecond control loops and necessitating faster algorithms.
Our efficient heuristic executes in the order of microseconds~(\Cref{subsec:sensitivityanalysis}), whereas running the RDC/OSSV algorithm with the \textit{same input size} and CPU yields runtimes three orders of magnitude higher~(\Cref{{app:control-eff-contd}}).
As \Lightpass/ operates in a single switch, it can be deployed directly in today's data centers without any changes to existing infrastructure, in contrast to optical edge proposals that require deploying optical circuit switches first.
%This context allows \Lightpass/ to rely on a centralized controller without the scalability issues and other usual downsides, while scaling only with the number of ports and ASICs, rather than the number of hosts and ToRs across the entire datacenter.
\lukas{I think the last two sentences need to be changed}

\section{Implementation}\label{implementation}
We extensively evaluate \Lightpass/ using both packet-level simulation and a hardware prototype.
First, we briefly discuss the simulation setup along with different switch architectures. 
Next, we discuss the prototype implementation details.

\subsection{Packet Level Simulation}
\begin{figure}[h]
    \centering
    \begin{tikzpicture}[x=0.1cm,y=0.1cm,scale=1]

    \newcommand{\xofideal}{64}
    \newcommand{\xoflight}{0}
    \newcommand{\xofbase}{-22}
    \newcommand{\xofstatic}{-44}
    
    % Ideal
    \node[draw=\archline,rectangle, fill=\archcol, rounded corners=\cornerrad, minimum height=15mm, minimum width=18mm] at (2+\xofideal,10) (idealback) {};
    \node [draw, rectangle, fill=white, rounded corners=\cornerrad, text height=2mm, inner sep=0.1cm, minimum height=10mm, minimum width=10mm] at (2+\xofideal,10) (Base) {};
    \node [] at (2+\xofideal,10) {\packetswitch[1.5]};
    
    \node[rectangle, inner sep=0mm, minimum height=2mm, color=myblue] at ([yshift=2mm] idealback.north) {\footnotesize \textbf{\Ideal/\strut}};

    % Baseline
    \node[draw=\archline, rectangle, fill=\archcol, rounded corners=\cornerrad, minimum height=15mm, minimum width=20mm] at (23+\xofbase,10) (baseback) {};
    \node [draw, rectangle,fill=white, rounded corners=\cornerrad, inner sep=0cm, minimum height=5mm,minimum width=5mm] at (18+\xofbase,8) (Base1) {};
    \node [] at (18+\xofbase,8) {\packetswitch[1]};
    
    \node [draw, rectangle, fill=white, rounded corners=\cornerrad, inner sep=0cm, minimum height=5mm, minimum width=5mm] at (28+\xofbase,8) (Base2) {};
    \node [] at (28+\xofbase,8) {\packetswitch[1]};
    \node [draw, rectangle, fill=white, rounded corners=\cornerrad, inner sep=0mm, minimum height=3mm, minimum width=15mm] at (23+\xofbase,15) (FabricBase) {\footnotesize Fabric\strut};
    
    \node[rectangle, inner sep=0mm, minimum height=2mm, color=myviolet] at ([yshift=2mm] baseback.north) {\footnotesize \textbf{\Baseline/\strut}};

    \node [rectangle, inner sep=0cm, minimum width=3mm] () at (23+\xofbase,8){\footnotesize \dots};

    \path [draw] ([xshift=1.5mm] Base1.north) -- ([xshift=-3.5mm] FabricBase.south);
    \path [draw] ([xshift=-1.5mm] Base1.north) -- ([xshift=-6.5mm] FabricBase.south);
    \path [draw] ([xshift=0.5mm] Base1.north) -- ([xshift=-4.5mm] FabricBase.south);
    \path [draw] ([xshift=-0.5mm] Base1.north) -- ([xshift=-5.5mm] FabricBase.south);

    \path [draw] ([xshift=1.5mm] Base2.north) -- ([xshift=6.5mm] FabricBase.south);
    \path [draw] ([xshift=-1.5mm] Base2.north) -- ([xshift=3.5mm] FabricBase.south);
    \path [draw] ([xshift=0.5mm] Base2.north) -- ([xshift=5.5mm] FabricBase.south);
    \path [draw] ([xshift=-0.5mm] Base2.north) -- ([xshift=4.5mm] FabricBase.south);

    % Lightpass
    \node[draw=\archline,rectangle, fill=\archcol, rounded corners=\cornerrad, minimum height=15mm, minimum width=20mm] at (45+\xoflight,10) (lightback) {};
    \node [draw, rectangle,fill=white, rounded corners=\cornerrad, text height=2mm, inner sep=0.1cm, minimum height=5mm,minimum width=5mm] at (40+\xoflight,10) (Lightpass1) {};
    \node [] at (40+\xoflight,10) {\packetswitch[1]};
    
    \node [draw, rectangle,fill=white, rounded corners=\cornerrad, text height=2mm, inner sep=0.1cm, minimum height=5mm,minimum width=5mm] at (50+\xoflight,10) (Lightpass2) {};
    \node [] at (50+\xoflight,10) {\packetswitch[1]};
    
    \node [draw, rectangle,fill=white, rounded corners=\cornerrad,text height=1.5mm, inner sep=0cm, minimum height=3mm, minimum width=1.6cm] at (45,5.25) (CircuitSwitch) {};
    \node [] at (45+\xoflight,5.25) {\circuitswitch[1]};
    
    \node [draw, rectangle,fill=white, rounded corners=\cornerrad, inner sep=0mm, minimum height=3mm, minimum width=15mm] at (45+\xoflight,15) (FabricLightpass) {\footnotesize Fabric\strut};
    
    \node[rectangle, inner sep=0mm, minimum height=2mm, color=myred] at ([yshift=2mm] lightback.north) {\footnotesize \textbf{\Lightpass/\strut}};
    
    \node [rectangle, inner sep=0cm, minimum width=3mm] () at (45+\xoflight,10){\footnotesize \dots};

    \path [draw] ([xshift=1mm] Lightpass1.north) -- ([xshift=-4mm] FabricLightpass.south);
    \path [draw] ([xshift=-1mm] Lightpass1.north) -- ([xshift=-6mm] FabricLightpass.south);
    \path [draw] ([xshift=1mm] Lightpass2.north) -- ([xshift=6mm] FabricLightpass.south);
    \path [draw] ([xshift=-1mm] Lightpass2.north) -- ([xshift=4mm] FabricLightpass.south);

    % Static
    \node[draw=\archline,rectangle, fill=\archcol, rounded corners=\cornerrad, minimum height=15mm, minimum width=20mm] at (67+\xofstatic,10) (staticback) {};
    \node [draw, rectangle,fill=white, rounded corners=\cornerrad, inner sep=0cm, minimum height=5mm,minimum width=5mm] at (62+\xofstatic,8) (Static1) {};
    \node [] at (62+\xofstatic,8) {\packetswitch[1]};
    
    \node [draw, rectangle,fill=white, rounded corners=\cornerrad, inner sep=0cm, minimum height=5mm,minimum width=5mm] at (72+\xofstatic,8) (Static2) {};
    \node [] at (72+\xofstatic,8) {\packetswitch[1]};
    \node [draw, rectangle,fill=white, rounded corners=\cornerrad, inner sep=0mm, minimum height=3mm, minimum width=15mm] at (67+\xofstatic,15) (FabricStatic) {\footnotesize Fabric\strut};
    
    \node[rectangle, inner sep=0mm, minimum height=2mm, color=mygreen] at ([yshift=2mm] staticback.north) {\footnotesize \textbf{\Static/\strut}};

    \node [rectangle, inner sep=0cm, minimum width=3mm] () at (67+\xofstatic,8){\footnotesize \dots};

    \path [draw] ([xshift=1mm] Static1.north) -- ([xshift=-4mm] FabricStatic.south);
    \path [draw] ([xshift=-1mm] Static1.north) -- ([xshift=-6mm] FabricStatic.south);

    \path [draw] ([xshift=1mm] Static2.north) -- ([xshift=6mm] FabricStatic.south);
    \path [draw] ([xshift=-1mm] Static2.north) -- ([xshift=4mm] FabricStatic.south);

    \node [rectangle, text width=28mm] at (27,-1) {\footnotesize Packet Switch ASIC};
    \node [rectangle] at (10,-1) {\packetswitch[1]};
    \node [rectangle,text width=28mm] at (60,-1) {\footnotesize Circuit Switch};    
    \node [rectangle] at (43,-1) {\circuitswitch[1]};

    \foreach \x in {0,...,7} {
        \pgfmathparse{\x - 3.5}
        \path [draw] ([xshift=\pgfmathresult mm] Base.south) -- (2 + \xofideal + \pgfmathresult, 4);
    }
    
    \foreach \x in {0,...,3}{
        \pgfmathparse{\x - 1.5}
        \path [draw] ([xshift=\pgfmathresult mm] Base1.south) -- (18 + \xofbase + \pgfmathresult, 4);
        \path [draw] ([xshift=\pgfmathresult mm] Base2.south) -- (28 + \xofbase + \pgfmathresult, 4);
        \path [draw] ([xshift=\pgfmathresult mm] Base1.south) -- (18 + \xofbase + \pgfmathresult, 4);
        \path [draw] ([xshift=\pgfmathresult mm] Base2.south) -- (28 + \xofbase + \pgfmathresult, 4);

        \path [draw] ([xshift=\pgfmathresult mm] Static1.south) -- (62 + \xofstatic + \pgfmathresult, 4);
        \path [draw] ([xshift=\pgfmathresult mm] Static2.south) -- (72 + \xofstatic + \pgfmathresult, 4);
        \path [draw] ([xshift=\pgfmathresult mm] Static1.south) -- (62 + \xofstatic + \pgfmathresult, 4);
        \path [draw] ([xshift=\pgfmathresult mm] Static2.south) -- (72 + \xofstatic + \pgfmathresult, 4);
        
        \path [draw] ([xshift=\pgfmathresult mm] Lightpass1.south) -- (40 + \xoflight + \pgfmathresult, 6.75);
        \path [draw] ([xshift=\pgfmathresult mm] Lightpass2.south) -- (50 + \xoflight + \pgfmathresult, 6.75);
        \path [draw] ([xshift=\pgfmathresult mm,yshift=-3.75 mm] Lightpass1.south) -- (40 + \xoflight + \pgfmathresult, 3);
        \path [draw] ([xshift=\pgfmathresult mm,yshift=-3.75 mm] Lightpass2.south) -- (50 + \xoflight + \pgfmathresult, 3);
    }
    
    \path (current bounding box.south west) +(0,-0.5) (current bounding box.north east) +(0,0);
    
\end{tikzpicture}%
    \caption[Switch architectures used in the simulation.]{Switch architectures used in the simulation. For the 2-ASIC case, no fabric is necessary.}
    \label{fig:architectures}
\end{figure}
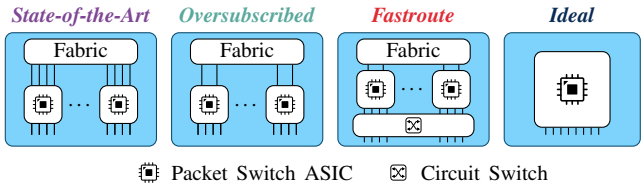

For the packet-level simulation, we leverage the Netbench simulator~\cite{kassing2025Netbench,kassing2017fattreesa}, which includes congestion control and queueing mechanisms.
We extend the simulator by implementing multi-ASIC switch architectures without and with an additional circuit-switch layer. 
Specifically, we evaluate the performance of four switch architectures~(\Cref{fig:architectures}).
The default parameters, if applicable for the architecture, are shown in \Cref{tab:default-values}.
The individual Packet Switch ASICs for all three multi-ASIC architectures have the \textit{same capacity}.

\begin{table}[b]
    \centering
    \footnotesize
    \caption{Default values used in the simulation.}
    \vspace{0.25em}
    \renewcommand{\arraystretch}{1.1}
    \begin{tabularx}{\columnwidth}{lXc}
        \toprule
        Parameters  &&  Default Values \\
        \midrule
        Reconfiguration Downtime (OCS) && $1\,\mu s$ \\
        Reconfiguration Computation Delay && $10\,\mu s$ \\
        Reconfiguration Interval && $100\,\mu s$ \\
        Inter-ASIC Delay && $10\,ns$ \\
        Inter-ASIC Buffer && $0.3\,MB\,/\,100\,Gbps^1$ \\
        \bottomrule
        \\[-2mm]
        \multicolumn{3}{l}{$^1$Buffer space per 100 Gbps in Tomahawk 5 chips}
    \end{tabularx}
    \label{tab:default-values}
\end{table}

\begin{enumerate}[leftmargin=*]
    \itemsep 0em 
    \item \textbf{\Baseline/:} A traditional multi-ASIC packet switch design, using multiple ASICs and a fully provisioned non-blocking inter-ASIC fabric~(\textit{not} oversubscribed).
    %This reduces the Packet Switch ASIC bandwidth available to the outside.
    
    \item \textbf{\Static/:} This architecture, unlike the \Baseline/, uses oversubscribed inter-ASIC links, providing more of the available Packet Switch ASIC bandwidth to the outside, but risking inter-ASIC congestion in the fabric.
    
    \item \textbf{\Lightpass/:} Our proposed architecture, just like \Static/, includes an oversubscribed inter-ASIC fabric, but in addition has a circuit switch indirection layer.
    The circuit switch has a $1$ $\mu$s reconfiguration downtime~(silicon photonics OCS), a delay of $10$ $\mu$s to compute the new mapping~(\Cref{subsec:sensitivityanalysis}) and a $100$ $\mu$s reconfiguration interval (\Cref{tab:default-values}). 
    
    \item \textbf{\Ideal/:} A packet switch composed of one big ASIC, that provides the theoretical performance upper bound.
\end{enumerate}

\subsubsection{Network Topology}
The default network topology is a two-stage fat-tree with 2048 hosts.
The \Ideal/ switch has 64 ports with 800 Gbps link speed~(51.2 Tbps per switch).
The switches use ECMP to load-balance across available paths.
The hosts use the DCTCP \cite{alizadeh2010Data} congestion control algorithm.

\subsubsection{Host Count}
To ensure a fair comparison, the number of hosts must remain constant across all architectures. 
Changing the host count alters the generated traffic trace, preventing an accurate apples-to-apples comparison of the traffic patterns.
%However, keeping the host count uniform is non-trivial since the evaluated architectures differ in switch radix.
We accomplish a constant host count by dividing each switch's capacity among a fixed number of ports.
%% safe option
%We evaluate the effect of this using the Web traffic trace; its uniform random traffic characteristics make it largely invariant to changes in the number of hosts, unlike more structured traffic such as Shuffle and Stride.
%The experiment in \Cref{app:host} presents the evaluation for varying numbers of hosts.
%The results are identical to those of the evaluation in \Cref{subsec:performance} with a fixed number of hosts, thereby validating our methodology.
%% aggressive version
%We evaluate this using the Web traffic trace; its uniform random traffic characteristics make it largely invariant to changes in the number of hosts, unlike more structured traffic such as Shuffle and Stride.

The experiment in \Cref{app:host} presents the evaluation with a variable host count instead.
Compared to our evaluation in \Cref{subsec:performance} of the same trace, the improvement of \Lightpass/ over the \Baseline/ is even better.
This is because \Lightpass/ improves not only bandwidth but also the radix compared to the \Baseline/.

%The experiment shown in \Cref{app:host} presents an evaluation where the host count is variable.
%Doing this slightly worsens the \Baseline/ performance metrics relative to the constant host count used in %\Cref{subsec:performance}.
%This is because \Lightpass/ improves not only bandwidth but also radix (scaling hosts quadratically in two-stage topologies) compared to the \Baseline/. 
%Keeping the host count constant neutralizes this additional disadvantage for the \Baseline/ architecture in our evaluation.

\lukas{needs polishing}

\subsubsection{Metrics}
We report the mean throughput normalized to the throughput of the \Ideal/ architecture, to show average performance~(higher is better).
To also include tail-end performance, we show the $99^{th}$ percentile flow completion time (lower is better).
For some experiments, we show the time series of the aggregate inter-ASIC bandwidth (the sum of all traffic sent to another ASIC) or other related values.

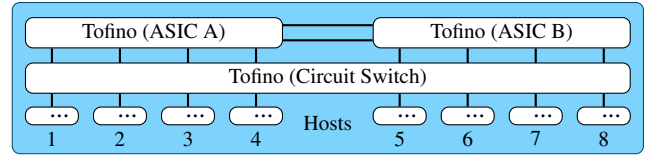
\begin{figure}
    \centering
    \begin{tikzpicture}[x=0.1cm,y=0.1cm,scale=1]

    \node[draw=\archline,fill=\archcol,rectangle, rounded corners=\cornerrad, minimum height=20mm,inner sep=0mm, minimum width=83.5mm] at (23,-6) {};
    \node [draw, rectangle, fill=white, rounded corners=\cornerrad, inner sep=0cm, minimum height=4mm, minimum width=3.4cm] at (0,0) (Lightpass1) {\footnotesize Tofino (ASIC A)};
    \node [draw, rectangle, fill=white, rounded corners=\cornerrad, inner sep=0cm, minimum height=4mm, minimum width=3.4cm] at (46,0) (Lightpass2) {\footnotesize Tofino (ASIC B)};
    \node [draw, rectangle, fill=white, rounded corners=\cornerrad, inner sep=0cm, minimum height=4mm, minimum width=8cm] at (23,-6) (CircuitSwitch){\footnotesize  Tofino (Circuit Switch)};

    \path [draw,thick] ([yshift=1mm] Lightpass1.east) -- ([yshift=1mm] Lightpass2.west);
    \path [draw,thick] ([yshift=-1mm] Lightpass1.east) -- ([yshift=-1mm] Lightpass2.west);
    
    \foreach \x in {1,...,4}{
        \pgfmathtruncatemacro{\offset}{\x * 9 - 23}
        \pgfmathtruncatemacro{\xtwo}{\x+4}
        
        \path [draw,thick] ([xshift=\offset mm + 0.5 mm] Lightpass1.south) -- (0.5 + \offset, -4);
        \path [draw,thick] ([xshift=\offset mm + 0.5 mm] Lightpass2.south) -- (46.5 + \offset, -4);
        
        \node[draw, rectangle, fill=white, rounded corners=\cornerrad, minimum width=7mm] at (0.5 + \offset, -11) (node\x) {};
        \node[draw, rectangle, fill=white, rounded corners=\cornerrad, minimum width=7mm] at (46.5 + \offset, -11) (node\xtwo) {};
        \node[minimum width=8mm] at (0.5 + \offset, -14) (node\x text) {\footnotesize\x};
        \node[minimum width=8mm] at (46.5 + \offset, -14) (node\xtwo text) {\footnotesize\xtwo}; 
        
        \node[minimum width=8mm] at (1.5 + \offset, -11) (node\x point) {...};
        \node[minimum width=8mm] at (47.5 + \offset, -11) (node\xtwo point) {...};

        \path [draw,thick] ([xshift=\offset mm + 0.5 mm,yshift=-6 mm] Lightpass1.south) -- (node\x);
        \path [draw,thick] ([xshift=\offset mm + 0.5 mm,yshift=-6 mm] Lightpass2.south) -- (node\xtwo);
    }
    %\node [text centered] at (23,2.5) (speed) {\footnotesize 2x100G};
    \node [text centered] at (23,-12) (hosts) {\footnotesize Hosts};
    \path (current bounding box.south west) +(0,-1) (current bounding box.north east) +(0,0);
    
\end{tikzpicture}%
    \caption{Hardware prototype of \Lightpass/ using programmable switches and 8 hosts (100 Gbps each).}
    \label{fig:testbed-overview}
\end{figure}

\subsection{Hardware Prototype}
\label{subsec:hardwaresetup}

The hardware prototype is built with P4-capable switches (Tofino 1), which enable us to record the required TM statistics and emulate the circuit-switch layer.
Note that programmable switches are not necessary for \Lightpass/ and are only used in this case to collect the needed statistics.
As shown in~\Cref{fig:testbed-overview}, the prototype consists of a) 8 hosts, with a 100 Gbps interface each, and b) 2 Tofinos, one acting as the two-ASIC (ASIC A and ASIC B) packet switch and another emulating the circuit switch.
The ASICs are connected via two 100 Gbps interfaces, resulting in a $2{:}1$ oversubscription.
The statistics are collected in the dataplane and read out from the control plane.
The new mapping is then computed using the heuristic explained in~\cref{subsec:efficient_heuristic_design}.
The ingress port-to-ASIC mapping is applied by the Tofino acting as the circuit switch and is updated every $500\,\mu s$.
The Tofinos, acting as the two-ASIC packet switch, have no additional logic implemented; they simply forward traffic based on the destination IP.

\lukas{remove 2x100G once the new hardware eval is added}

\section{Evaluation}
\label{evaluation}

Our evaluation addresses the central research question: 
\textit{Can a multi-ASIC switch with reduced inter-ASIC bandwidth deliver performance close to that of a single-ASIC switch?} 

We answer this through extensive evaluation with packet-level simulations and a hardware prototype.
We begin with application-oriented patterns \eg Shuffle, Stride, and All-Reduce, followed by trace-driven datacenter traffic.
Using those scenarios, we validate that \Lightpass/ works for an arbitrary number of ASICs and tolerates multiple smaller independent circuit switches.
We then analyze sensitivity to key parameters (traffic stability, buffer size, reconfiguration overhead), benchmark the control plane and evaluate the power savings enabled by \Lightpass/.
Here are the key findings:

\begin{enumerate}[leftmargin=*, topsep=2pt]
    \itemsep -0pt 
    \item Under application-oriented traffic patterns~(\Cref{subsec:special_patterns})  and realistic datacenter traces~(\Cref{subsec:performance}), \Lightpass/ sustains performance within $\approx1\,\%$ of an ideal single-ASIC switch, for 2-ASICs and $n$-ASICs~(\Cref{subsec:n-asic-case}) and outperforms both \Static/ and \Baseline/ architectures.
    \item On a hardware prototype, \Lightpass/ achieves a $3\times$ higher throughput than a \Static/ multi-ASIC switch under a LLM workloads~(\Cref{subsec:special_patterns}).
    %\item \Lightpass/ is robust against dynamic workloads~(\Cref{subsec:sensitivityanalysis}).
    \lukas{maybe change this}
    \item \Lightpass/ reduces the overall power consumption by hundreds of watts compared to a traditional design~(\Cref{subsec:powerandcost}). 
\end{enumerate}

\subsection{Application Oriented Traffic}
\label{subsec:special_patterns}
%Apart from trace-based DCN traffic, 
We evaluate the performance of \Lightpass/ under well-known application-oriented traffic patterns, emulating the behavior of HPC/ML applications. 
We consider widely-used traffic patterns such as Shuffle, Stride and Ring All-Reduce described below.
In this section, we consider 2 ASICs with an oversubscription ratio of $2{:}1$.
For the $n$-ASIC scenario, see~\Cref{subsec:n-asic-case}.

\begin{enumerate}[leftmargin=*,topsep=0pt]
    \setlength\abovedisplayskip{1pt}
    \setlength\belowdisplayskip{1pt}
    \itemsep -0pt
    \item \textbf{Shuffle:} Each host $i$ communicates with
$k$ other hosts, each with a different constant offset.  
\[ i \to (i + j \cdot s) \bmod H \quad j \in \{1, \dots, k\} \]
    \item \textbf{Stride:} Each host $i$ communicates only with one other host that has a constant offset $s$. 
\[ i \to (i + s) \bmod H \]
    \item \textbf{Ring All-Reduce:} Each host $i$ communicates with its two neighbors in a ring, until all values are aggregated. 
\[ i \to (i+1) \bmod H, \quad (i-1) \bmod H \to i \]

\if 0
    \item \textbf{Hierarchical All-Reduce:} Hosts are divided into groups ($M$). 
    During the first phase, each host $i$ in group $g$ ($i_g$) communicates within its own group, creating a local ring. In each subsequent phase, host $i_g$ communicates with its counterpart $i$ in another group $g'$ ($i_{g'}$). 
\begin{gather*} 
    i_g \to (i+1)_g \bmod |G|, \quad \forall g \in \{0, \dots, M-1\} \\
    i_g \leftrightarrow i_{g'} \quad \forall g' \neq g 
\end{gather*}
\fi

\end{enumerate}

\def\addlegend{11}
\begin{figure}
    \centering
    \begin{tikzpicture}
\pgfplotsset{every axis/.append style={font=\large,line width=1pt,tick style={line width=0.8pt}}, width=0.53\linewidth}
\begin{axis}[ 
 title=Mean Throughput, ylabel style={at={(axis description cs:-0.2,.5)},anchor=south,align=center},ylabel={\footnotesize Thrp. [\%]}, xtick pos=left, ybar,   axis on top, xtick=data, xtick align=inside, scaled ticks=false,symbolic x coords={Shuffle,Stride,ML},ticklabel style={/pgf/number format/fixed,/pgf/number format/.cd, 1000 sep = {}},bar width=1.2mm, xmin={[normalized]-0.6}, xmax={[normalized]+2.6},ymin=0, name=first plot, legend style={at={(1.1,1.25)}, draw=none, anchor=south}, legend columns=-1] 
\addplot [myviolet, fill=myviolet] table { 
 
Scenario "Mean Throughput [%]"

Shuffle 76.66011361601372
Stride 75.91166395717998
ML 65.9392058291405
};
\addplot [mygreen, fill=mygreen] table { 
 
Scenario "Mean Throughput [%]"

Shuffle 65.58331528317697
Stride 52.14386579506838
ML 53.97658870223787
};
\addplot [myred, fill=myred] table { 
 
Scenario "Mean Throughput [%]"

Shuffle 96.41696270365208
Stride 95.899068419282
ML 99.62443527389807
};
\addplot [myblue, fill=myblue] table { 
 
Scenario "Mean Throughput [%]"

Shuffle 100.0
Stride 100.0
ML 100.0
};
\addlegendimage{myviolet, ybar, ybar legend, fill=myviolet}
\addlegendimage{mygreen, ybar, ybar legend, fill=mygreen}
\addlegendimage{myred, ybar, ybar legend, fill=myred}
\addlegendimage{myblue, ybar, ybar legend, fill=myblue}
\if\addlegend 
 \legend{\BaselinePlot/,\StaticPlot/,\LightpassPlot/,\IdealPlot/}
 \fi 
\end{axis}
\begin{axis}[ 
 title=$99^{th}$ Perc. FCT, ylabel style={at={(axis description cs:-0.15,.5)},anchor=south, align=center},ylabel={\footnotesize FCT [ms]}, xtick pos=left, ybar, axis on top, xtick=data, xtick align=inside, scaled ticks=false,symbolic x coords={Shuffle,Stride,ML},ticklabel style={/pgf/number format/fixed,/pgf/number format/.cd, 1000 sep = {}},bar width=1.2mm, xmin={[normalized]-0.6}, xmax={[normalized]+2.6}, ymin=0, at=(first plot.east), anchor=west, xshift=14mm, xlabel style={yshift=1mm}, ytick distance=1] 
\addplot [myviolet, fill=myviolet] table { 
 
Scenario "All 99th FCT [ms]"

Shuffle 1.52647733
Stride 1.4387422399999998
ML 2.5073815
};
\addplot [mygreen, fill=mygreen] table { 
 
Scenario "All 99th FCT [ms]"

Shuffle 1.3677464
Stride 1.5293397699999995
ML 2.46669475
};
\addplot [myred, fill=myred] table { 
 
Scenario "All 99th FCT [ms]"

Shuffle 1.05847647
Stride 1.0206405899999995
ML 1.6286915
};
\addplot [myblue, fill=myblue] table { 
 
Scenario "All 99th FCT [ms]"

Shuffle 1.0521218100000003
Stride 1.03237396
ML 1.6351165
};
\end{axis}
\end{tikzpicture}%
    \caption{\Lightpass/ is able to adapt to different traffic patterns, providing near ideal performance.}
    \label{fig:special}
\end{figure}
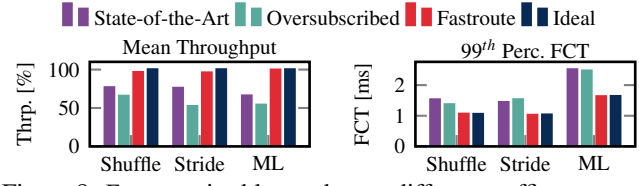

\label{subsec:shuffleandstride}
\subsubsection{Shuffle and Stride}
Shuffle and Stride are two traffic patterns prevalent in HPC workloads where a host communicates with other hosts at constant offsets.
\Cref{fig:special} shows the throughput and $99^{th}$ percentile FCT performance of different switch architectures for these patterns.
We observe that while \Lightpass/ can adapt to traffic patterns, the \Static/ and \Baseline/ switch architectures suffer significant performance degradation.
\Lightpass/ automatically remaps ingress ports to ASICs, minimizing inter-ASIC traffic and achieving near-single-ASIC performance even under oversubscription.

\subsubsection{Ring All-Reduce}
\label{subsec:mltraffic}
The Ring All-Reduce pattern is the heart of today's ML training workloads; known for high predictability and low entropy. 
To emulate ML traffic patterns, we randomly split up all hosts into groups of 16, with each group forming a logical ring, and repeat this every $0.5\,ms$.
\Cref{fig:special} shows the throughput and $99^{th}$ percentile FCT performance of different switch architectures for the ML Ring All-Reduce pattern.
We show that \Lightpass/ sustains performance within $<0.5\,\%$ of the \Ideal/, while both the \Baseline/ and \Static/ suffer from reduced throughput and higher FCT.

\if 0
\begin{table}
    \centering
    \footnotesize
    \caption{Prototype throughput performance.}
    \vspace{1mm}
    \renewcommand{\arraystretch}{1.1}
    \begin{tabularx}{\columnwidth}{lXXcXc}
        \toprule
        &&& \Static/ && \Lightpass/ \\
        \midrule
        Throughput Phase 1 &&& 92.37 Gbps &&  92.34 Gbps \\
        Throughput Phase 2 &&& \textbf{45.11} Gbps && \textbf{92.36} Gbps \\
        \bottomrule
    \end{tabularx}
    \label{tab:prototype-perf}
\end{table}
\fi

\begin{figure}
    \centering
    \begin{tikzpicture}
\pgfplotsset{every axis/.append style={font=\large,line width=1pt,tick style={line width=0.8pt}}, width=0.85\linewidth}
\begin{axis}[ 
 title={}, ylabel style={at={(axis description cs:-0.2,.5)},anchor=south,align=center},ylabel={\footnotesize Tokens \\ per Second}, xtick pos=left, ybar=2mm, axis on top, xtick=data, xtick align=inside, scaled ticks=false,symbolic x coords={LLM Training},bar width=8mm, xmin={[normalized]-0.6}, xmax={[normalized]+0.6},ymin=0, name=first plot, legend style={at={(0.4,1.25)}, draw=none, anchor=south}, legend columns=-1] 
\addplot [myviolet, fill=myviolet] table { 
Scenario Tokens/second
{LLM Training} 1878.754760514296
};
\addplot [mygreen, fill=mygreen] table { 
Scenario Tokens/second
{LLM Training} 2008.434530054712
};
\addplot [myred, fill=myred] table { 
Scenario Tokens/second
{LLM Training} 5970.1860321932045
};
\addplot [myblue, fill=myblue] table { 
Scenario Tokens/second
{LLM Training} 6145.73993351425
};
\addlegendimage{myviolet, ybar, ybar legend, fill=myviolet}
\addlegendimage{mygreen, ybar, ybar legend, fill=mygreen}
\addlegendimage{myred, ybar, ybar legend, fill=myred}
\addlegendimage{myblue, ybar, ybar legend, fill=myblue}
\if\addlegend
\legend{\BaselinePlot/,\StaticPlot/,\LightpassPlot/,\IdealPlot/}
\fi
\end{axis}
\end{tikzpicture}%
    \caption{The hardware prototype validates that \Lightpass/ performs well in real-world scenarios.}
    \label{fig:prototype}
\end{figure}
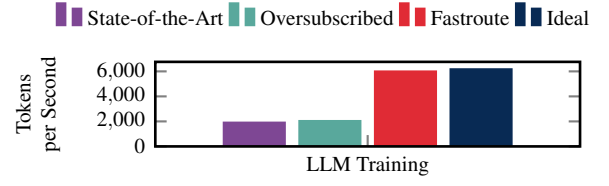

\begin{figure}
    \centering
    \input{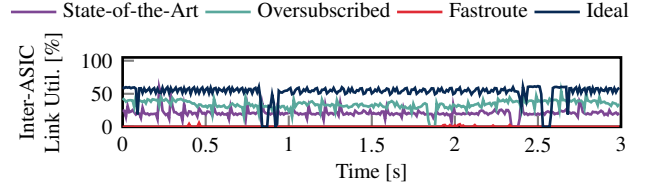}%
    \caption{The inter-ASIC link utilization for \Lightpass/ is reduced to almost zero, adapting to new traffic almost instantly.}
    \label{fig:prototype-bw}
\end{figure}

\subsubsection{Testbed Evaluation: LLM Training} % on the Hardware Prototype}
\label{subsec:hardwareprototype}

To practically evaluate \Lightpass/, we run an LLM training job on our hardware testbed described in \Cref{subsec:hardwaresetup}.
We use PyTorch FSDP to fine-tune the 3-billion-parameter Llama 3.2 model on 8 RTX 4000 Ada GPUs.
Each GPU is attached to the network via a dedicated 100G ConnectX-6 interface.
To create inter-ASIC traffic, we assign ranks such that the next rank ID is always on the other ASIC.
To compare the different topologies, we adjust the number of inter-ASIC links and their corresponding speeds on the hardware prototype accordingly.
We run the LLM training job for a few iterations to capture multiple collective operations.
\Cref{fig:prototype} shows the average training throughput (tokens/sec) during the run.
The performance of \Lightpass/ closely matches that of the \Ideal/ architecture, validating our simulation results. 
Compared to the \Baseline/ and \Static/ architectures, \Lightpass/ has a $3\times$ higher token throughput.

\Cref{fig:prototype-bw} shows the inter-ASIC link utilization measured in $10\,ms$ intervals during the LLM training.
We can clearly see that the indirection layer of \Lightpass/ is working as intended, since the link utilization is almost zero while the performance is on par with the \Ideal/ architecture.
The indirection layer is updated every $500\,\mu s$ and does not cause any noticeable reordering issues.
For the \Static/ and \Baseline/ architectures, the low link utilization might look a bit surprising, as it does not appear congested.
Looking at shorter timescales ($<1\,ms$) reveals microbursts that fully utilize the inter-ASIC links for a short time before congestion control drastically reduces the sending rate.
This results in the lower-than-expected average utilization at the millisecond timescale, with congestion only visible at the microsecond timescale.

\subsection{Datacenter Traffic}
\label{subsec:performance}

We generate flow-level Web and Hadoop traffic traces having inter-rack volume, inter-arrival time, and flow size distribution obtained from the Facebook datacenter~\cite{roy2015Social}. 
For both traffic scenarios, the flows are mostly inter-rack and small in size. For the Web trace, the inter-rack traffic is $82\,\%$ with flow sizes mostly below $10\,KB$. 
For the Hadoop trace, the inter-rack traffic is $86\,\%$; apart from small flows, it also includes a small number of large intra-rack flows. 
We scale up the flow inter-arrival time to generate different network loads.
In this section, we consider 2 ASICs with an oversubscription ratio of $2{:}1$. 
For the evaluation of the $n$-ASIC scenario, see~\Cref{subsec:n-asic-case}.

\def\addlegend{11}
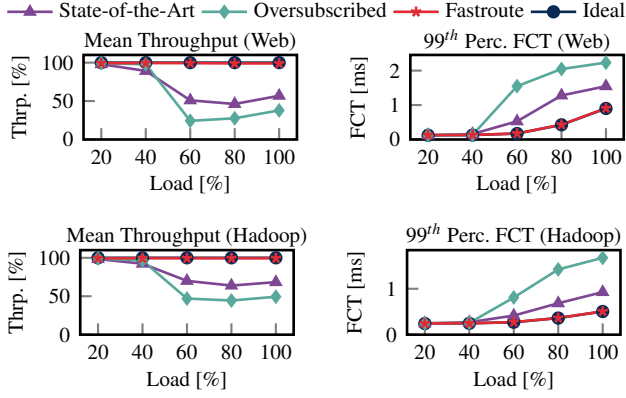
\begin{figure}
    \begin{subfigure}{\linewidth}
        \centering
        \begin{tikzpicture}
\pgfplotsset{every axis/.append style={font=\large, line width=1pt, tick style={line width=0.8pt}},width=0.52\linewidth}
\begin{axis}[ 
 title= Mean Throughput (Web), ylabel style={at={(axis description cs:-0.2,.5)}, anchor=south, align=center}, ylabel={\footnotesize Thrp. [\%]}, scaled ticks=false, ticklabel style={/pgf/number format/fixed,/pgf/number format/.cd, 1000 sep = {}}, tick pos=left,ymin=0,axis on top,xlabel={Load [\%]},name=first plot, legend style={at={(1.1,1.26)}, draw=none, anchor=south}, legend columns=-1, xlabel style={yshift=1mm}] 
\addplot [myblue, mark=*, forget plot] table { 
 
"Load [%]" "Mean Throughput [%]"

20 100.0
40 100.0
60 100.0
80 100.0
100 100.0
};
\addplot [myviolet, mark=triangle*, forget plot] table { 
 
"Load [%]" "Mean Throughput [%]"

20 97.99838627920612
40 89.21752839374784
60 50.96464678515738
80 46.20159313578303
100 56.911079065274286
};
\addplot [mygreen, mark=diamond*, forget plot] table { 
 
"Load [%]" "Mean Throughput [%]"

20 99.2652592892513
40 97.81529792134117
60 24.29328765100239
80 27.609032002068446
100 37.88225803839543
};
\addplot [myred, mark=star, forget plot] table { 
 
"Load [%]" "Mean Throughput [%]"

20 99.33064997684372
40 99.45468778021235
60 99.36063445035265
80 98.98655303359675
100 98.94247475978975
};
\addlegendimage{myviolet, mark=triangle*}
\addlegendimage{mygreen, mark=diamond*}
\addlegendimage{myred, mark=star}
\addlegendimage{myblue, mark=*}
\if\addlegend 
 \legend{\BaselinePlot/,\StaticPlot/,\LightpassPlot/,\IdealPlot/}
 \fi 
\end{axis}
\begin{axis}[ 
 title= $99^{th}$ Perc. FCT (Web), ylabel style={at={(axis description cs:-0.15,.5)}, anchor=south, align=center}, ylabel={\footnotesize FCT [ms]}, xlabel={Load [\%]}, scaled ticks=false,ytick distance=1,ticklabel style={/pgf/number format/fixed,/pgf/number format/.cd, 1000 sep = {}}, tick pos=left,ymin=0,axis on top,at=(first plot.east), anchor=west, xshift=15mm, xlabel style={yshift=1mm}] 
\addplot [myblue, mark=*, forget plot] table { 
 
"Load [%]" "All 99th FCT [ms]"

20 0.122524
40 0.128449
60 0.17228700999999977
80 0.426354
100 0.9009540499999988
};
\addplot [myviolet, mark=triangle*, forget plot] table { 
 
"Load [%]" "All 99th FCT [ms]"

20 0.125483
40 0.15986500999999978
60 0.528031
80 1.2786790199999996
100 1.5470351399999969
};
\addplot [mygreen, mark=diamond*, forget plot] table { 
 
"Load [%]" "All 99th FCT [ms]"

20 0.123534
40 0.132549
60 1.5475960299999993
80 2.0468070699999985
100 2.2315350199999995
};
\addplot [myred, mark=star, forget plot] table { 
 
"Load [%]" "All 99th FCT [ms]"

20 0.12542805999999865
40 0.1330650099999998
60 0.17417300999999977
80 0.4300781099999975
100 0.9077750199999995
};
\end{axis}
\end{tikzpicture}%
        \label{fig:load_web}
    \end{subfigure}
    \begin{subfigure}{\linewidth}
        \vspace{-3mm}
        \centering
        \def\addlegend{0}
        \begin{tikzpicture}
\pgfplotsset{every axis/.append style={font=\large, line width=1pt, tick style={line width=0.8pt}},width=0.52\linewidth}
\begin{axis}[ 
 title= Mean Throughput (Hadoop), ylabel style={at={(axis description cs:-0.2,.5)}, anchor=south, align=center}, ylabel={\footnotesize Thrp. [\%]}, scaled ticks=false, ticklabel style={/pgf/number format/fixed,/pgf/number format/.cd, 1000 sep = {}}, tick pos=left,ymin=0,axis on top,xlabel={Load [\%]},name=first plot, legend style={at={(1.1,1.26)}, draw=none, anchor=south}, legend columns=-1, xlabel style={yshift=1mm}] 
\addplot [myblue, mark=*, forget plot] table { 
 
"Load [%]" "Mean Throughput [%]"

20 100.0
40 100.0
60 100.0
80 100.0
100 100.0
};
\addplot [myviolet, mark=triangle*, forget plot] table { 
 
"Load [%]" "Mean Throughput [%]"

20 98.39396403735319
40 92.34859798637758
60 70.0104228761231
80 63.896003225465
100 68.35276954455392
};
\addplot [mygreen, mark=diamond*, forget plot] table { 
 
"Load [%]" "Mean Throughput [%]"

20 99.21570980228418
40 96.48362728521269
60 46.98246509067967
80 44.460391725320605
100 49.34989652739002
};
\addplot [myred, mark=star, forget plot] table { 
 
"Load [%]" "Mean Throughput [%]"

20 99.30353063606367
40 99.53316442541173
60 99.46304866915422
80 99.42234846453132
100 99.48510578036401
};
\addlegendimage{myviolet, mark=triangle*}
\addlegendimage{mygreen, mark=diamond*}
\addlegendimage{myred, mark=star}
\addlegendimage{myblue, mark=*}
\if\addlegend 
 \legend{\BaselinePlot/,\StaticPlot/,\LightpassPlot/,\IdealPlot/}
 \fi 
\end{axis}
\begin{axis}[ 
 title= $99^{th}$ Perc. FCT (Hadoop), ylabel style={at={(axis description cs:-0.15,.5)}, anchor=south, align=center}, ylabel={\footnotesize FCT [ms]}, xlabel={Load [\%]}, scaled ticks=false,ytick distance=1,ticklabel style={/pgf/number format/fixed,/pgf/number format/.cd, 1000 sep = {}}, tick pos=left,ymin=0,axis on top,at=(first plot.east), anchor=west, xshift=15mm, xlabel style={yshift=1mm}] 
\addplot [myblue, mark=*, forget plot] table { 
 
"Load [%]" "All 99th FCT [ms]"

20 0.24324301999999956
40 0.24682902999999934
60 0.2712800099999998
80 0.3614870299999993
100 0.5045830199999995
};
\addplot [myviolet, mark=triangle*, forget plot] table { 
 
"Load [%]" "All 99th FCT [ms]"

20 0.24728700999999978
40 0.27128001999999957
60 0.4158070199999995
80 0.684348
100 0.9276360199999996
};
\addplot [mygreen, mark=diamond*, forget plot] table { 
 
"Load [%]" "All 99th FCT [ms]"

20 0.245268
40 0.25617100999999975
60 0.8094950799999983
80 1.421185049999999
100 1.6739700599999987
};
\addplot [myred, mark=star, forget plot] table { 
 
"Load [%]" "All 99th FCT [ms]"

20 0.24510501999999956
40 0.248444
60 0.273496
80 0.3640820399999991
100 0.5083510299999994
};
\end{axis}
\end{tikzpicture}%
        \label{fig:load_hadoop}
    \end{subfigure}
    \caption{For network load above $60\,\%$, the limited inter-ASIC bandwidth becomes the bottleneck.}
    \label{fig:load}
\end{figure}

\subsubsection{Impact of Network Load}
\label{subsec:networkload}
\Cref{fig:load} shows the mean throughput and $99^{th}$ percentile FCTs of the various architectures across different network loads for Web and Hadoop traces.
In both scenarios, we observe that the \Static/ architecture is not able to handle loads above $60\,\%$ due to congestion at the inter-ASIC link; performance degrades significantly at higher loads, e.g., the throughput is reduced by more than half.
The \Baseline/ does not suffer from congestion at the inter-ASIC link, but is limited by the reduction in usable capacity as more of it is allocated to the inter-ASIC links.
\Lightpass/, on the other hand, closely matches the performance of the \Ideal/ architecture, across the whole load range. 
There is a small throughput penalty for \Lightpass/, i.e., around $1\,\%$ compared to the \Ideal/, primarily due to circuit reconfiguration downtime and increased latency when traversing inter-ASIC links. 
For the rest of the evaluation, we use a high network load ($80\,\%$) sufficient to create an inter-ASIC bottleneck.

\begin{figure}
        \centering
        \begin{tikzpicture}
\pgfplotsset{every axis/.append style={font=\large, line width=1pt, tick style={line width=0.8pt}},width=0.52\linewidth}
\begin{axis}[ 
 title= Mean Throughput (Web), ylabel style={at={(axis description cs:-0.2,.5)}, anchor=south, align=center}, ylabel={\footnotesize Thrp. [\%]}, scaled ticks=false, ticklabel style={/pgf/number format/fixed,/pgf/number format/.cd, 1000 sep = {}}, tick pos=left,ymin=0,axis on top,xlabel={Intra-Rack [\%]},name=first plot, legend style={at={(1.1,1.26)}, draw=none, anchor=south}, legend columns=-1, xlabel style={yshift=1mm}] 
\addplot [myblue, mark=*, forget plot] table { 
 
"Intra-Rack [%]" "Mean Throughput [%]"

0 100.0
25 100.0
50 100.0
75 100.0
100 100.0
};
\addplot [myviolet, mark=triangle*, forget plot] table { 
 
"Intra-Rack [%]" "Mean Throughput [%]"

0 47.3497677589819
25 47.36505925773009
50 55.477509081988295
75 62.18220567499135
100 64.1666389115614
};
\addplot [mygreen, mark=diamond*, forget plot] table { 
 
"Intra-Rack [%]" "Mean Throughput [%]"

0 23.733102131985053
25 31.554350580820195
50 89.0382591989631
75 99.90764392270552
100 100.0
};
\addplot [myred, mark=star, forget plot] table { 
 
"Intra-Rack [%]" "Mean Throughput [%]"

0 97.85420964059487
25 99.28328482903522
50 99.92185086877208
75 99.34261071518168
100 100.0
};
\addlegendimage{myviolet, mark=triangle*}
\addlegendimage{mygreen, mark=diamond*}
\addlegendimage{myred, mark=star}
\addlegendimage{myblue, mark=*}
\if\addlegend 
 \legend{\BaselinePlot/,\StaticPlot/,\LightpassPlot/,\IdealPlot/}
 \fi 
\end{axis}
\begin{axis}[ 
 title= $99^{th}$ Perc. FCT (Web), ylabel style={at={(axis description cs:-0.15,.5)}, anchor=south, align=center}, ylabel={\footnotesize FCT [ms]}, xlabel={Intra-Rack [\%]}, scaled ticks=false,ytick distance=1,ticklabel style={/pgf/number format/fixed,/pgf/number format/.cd, 1000 sep = {}}, tick pos=left,ymin=0,axis on top,at=(first plot.east), anchor=west, xshift=15mm, xlabel style={yshift=1mm}] 
\addplot [myblue, mark=*, forget plot] table { 
 
"Intra-Rack [%]" "All 99th FCT [ms]"

0 0.6247880499999989
25 0.380857
50 0.3132280099999998
75 0.287898
100 0.2558650399999991
};
\addplot [myviolet, mark=triangle*, forget plot] table { 
 
"Intra-Rack [%]" "All 99th FCT [ms]"

0 1.5347113099999932
25 1.174732129999997
50 0.8424364099999908
75 0.6198310699999985
100 0.4902780599999987
};
\addplot [mygreen, mark=diamond*, forget plot] table { 
 
"Intra-Rack [%]" "All 99th FCT [ms]"

0 2.6920300699999986
25 1.761623
50 0.36717210999999755
75 0.28819901999999953
100 0.2558650399999991
};
\addplot [myred, mark=star, forget plot] table { 
 
"Intra-Rack [%]" "All 99th FCT [ms]"

0 0.6384510499999989
25 0.3843970799999982
50 0.31132800999999977
75 0.2894920099999998
100 0.2558650399999991
};
\end{axis}
\end{tikzpicture}%
        \caption{For high intra-rack traffic, the switch is underutilized due to the uplink ports being idle.}
        \label{fig:inrack_web}
\end{figure}
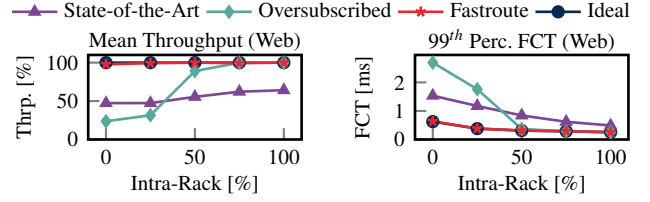

\subsubsection{Impact of Intra-Rack Traffic}
\label{subsec:inrack_traffic}
Measurements from Meta’s datacenters show that both scenarios are dominated by inter-rack traffic, with intra-rack traffic around $18\,\%$ for Web and $14\,\%$ for Hadoop~\cite{roy2015Social}.
In this experiment, we synthesize traffic traces with varying intra-rack traffic fractions and observe their performance impact. \Cref{fig:inrack_web} shows the mean throughput and $99^{th}$ percentile FCTs of different switch architectures for Web traffic across different intra-rack traffic fractions.
As observed, the \Static/ architecture performs worse at low intra-rack traffic; i.e., the inter-ASIC links are not a bottleneck when intra-rack traffic is higher.
The underlying reason is that the ToR uplinks towards the spine layer are underutilized in those scenarios. 
However, the performance gap between \Lightpass/ and \Ideal/ is small even at very low intra-rack traffic, demonstrating the design's efficacy.
The \Baseline/ struggles across the entire range equally, since the non-oversubscribed inter-ASIC link makes it independent of traffic patterns but reduces overall available capacity.

\subsection{\Lightpass/ with \texorpdfstring{$n$-ASICs}{Lg}}
\label{subsec:n-asic-case}
Increasing the number of ASICs might be unavoidable to meet performance targets, where \Lightpass/ can reduce the overhead associated with multi-ASIC designs.
While the indirection layer's efficiency is reduced, \Lightpass/ still performs better than the \Baseline/.
\lukas{need to adapt this to show that targeting higher than 4 ASIC is not really the point}
\lukas{rewrite this part}

\def\addlegend{11}
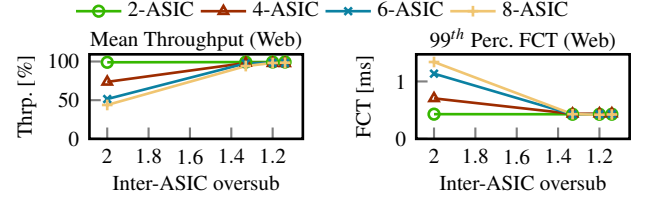
\begin{figure}
    \centering
    \begin{tikzpicture}
\pgfplotsset{every axis/.append style={font=\large, line width=1pt, tick style={line width=0.8pt}},width=0.52\linewidth}
\begin{axis}[ 
 title= Mean Throughput (Web), x dir=reverse,ylabel style={at={(axis description cs:-0.2,.5)}, anchor=south, align=center}, ylabel={\footnotesize Thrp. [\%]}, scaled ticks=false, ticklabel style={/pgf/number format/fixed,/pgf/number format/.cd, 1000 sep = {}}, tick pos=left,ymin=0,axis on top,xlabel={Inter-ASIC oversub},name=first plot, legend style={at={(1.1,1.26)}, draw=none, anchor=south}, legend columns=-1, xlabel style={yshift=1mm}] 
\addplot [mylightgreen, mark=o, forget plot] table { 
 
"Inter-ASIC oversub" "Mean Throughput [%]"

2.0 98.98655303359675
1.33 99.24839173446772
1.2 99.33255830367838
1.14 99.34958323120466
};
\addplot [mydarkred, mark=triangle, forget plot] table { 
 
"Inter-ASIC oversub" "Mean Throughput [%]"

2.0 73.65994404448965
1.33 98.07348992139346
1.2 97.83741997835719
1.14 97.6391373029892
};
\addplot [mylightblue, mark=x, forget plot] table { 
 
"Inter-ASIC oversub" "Mean Throughput [%]"

2.0 51.48567437178252
1.33 97.55627591879782
1.2 98.17434485689716
1.14 98.0293335446316
};
\addplot [myyellow, mark=+, forget plot] table { 
 
"Inter-ASIC oversub" "Mean Throughput [%]"

2.0 43.808398075722266
1.33 94.09162708148986
1.2 98.12322661906563
1.14 98.44001348260886
};
\addlegendimage{mylightgreen, mark=o}
\addlegendimage{mydarkred, mark=triangle}
\addlegendimage{mylightblue, mark=x}
\addlegendimage{myyellow, mark=+}
\if\addlegend 
 \legend{2-ASIC,4-ASIC,6-ASIC,8-ASIC}
 \fi 
\end{axis}
\begin{axis}[ 
 title= $99^{th}$ Perc. FCT (Web), x dir=reverse,ylabel style={at={(axis description cs:-0.15,.5)}, anchor=south, align=center}, ylabel={\footnotesize FCT [ms]}, xlabel={Inter-ASIC oversub}, scaled ticks=false,ytick distance=1,ticklabel style={/pgf/number format/fixed,/pgf/number format/.cd, 1000 sep = {}}, tick pos=left,ymin=0,axis on top,at=(first plot.east), anchor=west, xshift=15mm, xlabel style={yshift=1mm}] 
\addplot [mylightgreen, mark=o, forget plot] table { 
 
"Inter-ASIC oversub" "All 99th FCT [ms]"

2.0 0.4300781099999975
1.33 0.42859206999999844
1.2 0.4282880399999991
1.14 0.4292740599999987
};
\addplot [mydarkred, mark=triangle, forget plot] table { 
 
"Inter-ASIC oversub" "All 99th FCT [ms]"

2.0 0.705165
1.33 0.43547100999999977
1.2 0.43865401999999953
1.14 0.4406780099999998
};
\addplot [mylightblue, mark=x, forget plot] table { 
 
"Inter-ASIC oversub" "All 99th FCT [ms]"

2.0 1.1355930399999992
1.33 0.42591601999999956
1.2 0.4286860099999998
1.14 0.432319
};
\addplot [myyellow, mark=+, forget plot] table { 
 
"Inter-ASIC oversub" "All 99th FCT [ms]"

2.0 1.3386000099999997
1.33 0.42980501999999954
1.2 0.42026106999999846
1.14 0.4243720199999996
};
\end{axis}
\end{tikzpicture}%
    \caption{Decreasing the oversubscription restores the single-ASIC-like performance of \Lightpass/. }
    \label{fig:n-asic_interos}
\end{figure}

\subsubsection{Oversubscription Ratio}
\label{subsec:oversubratio} 
With more ASICs, it becomes harder for the circuit switch to localize traffic.
In the worst-case scenario, an equal number of packets per port needs to be forwarded to each ASIC.
Due to the circuit-switched behaviour, traffic can only be redirected on a per-port basis, not per packet.
This means per port, we can only pick one ASIC to redirect traffic to, with everything else needing to cross between ASICs.
Hence, we can reduce the inter-ASIC bandwidth by at most $1 - (\frac{1}{\#ASIC})$. 
%\gf{connect the latter to oversubsciption ratio}
\Cref{fig:n-asic_interos} experimentally confirms this, where we observe that a higher number of ASICs needs a lower oversubscription to achieve high performance.
For the following experiments, we adapt the inter-ASIC oversubscription according to the formula.

\def\addlegend{11}
\begin{figure}
    \centering
    \begin{tikzpicture}
\pgfplotsset{every axis/.append style={font=\large, line width=1pt, tick style={line width=0.8pt}},width=0.52\linewidth}
\begin{axis}[ 
 title= Mean Throughput (Web), ylabel style={at={(axis description cs:-0.2,.5)}, anchor=south, align=center}, ylabel={\footnotesize Thrp. [\%]}, scaled ticks=false, symbolic x coords={2,4,6,8}, xtick=data,ticklabel style={/pgf/number format/fixed,/pgf/number format/.cd, 1000 sep = {}}, tick pos=left,ymin=0,axis on top,xlabel={No. ASIC},name=first plot, legend style={at={(1.1,1.26)}, draw=none, anchor=south}, legend columns=-1, xlabel style={yshift=1mm}] 
\addplot [myblue, mark=*, forget plot] table { 
 
"No. ASIC" "Mean Throughput [%]"

2 100.0
4 100.0
6 100.0
8 100.0
};
\addplot [myviolet, mark=triangle*, forget plot] table { 
 
"No. ASIC" "Mean Throughput [%]"

2 46.20159313578303
4 67.90936314127369
6 77.21664849937207
8 82.51824974974366
};
\addplot [mygreen, mark=diamond*, forget plot] table { 
 
"No. ASIC" "Mean Throughput [%]"

2 27.609032002068446
4 66.13223436731876
6 74.65392289107258
8 87.12814699280094
};
\addplot [myred, mark=star, forget plot] table { 
 
"No. ASIC" "Mean Throughput [%]"

2 98.98655303359675
4 98.07219568498527
6 98.17434485689716
8 98.4047159445231
};
\addlegendimage{myviolet, mark=triangle*}
\addlegendimage{mygreen, mark=diamond*}
\addlegendimage{myred, mark=star}
\addlegendimage{myblue, mark=*}
\if\addlegend 
 \legend{\BaselinePlot/,\StaticPlot/,\LightpassPlot/,\IdealPlot/}
 \fi 
\end{axis}
\begin{axis}[ 
 title= $99^{th}$ Perc. FCT (Web), ylabel style={at={(axis description cs:-0.15,.5)}, anchor=south, align=center}, ylabel={\footnotesize FCT [ms]}, xlabel={No. ASIC}, scaled ticks=false,ytick distance=1,symbolic x coords={2,4,6,8}, xtick=data,ticklabel style={/pgf/number format/fixed,/pgf/number format/.cd, 1000 sep = {}}, tick pos=left,ymin=0,axis on top,at=(first plot.east), anchor=west, xshift=15mm, xlabel style={yshift=1mm}] 
\addplot [myblue, mark=*, forget plot] table { 
 
"No. ASIC" "All 99th FCT [ms]"

2 0.426354
4 0.426354
6 0.426354
8 0.426354
};
\addplot [myviolet, mark=triangle*, forget plot] table { 
 
"No. ASIC" "All 99th FCT [ms]"

2 1.2786790199999996
4 0.811674
6 0.6680971599999964
8 0.5970840299999993
};
\addplot [mygreen, mark=diamond*, forget plot] table { 
 
"No. ASIC" "All 99th FCT [ms]"

2 2.0468070699999985
4 0.8728420099999997
6 0.6671340699999985
8 0.48882600999999976
};
\addplot [myred, mark=star, forget plot] table { 
 
"No. ASIC" "All 99th FCT [ms]"

2 0.4300781099999975
4 0.4361230799999982
6 0.4286860099999998
8 0.4242250799999982
};
\end{axis}
\end{tikzpicture}%
    \caption{Scaling to more than 2-ASICs is possible while keeping single-ASIC-like performance.}
    \label{fig:n-asic_dcn}
\end{figure}
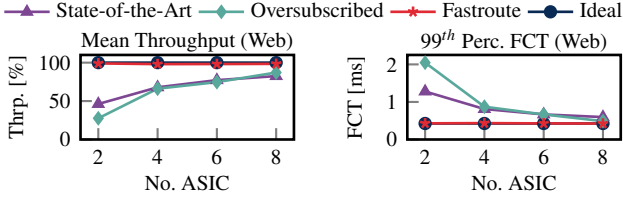

\subsubsection{DCN Traffic}
\label{subsec:DCN-traffic} 
\Cref{fig:n-asic_dcn} shows the mean throughput and tail FCT for the Web traffic traces for different numbers of ASICs.
We see that by adapting the inter-ASIC oversubscription, single-ASIC-like performance can be maintained even with 8 ASICs.
There \Lightpass/ is within $2\,\%$ of the throughput of the \Ideal/ architecture and within $1\,\%$ of the tail FCT.
Note that as the inter-ASIC oversubscription decreases with more ASICs (\Cref{fig:n-asic_interos}), both the \Baseline/ and \Static/ architectures also approach single-ASIC performance.

\subsubsection{ML Traffic}
\label{subsec:ML-traffic}
\Cref{fig:n-asic_ml} shows the mean throughput and tail FCT for ML traffic, where \Lightpass/'s performance is close to that of a single ASIC.
Overall, \Lightpass/ still works even at 8 ASICs across a diverse set of traffic scenarios, e.g., Web~(small flows, high entropy) and ML Ring All-Reduce~(large flows, low entropy).
The gap compared to more traditional designs narrows, indicating that \Lightpass/ is highly effective for the $2\mbox{ -- }4$ ASIC range.
Since integrating even 4 large-scale ASICs into a single package is a significant hurdle~\cite{2026NVIDIAs}, we believe that \Lightpass/ can cover the most relevant use cases.
%\Cref{fig:entropy} shows the ratio of traffic that needs to cross between ASICs for the 8-ASIC scenario.
%Because ML traffic has lower entropy, it is easier to localize, allowing inter-ASIC bandwidth to be reduced beyond the worst case.
%As a result, when workloads are known \emph{a priori} to have low entropy, an 8-ASIC switch can tolerate a higher inter-ASIC oversubscription while maintaining near-single-ASIC performance.

\def\addlegend{11}
\begin{figure}
    \centering
    \begin{tikzpicture}
\pgfplotsset{every axis/.append style={font=\large, line width=1pt, tick style={line width=0.8pt}},width=0.52\linewidth}
\begin{axis}[ 
 title= Mean Throughput (ML), ylabel style={at={(axis description cs:-0.2,.5)}, anchor=south, align=center}, ylabel={\footnotesize Thrp. [\%]}, scaled ticks=false, symbolic x coords={2,4,6,8}, xtick=data,ticklabel style={/pgf/number format/fixed,/pgf/number format/.cd, 1000 sep = {}}, tick pos=left,ymin=0,axis on top,xlabel={No. ASIC},name=first plot, legend style={at={(1.1,1.26)}, draw=none, anchor=south}, legend columns=-1, xlabel style={yshift=1mm}] 
\addplot [myblue, mark=*, forget plot] table { 
 
"No. ASIC" "Mean Throughput [%]"

2 100.0
4 100.0
6 100.0
8 100.0
};
\addplot [myviolet, mark=triangle*, forget plot] table { 
 
"No. ASIC" "Mean Throughput [%]"

2 65.9392058291405
4 82.49615723259399
6 88.12362336504899
8 91.10981584476332
};
\addplot [mygreen, mark=diamond*, forget plot] table { 
 
"No. ASIC" "Mean Throughput [%]"

2 53.97658870223787
4 90.12131725022485
6 89.61983607460726
8 91.6819364741201
};
\addplot [myred, mark=star, forget plot] table { 
 
"No. ASIC" "Mean Throughput [%]"

2 99.62443527389807
4 99.22824238439627
6 98.9102266858102
8 98.50591989332082
};
\addlegendimage{myviolet, mark=triangle*}
\addlegendimage{mygreen, mark=diamond*}
\addlegendimage{myred, mark=star}
\addlegendimage{myblue, mark=*}
\if\addlegend 
 \legend{\BaselinePlot/,\StaticPlot/,\LightpassPlot/,\IdealPlot/}
 \fi 
\end{axis}
\begin{axis}[ 
 title= $99^{th}$ Perc. FCT (ML), ylabel style={at={(axis description cs:-0.15,.5)}, anchor=south, align=center}, ylabel={\footnotesize FCT [ms]}, xlabel={No. ASIC}, scaled ticks=false,ytick distance=1,symbolic x coords={2,4,6,8}, xtick=data,ticklabel style={/pgf/number format/fixed,/pgf/number format/.cd, 1000 sep = {}}, tick pos=left,ymin=0,axis on top,at=(first plot.east), anchor=west, xshift=15mm, xlabel style={yshift=1mm}] 
\addplot [myblue, mark=*, forget plot] table { 
 
"No. ASIC" "All 99th FCT [ms]"

2 1.6351165
4 1.6351165
6 1.6351165
8 1.6351165
};
\addplot [myviolet, mark=triangle*, forget plot] table { 
 
"No. ASIC" "All 99th FCT [ms]"

2 2.5073815
4 1.98263975
6 1.857076
8 1.79802325
};
\addplot [mygreen, mark=diamond*, forget plot] table { 
 
"No. ASIC" "All 99th FCT [ms]"

2 2.46669475
4 1.71439975
6 1.7659255
8 1.6925205
};
\addplot [myred, mark=star, forget plot] table { 
 
"No. ASIC" "All 99th FCT [ms]"

2 1.6286915
4 1.63845925
6 1.64355025
8 1.65681075
};
\end{axis}
\end{tikzpicture}%
    \caption{\Lightpass/ can be scaled to more ASICs, at the cost of less oversubscription.}
    \label{fig:n-asic_ml}
\end{figure}
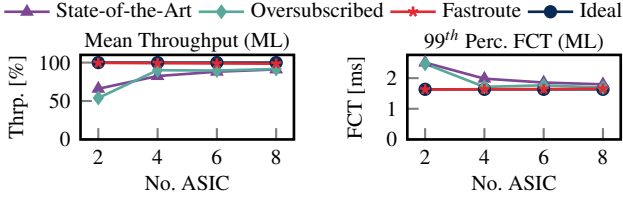

\subsubsection{Number of Circuit Switches}
\label{subsec:multiple-circuits}
With more ASICs and higher port counts, the question of building a super-high-radix circuit switch arises.
The natural solution is to split the monolithic circuit switch into multiple smaller radix circuit switches running in parallel; of course, this reduces flexibility.
\Cref{fig:n-asic-circuit} shows the mean throughput and $99^{th}$ percentile FCT for both the ML and Web traffic traces while using 8-ASICs.
We see that the performance delta between the number of independent circuit switches is $<1\,\%$.
\Lightpass/, therefore, operates well even with reduced reconfiguration flexibility by splitting the monolithic circuit switch into multiple smaller ones.

\if 0
\def\addlegend{11}
\begin{figure}
    \centering
    \input{plots/Entropy_TS.tikz}%
    \caption{Low entropy improves traffic localization.}
    \label{fig:entropy}
\end{figure}
\fi

\def\addlegend{11}
\begin{figure}
    \centering
    \begin{tikzpicture}
\pgfplotsset{every axis/.append style={font=\large,line width=1pt,tick style={line width=0.8pt}}, width=0.53\linewidth}
\begin{axis}[ 
 title=Mean Throughput, ylabel style={at={(axis description cs:-0.2,.5)},anchor=south,align=center},ylabel={\footnotesize Thrp. [\%]}, xtick pos=left, ybar,   axis on top, xtick=data, xtick align=inside, scaled ticks=false,symbolic x coords={Web,ML},ticklabel style={/pgf/number format/fixed,/pgf/number format/.cd, 1000 sep = {}},bar width=1.8mm, xmin={[normalized]-0.6}, xmax={[normalized]+1.6},ymin=0, name=first plot, legend style={at={(1.1,1.25)}, draw=none, anchor=south}, legend columns=-1] 
\addplot [mylightgreen, fill=mylightgreen] table { 
 
Scenario "Mean Throughput [%]"

Web 98.4047159445231
ML 98.50591989332082
};
\addplot [mydarkred, fill=mydarkred] table { 
 
Scenario "Mean Throughput [%]"

Web 98.35980710035906
ML 98.3693412796276
};
\addplot [mylightblue, fill=mylightblue] table { 
 
Scenario "Mean Throughput [%]"

Web 98.35191547271579
ML 98.32201672340251
};
\addplot [myyellow, fill=myyellow] table { 
 
Scenario "Mean Throughput [%]"

Web 98.252672259125
ML 98.19670570817507
};
\addlegendimage{mylightgreen, ybar, ybar legend, fill=mylightgreen}
\addlegendimage{mydarkred, ybar, ybar legend, fill=mydarkred}
\addlegendimage{mylightblue, ybar, ybar legend, fill=mylightblue}
\addlegendimage{myyellow, ybar, ybar legend, fill=myyellow}
\if\addlegend 
 \legend{1-CS,2-CS,4-CS,8-CS}
 \fi 
\end{axis}
\begin{axis}[ 
 title=$99^{th}$ Perc. FCT, ylabel style={at={(axis description cs:-0.15,.5)},anchor=south, align=center},ylabel={\footnotesize FCT [ms]}, xtick pos=left, ybar, axis on top, xtick=data, xtick align=inside, scaled ticks=false,symbolic x coords={Web,ML},ticklabel style={/pgf/number format/fixed,/pgf/number format/.cd, 1000 sep = {}},bar width=1.8mm,xmin={[normalized]-0.6}, xmax={[normalized]+1.6}, ymin=0, at=(first plot.east), anchor=west, xshift=14mm, xlabel style={yshift=1mm}, ytick distance=1] 
\addplot [mylightgreen, fill=mylightgreen] table { 
 
Scenario "All 99th FCT [ms]"

Web 0.4242250799999982
ML 1.65681075
};
\addplot [mydarkred, fill=mydarkred] table { 
 
Scenario "All 99th FCT [ms]"

Web 0.42425208999999797
ML 1.64912825
};
\addplot [mylightblue, fill=mylightblue] table { 
 
Scenario "All 99th FCT [ms]"

Web 0.42267806999999846
ML 1.64578025
};
\addplot [myyellow, fill=myyellow] table { 
 
Scenario "All 99th FCT [ms]"

Web 0.42390109999999775
ML 1.6471455
};
\end{axis}
\end{tikzpicture}%
    \caption{\Lightpass/ works even with reduced flexibility of multiple independent circuit switches.}
    \label{fig:n-asic-circuit}
\end{figure}
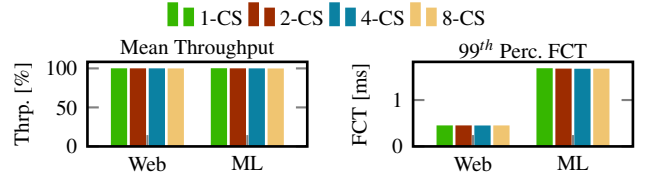

\subsection{Sensitivity Analysis}
\label{subsec:sensitivityanalysis}
We analyze the \Lightpass/ performance sensitivity to several parameters, e.g., traffic stability, buffer size, and reconfiguration overhead; also benchmark the control-plane efficiency. 

\subsubsection{Impact of Traffic Stability}
\label{subsec:dynamic_patterns}
To evaluate the robustness of \Lightpass/ w.r.t. the traffic stability, we synthesize a traffic scenario (\textbf{Fast}) that presents the worst-case scenario, consisting of two periodic phases, i.e., a) fully inter-rack in the first phase, and b) fully intra-rack in the second phase. 
We adjust the interval between these two phases to vary traffic stability.
\Cref{fig:fast} shows the mean throughput and $99^{th}$ FCTs of different switch architectures across a range of traffic phase-switching intervals, while keeping the \Lightpass/ circuit reconfiguration interval as $100\,\mu s$. 
The insight is that \Lightpass/ closely matches the performance of the \Ideal/ if it can adapt at least $4\times$ faster than the rate of worst-case traffic variability. 
We also observe that, when the traffic phase-switching interval is faster than $200\,\mu s$ (i.e., within $2\times$ of the reconfiguration interval), the performance of \Lightpass/ starts degrading gradually, but still outperforms both the \Static/ and \Baseline/ architectures over the whole range.
The effectiveness of \Lightpass/ depends on the relative frequency of reconfiguration vs. traffic variability.
Luckily, DCN workloads are reasonably stable over short time scales~\cite {benson2011MicroTE,roy2015Social}, making \Lightpass/ perform well in practice (\Cref{subsec:performance}).
We show this in our evaluation, which includes highly bursty traffic across a wide range of workloads, including Web/Hadoop and ML.

\def\addlegend{11}
\begin{figure}
        \centering
        \begin{tikzpicture}
\pgfplotsset{every axis/.append style={font=\large, line width=1pt, tick style={line width=0.8pt}},width=0.52\linewidth}
\begin{axis}[ 
 title= Mean Throughput (Fast), ylabel style={at={(axis description cs:-0.2,.5)}, anchor=south, align=center}, ylabel={\footnotesize Thrp. [\%]}, scaled ticks=false, ticklabel style={/pgf/number format/fixed,/pgf/number format/.cd, 1000 sep = {}}, tick pos=left,ymin=0,axis on top,xlabel={Traffic Phase Interval [$\mu$s]},name=first plot, legend style={at={(1.1,1.26)}, draw=none, anchor=south}, legend columns=-1, xlabel style={yshift=1mm}] 
\addplot [myblue, mark=*, forget plot] table { 
 
"Traffic Phase Interval [us]" "Mean Throughput [%]"

50 100.0
100 100.0
150 100.0
200 100.0
400 100.0
800 100.0
};
\addplot [myviolet, mark=triangle*, forget plot] table { 
 
"Traffic Phase Interval [us]" "Mean Throughput [%]"

50 65.65504521428791
100 46.46343544856247
150 45.307444702031304
200 45.39003961786593
400 48.73185278060491
800 54.13412573217609
};
\addplot [mygreen, mark=diamond*, forget plot] table { 
 
"Traffic Phase Interval [us]" "Mean Throughput [%]"

50 65.68628525328322
100 53.415799724891585
150 57.9175768250139
200 56.12154316907502
400 50.08105195548508
800 47.59227420757096
};
\addplot [myred, mark=star, forget plot] table { 
 
"Traffic Phase Interval [us]" "Mean Throughput [%]"

50 69.73674871816557
100 61.08842268858298
150 64.73946379474705
200 69.96940464475995
400 89.43881863306106
800 95.48281342841344
};
\addlegendimage{myviolet, mark=triangle*}
\addlegendimage{mygreen, mark=diamond*}
\addlegendimage{myred, mark=star}
\addlegendimage{myblue, mark=*}
\if\addlegend 
 \legend{\BaselinePlot/,\StaticPlot/,\LightpassPlot/,\IdealPlot/}
 \fi 
\end{axis}
\begin{axis}[ 
 title= $99^{th}$ Perc. FCT (Fast), ylabel style={at={(axis description cs:-0.15,.5)}, anchor=south, align=center}, ylabel={\footnotesize FCT [ms]}, xlabel={Traffic Phase Interval [$\mu$s]}, scaled ticks=false,ytick distance=1,ticklabel style={/pgf/number format/fixed,/pgf/number format/.cd, 1000 sep = {}}, tick pos=left,ymin=0,axis on top,at=(first plot.east), anchor=west, xshift=15mm, xlabel style={yshift=1mm}] 
\addplot [myblue, mark=*, forget plot] table { 
 
"Traffic Phase Interval [us]" "All 99th FCT [ms]"

50 0.081515
100 0.16052675
150 0.19248356999999983
200 0.24438878000000003
400 0.43104942
800 0.8312670199999997
};
\addplot [myviolet, mark=triangle*, forget plot] table { 
 
"Traffic Phase Interval [us]" "All 99th FCT [ms]"

50 0.139168
100 0.378994
150 0.51849492
200 0.6430201100000007
400 1.00818384
800 1.67139205
};
\addplot [mygreen, mark=diamond*, forget plot] table { 
 
"Traffic Phase Interval [us]" "All 99th FCT [ms]"

50 0.113779
100 0.3330935
150 0.42815273
200 0.53512613
400 0.91920571
800 1.6640429899999991
};
\addplot [myred, mark=star, forget plot] table { 
 
"Traffic Phase Interval [us]" "All 99th FCT [ms]"

50 0.09231448999999999
100 0.198656
150 0.26816046
200 0.29905304000000005
400 0.436739
800 0.8388932399999991
};
\end{axis}
\end{tikzpicture}%
        \caption{\Lightpass/ remains robust as long as the reconfiguration is $3$–$4\times$ faster than large traffic matrix changes.}
        \label{fig:fast}
\end{figure}
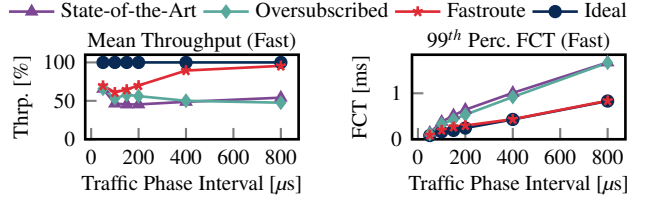

\subsubsection{Impact of Inter-ASIC Buffer Size}
\label{subsec:buffersize}
\Cref{fig:buffer} shows the mean throughput and $99^{th}$ percentile FCTs for Hadoop traffic, while varying the buffer size of the inter-ASIC links.
As observed, \Lightpass/ and \Baseline/ performance are mostly agnostic to buffer size, whereas the \Static/ architecture is impacted as the buffer builds up due to high inter-ASIC congestion. 
A smaller buffer size slightly improves the mean throughput but comes at the expense of a higher tail FCT. 
Dropping packets for longer flows degrades performance by forcing the congestion control algorithm to reduce the sending rate, thereby increasing tail latency.

\def\addlegend{11}
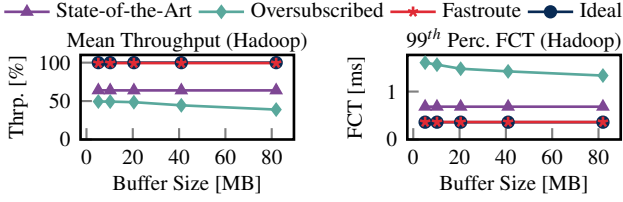
\begin{figure}
    \centering
    \begin{tikzpicture}
\pgfplotsset{every axis/.append style={font=\large, line width=1pt, tick style={line width=0.8pt}},width=0.52\linewidth}
\begin{axis}[ 
 title= Mean Throughput (Hadoop), ylabel style={at={(axis description cs:-0.2,.5)}, anchor=south, align=center}, ylabel={\footnotesize Thrp. [\%]}, scaled ticks=false, ticklabel style={/pgf/number format/fixed,/pgf/number format/.cd, 1000 sep = {}}, tick pos=left,ymin=0,axis on top,xlabel={Buffer Size [MB]},name=first plot, legend style={at={(1.1,1.26)}, draw=none, anchor=south}, legend columns=-1, xlabel style={yshift=1mm}] 
\addplot [myblue, mark=*, forget plot] table { 
 
"Buffer Size [MB]" "Mean Throughput [%]"

5.12 100.0
10.24 100.0
20.48 100.0
40.96 100.0
81.92 100.0
};
\addplot [myviolet, mark=triangle*, forget plot] table { 
 
"Buffer Size [MB]" "Mean Throughput [%]"

5.12 63.994195515397955
10.24 63.91432319753329
20.48 63.896003225465
40.96 63.896003225465
81.92 63.896003225465
};
\addplot [mygreen, mark=diamond*, forget plot] table { 
 
"Buffer Size [MB]" "Mean Throughput [%]"

5.12 49.38000120844325
10.24 49.31281182576589
20.48 48.53866048078342
40.96 44.460391725320605
81.92 38.91369782486962
};
\addplot [myred, mark=star, forget plot] table { 
 
"Buffer Size [MB]" "Mean Throughput [%]"

5.12 99.40789084605194
10.24 99.42234846453132
20.48 99.42234846453132
40.96 99.42234846453132
81.92 99.42234846453132
};
\addlegendimage{myviolet, mark=triangle*}
\addlegendimage{mygreen, mark=diamond*}
\addlegendimage{myred, mark=star}
\addlegendimage{myblue, mark=*}
\if\addlegend 
 \legend{\BaselinePlot/,\StaticPlot/,\LightpassPlot/,\IdealPlot/}
 \fi 
\end{axis}
\begin{axis}[ 
 title= $99^{th}$ Perc. FCT (Hadoop), ylabel style={at={(axis description cs:-0.15,.5)}, anchor=south, align=center}, ylabel={\footnotesize FCT [ms]}, xlabel={Buffer Size [MB]}, scaled ticks=false,ytick distance=1,ticklabel style={/pgf/number format/fixed,/pgf/number format/.cd, 1000 sep = {}}, tick pos=left,ymin=0,axis on top,at=(first plot.east), anchor=west, xshift=15mm, xlabel style={yshift=1mm}] 
\addplot [myblue, mark=*, forget plot] table { 
 
"Buffer Size [MB]" "All 99th FCT [ms]"

5.12 0.3614870299999993
10.24 0.3614870299999993
20.48 0.3614870299999993
40.96 0.3614870299999993
81.92 0.3614870299999993
};
\addplot [myviolet, mark=triangle*, forget plot] table { 
 
"Buffer Size [MB]" "All 99th FCT [ms]"

5.12 0.684871089999998
10.24 0.6845770999999977
20.48 0.684348
40.96 0.684348
81.92 0.684348
};
\addplot [mygreen, mark=diamond*, forget plot] table { 
 
"Buffer Size [MB]" "All 99th FCT [ms]"

5.12 1.6035580399999991
10.24 1.5609490099999999
20.48 1.4762591499999966
40.96 1.421185049999999
81.92 1.334040399999991
};
\addplot [myred, mark=star, forget plot] table { 
 
"Buffer Size [MB]" "All 99th FCT [ms]"

5.12 0.364647
10.24 0.3640820399999991
20.48 0.3640820399999991
40.96 0.3640820399999991
81.92 0.3640820399999991
};
\end{axis}
\end{tikzpicture}%
    \caption{Smaller buffers slightly improve the \Static/ architecture performance due to less buffer buildup.}
    \label{fig:buffer}
\end{figure}

\subsubsection{Impact of Reconfiguration Overhead}
\label{subsec:reconfigurationinterval}
\Cref{fig:reconf} profiles the mean throughput and $99^{th}$ percentile FCT performance of \Lightpass/ w.r.t. the ``reconfiguration overhead'' defined as the ratio of circuit switch downtime and the reconfiguration interval. 
For example, if the downtime is $1\,\mu s$ and the reconfiguration interval is $100\,\mu s$ (\Cref{tab:default-values}), the overhead is $1\,\%$; leading up to $1\,\%$ bandwidth wastage on average. 
In this experiment, we tune this overhead by only varying the reconfiguration downtime. 
As observed, the performance degrades at higher reconfiguration overhead. 
However, the performance impact is negligible when the overhead is around $1\,\%$.
Given more stable traffic~\cite {benson2011MicroTE,roy2015Social}, this overhead is further reduced.

\def\addlegend{11}
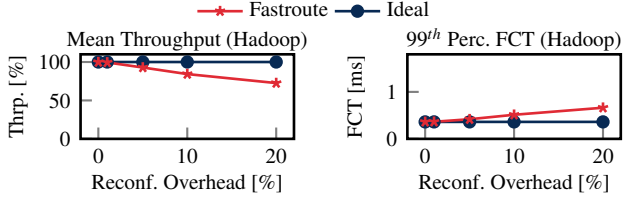
\begin{figure}
        \centering
        \begin{tikzpicture}
\pgfplotsset{every axis/.append style={font=\large, line width=1pt, tick style={line width=0.8pt}},width=0.52\linewidth}
\begin{axis}[ 
 title= Mean Throughput (Hadoop), ylabel style={at={(axis description cs:-0.2,.5)}, anchor=south, align=center}, ylabel={\footnotesize Thrp. [\%]}, scaled ticks=false, ticklabel style={/pgf/number format/fixed,/pgf/number format/.cd, 1000 sep = {}}, tick pos=left,ymin=0,axis on top,xlabel={Reconf. Overhead [\%]},name=first plot, legend style={at={(1.1,1.26)}, draw=none, anchor=south}, legend columns=-1, xlabel style={yshift=1mm}] 
\addplot [myblue, mark=*, forget plot] table { 
 
"Reconf. Overhead [%]" "Mean Throughput [%]"

0.0 100.0
1.0 100.0
5.0 100.0
10.0 100.0
20.0 100.0
};
\addplot [myred, mark=star, forget plot] table { 
 
"Reconf. Overhead [%]" "Mean Throughput [%]"

0.0 99.88990796414059
1.0 99.42234846453132
5.0 92.84221158888984
10.0 84.29445501235737
20.0 72.60504666467867
};
\addlegendimage{myred, mark=star}
\addlegendimage{myblue, mark=*}
\if\addlegend 
 \legend{\LightpassPlot/,\IdealPlot/}
 \fi 
\end{axis}
\begin{axis}[ 
 title= $99^{th}$ Perc. FCT (Hadoop), ylabel style={at={(axis description cs:-0.15,.5)}, anchor=south, align=center}, ylabel={\footnotesize FCT [ms]}, xlabel={Reconf. Overhead [\%]}, scaled ticks=false,ytick distance=1,ymax=1.8,ticklabel style={/pgf/number format/fixed,/pgf/number format/.cd, 1000 sep = {}}, tick pos=left,ymin=0,axis on top,at=(first plot.east), anchor=west, xshift=15mm, xlabel style={yshift=1mm}] 
\addplot [myblue, mark=*, forget plot] table { 
 
"Reconf. Overhead [%]" "All 99th FCT [ms]"

0.0 0.3614870299999993
1.0 0.3614870299999993
5.0 0.3614870299999993
10.0 0.3614870299999993
20.0 0.3614870299999993
};
\addplot [myred, mark=star, forget plot] table { 
 
"Reconf. Overhead [%]" "All 99th FCT [ms]"

0.0 0.361569
1.0 0.3640820399999991
5.0 0.418218
10.0 0.512762
20.0 0.6628610199999996
};
\end{axis}
\end{tikzpicture}%
        \caption{If the reconfiguration overhead is high, then too much bandwidth is lost during the downtime.}
        \label{fig:reconf}
\end{figure}

\subsubsection{Control Plane Efficiency}
\label{subsec:controlplaneperformance}
\begin{figure}
    \centering
    \begin{tikzpicture}
\pgfplotsset{every axis/.append style={font=\large,line width=1pt,tick style={line width=0.8pt}},height=25mm}
\begin{axis}[ 
 title=,ylabel style={at={(axis description cs:-0.1,.5)},anchor=south,align=center}, tick pos=left, ylabel={\footnotesize Runtime [$\mu$s]}, xlabel={\# Packet Switch Ports}, scaled ticks=false,ticklabel style={/pgf/number format/fixed,/pgf/number format/.cd, 1000 sep = {}},ymin=-100, ymax=1000,axis on top, name=first plot, legend style={at={(0.45,1.1)},draw=none,anchor=south},legend columns=-1, xlabel style={yshift=1mm}] 
\addplot [mylightgreen,mark=*] table { 
 
Ports "Time [us]"

512 2128.811429
480 1822.088982
448 1561.589355
416 1335.452084
384 1127.269162
352 922.838604
320 750.945568
288 601.626727
256 472.09223599999996
224 353.314013
192 257.724056
160 178.06045600000002
128 111.29206500000001
96 61.643663999999994
64 27.603562
};
\addplot [mydarkred,mark=*] table { 
 
Ports "Time [us]"

512 2967.7850630000003
480 2557.67844
448 2171.5446699999998
416 1828.781636
384 1513.016089
352 1243.522439
320 1016.884581
288 807.341996
256 624.015069
224 466.634372
192 341.171719
160 232.52425399999998
128 144.309236
96 82.41589599999999
64 38.26601
};
\addplot [mylightblue,mark=*] table { 
 
Ports "Time [us]"

512 3551.8102949999998
480 3012.499446
448 2568.02501
416 2168.679281
384 1824.480037
352 1536.246206
320 1215.17805
288 972.455708
256 745.744042
224 571.157153
192 415.945072
160 285.475306
128 180.379482
96 101.691642
64 48.92482
};
\addplot [mylightgreen,mark=x] table { 
 
Ports "Time [us]"

512 26.966873
480 25.377641
448 23.558358000000002
416 21.996596
384 20.139525000000003
352 18.741952
320 17.243809000000002
288 15.609706000000001
256 13.883294
224 12.052634
192 9.856334
160 8.52978
128 6.001991
96 5.828282
64 5.766323000000001
};
\addplot [mydarkred,mark=x] table { 
 
Ports "Time [us]"

512 59.956807999999995
480 55.716817000000006
448 51.268497
416 47.326616
384 43.249365
352 39.393964
320 35.501112
288 31.822049
256 28.088967
224 24.190226
192 20.840791
160 17.550276999999998
128 14.493191999999999
96 11.336316
64 6.625228
};
\addplot [mylightblue,mark=x] table { 
 
Ports "Time [us]"

512 127.323994
480 117.71757000000001
448 109.13928
416 100.41654
384 92.394908
352 83.470832
320 74.724402
288 66.428341
256 58.23838
224 51.228274
192 42.165218
160 34.588193
128 27.259138
96 20.61468
64 14.305301
};
\legend{2,4,8,Heur. 2,Heur. 4,Heur. 8}
\end{axis}
\end{tikzpicture}%
    \caption{The heuristic is much faster for high port counts.}
    \label{fig:algo_perf}
\end{figure}
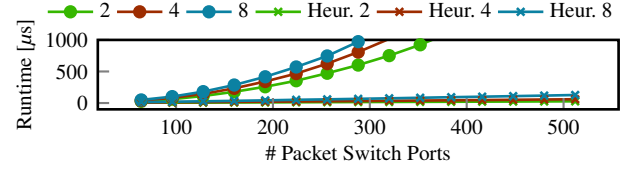

\Cref{fig:algo_perf} shows the runtime of the \Lightpass/ control plane while varying the number of ports and ASICs. 
We consider both the optimal matching algorithm (\Cref{subsec:matching}) and the heuristic (\Cref{subsec:efficient_heuristic_design}), abbreviated as \textbf{Heur. $N$}. 
The experiments run single-threaded on one CPU core (AMD Zen~3) and are averaged over 1k iterations.  
As expected, the optimal algorithm scales poorly, reaching $3.5\,ms$ for 512 ports and 8 ASICs. 
In contrast, the heuristic scales almost linearly.
For 512 ports, the 2-ASIC and 8-ASIC cases require only $28\,\mu s$ and $127\,\mu s$ respectively. 
%The 2-ASIC case with 512 ports finishes in only $28\,\mu s$ with the 8-ASIC case taking $127\,\mu s$ to complete.
%For 4 and 8 ASICs
For more than 2 ASICs, the heuristic is not strictly optimal, but only incurs $<1\,\%$ additional inter-ASIC traffic on average across 1k random traffic matrices. 
Runtime can be reduced further through port bundling or hardware acceleration~(\Cref{app:control-eff-contd}). 
%For instance, bundling 4 ports yields $<0.1\,\%$ performance penalty (\Cref{fig:bundle_median}), while specialized hardware could parallelize the mapping computation~\cite{song2016Parallel,abdelrasoul2021FPGA,jalilvand2025Sorting}. 

\subsection{Power and Overhead Analysis}
\label{subsec:powerandcost}

We estimate the total power savings from  \Lightpass/ and show the negligible overhead of adding a circuit-switched indirection layer.
\Cref{tab:fabric} shows the power savings of \Lightpass/ compared to the non-oversubscribed $n$-ASIC switch.
The savings are due to bandwidth reductions from inter-ASIC links, the fabric, and buffers, as discussed below.

\subsubsection{Baseline ASIC Chip}
For this evaluation, we consider the most cutting-edge baseline chip~($\approx$ Tomahawk 6; 102.4 Tbps) as the ASIC building block to create the $n$-ASIC switches (\Cref{fig:asic}).
We assume each Tomahawk 6 ASIC consumes $\approx750\,W$~\cite{michael2025Tomahawk} and occupies 
%an area of 
$\approx800\,mm^2$~\cite{yeluri2023Chiplets}.
\susdas{update this last phrase a little}
% this sentence needs to be clarified
%The Tomahawk 6 is the most cutting-edge chip based on a chiplet design.
%We use this building block to create $n$-ASIC switches as shown in~\Cref{fig:asic}.

\subsubsection{Inter-ASIC Throughput Reduction Factor}
\Lightpass/ reduces the required inter-ASIC throughput by a factor of $\frac{1}{\#ASIC}$ without sacrificing performance~(\Cref{evaluation}).
We define the inter-ASIC throughput reduction factor as $\alpha = \frac{Throughput}{\#ASIC}$, which is used to compute the following power savings. 

\subsubsection{Power Savings from Inter-ASIC Links}
To estimate the power required to transmit data between ASICs, we use the energy per bit value from a UCIe-compliant interface fabricated in $3\,nm$~\cite{lin2025361}. 
This interface consumes $0.6\,pJ/bit$ and results in a power reduction of $\alpha \cdot 0.6\,pJ/bit  \approx 41\,W$ for the 2-ASIC case and $\approx 33\,W$ for the 8-ASIC case.

%In terms of area, even when assuming the highest possible density for UCIe-A~\cite{dassharma2024Highperformance}, the reduction in inter-ASIC bandwidth frees up $5\,mm^2$ on the ASIC chiplets.
%UCIe-A requires expensive packaging and has limited reach; instead, one might use the UCIe-S standard, freeing up $51\,mm^2$ on the ASICs.

\subsubsection{Power Savings from Fabric}
For more than two ASICs, a fabric is needed to interconnect them. 
By reducing the inter-ASIC bandwidth, we can also reduce the fabric bandwidth.
We estimate that fabric chips consume $\approx 70\,\%$ of the power of a packet switch ASIC, based on the power consumption breakdown of a multi-ASIC chassis switch~\cite{fevrier2023PTX10000}.
This results in $5.13\,pJ/bit$ and provides savings of $\alpha \cdot 5.13\,pJ/bit \approx 300\,W$ for the 4-ASIC case and $\approx 280\,W$ for the 8-ASIC case.
%Note that with 2 ASICs, no fabric is needed.

%For more details, check~(\Cref{app:fabricpowerformula}).

\subsubsection{Power Savings From Buffer}
Due to localized traffic, packets might need to be buffered only at a single ASIC rather than at two. 
Buffering twice can be avoided, but requires complex buffering logic instead. 
Examining similar-sized SRAM modules, as found in network ASIC buffers, reveals an energy consumption of $\approx2\,pJ/bit$ for one read access followed by one write access~\cite{chang2013Technology} when scaled to modern process nodes.
This results in savings of around $\alpha \cdot 2\,pJ/bit \approx 137\,W$ for the 2-ASIC case and $\approx 109\,W$ for the 8-ASIC case.

\subsubsection{Overhead from Indirection Layer}
For this, we consider an electrical circuit-switch~(\Cref{subsec:dataplane}), which can be implemented without relying on future co-packaged optics switches enabling OCS indirection layers.
Here we consider the specific electrical crossbar switch~\cite{cakir2015Modelinga}, implemented in a $40\,nm$ technology. 
We can approximate the power/area consumption of the design with a newer $5\,nm$ technology.
Looking at historical data from TSMC technology nodes, the performance per watt has increased by $\approx 3\times$ from $40\,nm$ to $5\,nm$ technology, while the area was reduced by $\approx 6\times$~\cite{dang2011TSMC,vlsi2021TSMC,schor2023TSMC}.
The electrical crossbar switch has a capacity of 30 Tbps while consuming $0.95\,W$ and occupying $0.6\,mm^2$.
This results in an overhead of $0.035\,pJ/bit$ and an area of $0.02\,mm^2/Tbps$.
For the throughput of the baseline chip~(102.4 Tbps) this results in $3.6\,W$ of excess power and $2.4\,mm^2$ of excess area, which is $<0.5\,\%$ of the power and area w.r.t. the baseline chip.

\subsubsection{Overall Power Savings}

Finally, we get total power savings of $173\,W$ for the 2-ASIC case and $407\,W$ for the 8-ASIC case~(\Cref{tab:fabric}).
Note that the peak power savings occur at 4 ASICs, due to large savings from both fabric and inter-ASIC links, while still allowing decent oversubscription.
With more ASICs, the efficiency of the indirection layer reduces.
%The effectiveness of the indirection layer decreases with a larger number of ASICs.
%Using more than $4$ ASICs is impractical due to the devices' total power consumption, even with larger savings, and should be avoided unless no other option exists.
%While the relative savings get smaller with more ASICs, the absolute savings remain large.

\begin{table}
    \centering
    \caption{\Lightpass/ saves hundreds of watts per device.}
    \vspace{0.25em}
    \setlength{\tabcolsep}{2mm}
    \renewcommand{\arraystretch}{1.1}
    \footnotesize
    \tikzmarknode{tab1}{
    \begin{tabularx}{\columnwidth}{Xcccc}
         \toprule
         \textbf{Number of ASIC} & 2 & 4 & 6 & 8 \\
         \textbf{Throughput [Tbps]}              & 136.5   & 234.1     & 335.1  & 436.9 \\
         \midrule
         Inter-ASIC Link Saving  [$W$]      &  41.0    &   35.1    &  33.5  &  32.8 \\
         Fabric Saving [$W$]                     & 0       &  300.2     & 286.5   & 280.2   \\
         Buffer Saving [$W$]                     & 136.5      &    117.0    & 111.7    & 109.2   \\
         Indirection Layer Overhead [$W$]        & \color{myred} -4.8     & \color{myred}  -8.2    & \color{myred} -11.7     & \color{myred} -15.3  \\
         \bottomrule
         Total Savings [$W$]              & 172.7   & 444.1  & 420.0    & 406.9 \\
         %Switch w/o Oversubscription  [$W$]         && 1354     & 3523 & 5044 & 6575 \\
         %Relative Savings             && $13\,\%$   & $13\,\%$   & $8\,\%$     & $6\,\%$ \\
         %Largest Switch Savings [$W$]   && 255 & 596 & 586 & 577 \\
    \end{tabularx}}
    \vspace{0.5em}
    \label{tab:fabric}
\end{table}

\subsubsection{Discussion}
Since \Lightpass/ frees up capacity from the inter-ASIC links, it achieves a higher total throughput~(\Cref{tab:fabric}) while
maintaining $102.4\,Tbps$ per ASIC.
This is because more of the throughput can be directed towards the physical ports rather than to the fabric~(\Cref{fig:asic}).
Achieving the same total throughput with a traditional $n$-ASIC switch without oversubscription requires ASICs with more than $102.4\,Tbps$ throughput, which might not be realizable today.
Even if such larger ASICs could be manufactured, \Lightpass/ still has an edge by saving hundreds of watts of power per switch.
The total power draw of such switches is shown in~\Cref{app:fabricpowerformula}.

%the non-oversubscribed switch would still draw hundreds of watts more power than one based on the \Lightpass/ architecture.
\lukas{make this clearer}
% would it make sense to add a sentence like this and fuse smoothly with your last sentence?
%Note that, rather than assuming a jump from single ASIC to $8$-ASICs, \Lightpass/ targets a more realistic transition regime, i.e., $2$- and $4$-ASICs with the most savings (highlighted in~\Cref{tab:fabric}).

\section{Related Work}
\subsubsection{Circuit Switching in Datacenters}
Circuit switches have been widely explored in datacenters to enable reconfigurable topologies. Most proposals employ OCS~\cite{urata2023Apollo,ballani2020Sirius,wang2023TopoOpt,zu2024Resiliency,mellette2024Realizinga,mellette2017RotorNet,hamedazimi2014FireFly,zhou2012Mirror,poutievski2022Jupitera}, while Larry~\cite{chatzieleftheriou2018Larrya} and Shoal~\cite{shrivastav2019Shoal} use ECS. Another line of work leverages reconfiguration to improve traffic locality: RDC~\cite{wang2022RDC} dynamically assigns frequently communicating hosts to the same ToR switch, and OSSV~\cite{das2024Rearchitecting} extends this idea to jointly minimize traffic skewness and inter-rack volume.
\Lightpass/ differs from those approaches in three important aspects.
First, the context of running within a single network switch; second, making the problem tractable by allowing flexibility only at the ingress; and third, designing an efficient heuristic having three orders of magnitude faster runtime~(\Cref{app:control-eff-contd}).
Splitting ingress and egress is enabled by the context of a single network switch and enables new algorithms to be used, distinguishing this work from other algorithmic proposals for circuit switching in datacenters.
More details can be found in~\Cref{subsec:discussion}.

\subsubsection{Scaling Packet Switches}
The end of Dennard scaling has exposed fundamental challenges in scaling packet switches. 
Prior work has explored several directions, e.g., improving packet parser~\cite{liu2022HyperParser}, combining cell and packet switching~\cite{zilberman2019Stardust}, and introducing multiple parallel dataplanes~\cite{guo2022Scaling}. 
Additionally, \cite{keslassy2025Petabita} uses wafer-scale packaging technology and fiber ribbons to increase the bandwidth of network switches.
Their approach targets a different class of devices, such as Cerbras~\cite{cerebras2026Cerebras}, stitching together a chip that is an order of magnitude larger.
Our work takes a complementary path, enabling packet switching to scale more efficiently to multiple ASICs.

\section{Future Research Directions}
The flexibility of the indirection layer opens up opportunities beyond reducing inter-ASIC bandwidth.
First, it could be leveraged for energy proportionality. 
By dynamically localizing traffic to a subset of ASICs, unused portions of the switch could be powered down at low load, improving energy efficiency without compromising performance.

Second, the same flexibility enables heterogeneous chiplets with specialized capabilities, e.g., a chiplet optimized for complex protocol handling. 
Ports could be dynamically reassigned to chiplets without manual recabling. 
This reduces the cost of adding the functionality to every chiplet while still supporting diverse use cases.
Finally, the indirection layer could enhance hardware resiliency. 
For an ASIC-side port failure, traffic can be rapidly remapped to spare healthy ASIC-ports, improving fault tolerance without manual intervention.

\section{Conclusion}
Multi-ASIC switches are the inevitable path forward as single-ASIC switch designs reach their physical and economic limits. 
We present \Lightpass/, a practical enabler of high-performance yet feasible multi-ASIC switch design. 
By leveraging fast circuit switches, \Lightpass/ remaps ingress ports to ASICs 
to minimize inter-ASIC traffic. 
Our evaluation shows that \Lightpass/ delivers performance within $1\,\%$ of a single-ASIC switch across a diverse set of traffic workloads, outperforming a traditional non-blocking design while enabling hundreds of watts of power savings.
%while requiring half the inter-ASIC bandwidth, with negligible overhead in chip area and power. 
%Moreover, the hardware prototype demonstrates the practical viability of our proposal.

\newpage

\bibliographystyle{plain}
\begin{small}
\bibliography{bibliography.bib}

@misc{2026NVIDIAs,
  title = {{{NVIDIA}}'s {{Rubin Ultra}} Reportedly Sticking to a Dual-Die Design Instead of a Four-Die Plan},
  year = 2026,
  month = apr,
  journal = {TweakTown},
  urldate = {2026-05-28},
  chapter = {Graphics Cards},
  langid = {american},
  howpublished = {\url{https://www.tweaktown.com/news/110819/nvidias-rubin-ultra-reportedly-sticking-to-a-dual-die-design-instead-of-a-four-die-plan/index.html}},
  key = {NVIDIA's Rubin Ultra reportedly sticking to a dual-die design instead of a four-die plan}
}

@inproceedings{abadi2016TensorFlow,
  title = {{{TensorFlow}}: {{A System}} for {{Large-Scale Machine Learning}}},
  booktitle = {12th {{USENIX}} Symposium on Operating Systems Design and Implementation {{OSDI}} 16},
  author = {Abadi, Mart{\'i}n and Barham, Paul and Chen, Jianmin and Chen, Zhifeng and Davis, Andy and Dean, Jeffrey and Devin, Matthieu and Ghemawat, Sanjay and Irving, Geoffrey and Isard, Michael and Kudlur, Manjunath and Levenberg, Josh and Monga, Rajat and Moore, Sherry and Murray, Derek G and Steiner, Benoit and Tucker, Paul and Vasudevan, Vijay and Warden, Pete and Wicke, Martin and Yu, Yuan and Zheng, Xiaoqiang},
  year = 2016,
  pages = {265--283},
  publisher = {USENIX Association},
  address = {Savannah, GA, USA},
  langid = {english}
}

@inproceedings{abdelrasoul2021FPGA,
  title = {{{FPGA Based Hardware Accelerator}} for {{Sorting Data}}},
  booktitle = {2021 9th {{International Japan-Africa Conference}} on {{Electronics}}, {{Communications}}, and {{Computations}} ({{JAC-ECC}})},
  author = {Abdelrasoul, Maher and Shaban, Ahmed Sayed and {Abdel-Kader}, Hala},
  year = 2021,
  month = dec,
  pages = {57--60},
  publisher = {IEEE},
  address = {Virtual},
  doi = {10.1109/JAC-ECC54461.2021.9691432},
  urldate = {2025-09-02}
}

@inproceedings{alizadeh2010Data,
  title = {Data Center {{TCP}} ({{DCTCP}})},
  booktitle = {Proceedings of the {{ACM SIGCOMM}} 2010 Conference},
  author = {Alizadeh, Mohammad and Greenberg, Albert and Maltz, David A. and Padhye, Jitendra and Patel, Parveen and Prabhakar, Balaji and Sengupta, Sudipta and Sridharan, Murari},
  year = 2010,
  month = aug,
  series = {{{SIGCOMM}} '10},
  pages = {63--74},
  publisher = {Association for Computing Machinery},
  address = {New York, NY, USA},
  doi = {10.1145/1851182.1851192},
  urldate = {2026-02-03},
  isbn = {978-1-4503-0201-2}
}

@article{amaru2012High,
  title = {High {{Speed Architectures}} for {{Finding}} the {{First}} Two {{Maximum}}/{{Minimum Values}}},
  author = {Amaru, Luca G. and Martina, Maurizio and Masera, Guido},
  year = 2012,
  month = dec,
  journal = {IEEE Transactions on Very Large Scale Integration (VLSI) Systems},
  volume = {20},
  number = {12},
  pages = {2342--2346},
  issn = {1557-9999},
  doi = {10.1109/TVLSI.2011.2174166},
  urldate = {2025-09-02}
}

@misc{bailey2020Designs,
  title = {Designs {{Beyond The Reticle Limit}}},
  author = {Bailey, Brian},
  year = 2020,
  month = nov,
  journal = {Semiconductor Engineering},
  urldate = {2026-02-03},
  langid = {american},
  howpublished = {\url{https://semiengineering.com/designs-beyond-the-reticle-limit/}},
  key = {Designs Beyond The Reticle Limit}
}

@inproceedings{ballani2020Sirius,
  title = {Sirius: {{A Flat Datacenter Network}} with {{Nanosecond Optical Switching}}},
  shorttitle = {Sirius},
  booktitle = {Proceedings of the {{Annual}} Conference of the {{ACM Special Interest Group}} on {{Data Communication}} on the Applications, Technologies, Architectures, and Protocols for Computer Communication},
  author = {Ballani, Hitesh and Costa, Paolo and Behrendt, Raphael and Cletheroe, Daniel and Haller, Istvan and Jozwik, Krzysztof and Karinou, Fotini and Lange, Sophie and Shi, Kai and Thomsen, Benn and Williams, Hugh},
  year = 2020,
  month = jul,
  pages = {782--797},
  publisher = {ACM},
  address = {Virtual Event USA},
  doi = {10.1145/3387514.3406221},
  urldate = {2024-05-28},
  isbn = {978-1-4503-7955-7},
  langid = {english}
}

@inproceedings{benson2011MicroTE,
  title = {{{MicroTE}}: Fine Grained Traffic Engineering for Data Centers},
  shorttitle = {{{MicroTE}}},
  booktitle = {Proceedings of the {{Seventh COnference}} on Emerging {{Networking EXperiments}} and {{Technologies}}},
  author = {Benson, Theophilus and Anand, Ashok and Akella, Aditya and Zhang, Ming},
  year = 2011,
  month = dec,
  pages = {1--12},
  publisher = {ACM},
  address = {Tokyo Japan},
  doi = {10.1145/2079296.2079304},
  urldate = {2026-02-03},
  isbn = {978-1-4503-1041-3},
  langid = {english}
}

@inproceedings{bohr2011evolutiona,
  title = {The Evolution of Scaling from the Homogeneous Era to the Heterogeneous Era},
  booktitle = {2011 {{International Electron Devices Meeting}}},
  author = {Bohr, Mark},
  year = 2011,
  month = dec,
  pages = {1.1.1-1.1.6},
  publisher = {IEEE},
  address = {Washington, D.C, USA},
  issn = {2156-017X},
  doi = {10.1109/IEDM.2011.6131469},
  urldate = {2026-02-01}
}

@inproceedings{borkar2007Thousand,
  title = {Thousand {{Core ChipsA Technology Perspective}}},
  booktitle = {2007 44th {{ACM}}/{{IEEE Design Automation Conference}}},
  author = {Borkar, Shekhar},
  year = 2007,
  month = jun,
  pages = {746--749},
  publisher = {Association for Computing Machinery},
  address = {San Diego, CA, USA},
  issn = {0738-100X},
  doi = {10.1109/DAC.2007.375263},
  urldate = {2026-02-03}
}

@misc{broadcom2026News,
  title = {News {{Releases}}},
  author = {Broadcom},
  year = 2026,
  journal = {Broadcom Newsroom},
  urldate = {2026-08-11},
  langid = {english},
  howpublished = {\url{https://www.broadcom.com/company/news/releases}},
  key = {News Releases}
}

@inproceedings{cakir2015Modelinga,
  title = {Modeling and {{Design}} of {{High-Radix On-Chip Crossbar Switches}}},
  booktitle = {Proceedings of the 9th {{International Symposium}} on {{Networks-on-Chip}}},
  author = {Cakir, Cagla and Ho, Ron and Lexau, Jon and Mai, Ken},
  year = 2015,
  month = sep,
  pages = {1--8},
  publisher = {ACM},
  address = {Vancouver BC Canada},
  doi = {10.1145/2786572.2786579},
  urldate = {2025-08-13},
  isbn = {978-1-4503-3396-2},
  langid = {english}
}

@misc{cerebras2026Cerebras,
  title = {Cerebras},
  author = {Cerebras},
  year = 2026,
  month = jan,
  journal = {Cerebras},
  urldate = {2026-01-29},
  langid = {english},
  howpublished = {\url{https://www.cerebras.ai}},
  key = {Cerebras}
}

@inproceedings{chang2013Technology,
  title = {Technology Comparison for Large Last-Level Caches ({{L3Cs}}): {{Low-leakage SRAM}}, Low Write-Energy {{STT-RAM}}, and Refresh-Optimized {{eDRAM}}},
  shorttitle = {Technology Comparison for Large Last-Level Caches ({{L3Cs}})},
  booktitle = {2013 {{IEEE}} 19th {{International Symposium}} on {{High Performance Computer Architecture}} ({{HPCA}})},
  author = {Chang, Mu-Tien and Rosenfeld, Paul and Lu, Shih-Lien and Jacob, Bruce},
  year = 2013,
  month = feb,
  pages = {143--154},
  publisher = {IEEE},
  address = {Shenzhen, China},
  issn = {1530-0897},
  doi = {10.1109/HPCA.2013.6522314},
  urldate = {2026-01-06}
}

@inproceedings{chatzieleftheriou2018Larrya,
  title = {Larry: {{Practical Network Reconfigurability}} in the {{Data Center}}},
  shorttitle = {Larry},
  booktitle = {15th {{USENIX Symposium}} on {{Networked Systems Design}} and {{Implementation}} ({{NSDI}} 18)},
  author = {Chatzieleftheriou, Andromachi and Legtchenko, Sergey and Williams, Hugh and Rowstron, Antony},
  year = 2018,
  pages = {141--156},
  publisher = {USENIX Association},
  address = {Renton, WA, USA},
  urldate = {2025-08-26},
  isbn = {978-1-939133-01-4},
  langid = {english}
}

@misc{cisco2026Cisco,
  title = {Cisco {{Catalyst}} 9500 {{Series Architecture White Paper}}},
  author = {Cisco},
  year = 2026,
  month = sep,
  journal = {Cisco},
  urldate = {2026-01-09},
  langid = {english},
  howpublished = {\url{https://www.cisco.com/c/en/us/products/collateral/switches/catalyst-9500-series-switches/nb-06-cat9500-architecture-cte-en.html}},
  key = {Cisco Catalyst 9500 Series Architecture White Paper}
}

@misc{dang2011TSMC,
  title = {{{TSMC}} Turns on the "Volume" for Its 28nm Process},
  author = {Dang, Annie},
  year = 2011,
  month = oct,
  journal = {Manufacturers' Monthly},
  urldate = {2025-09-12},
  langid = {australian},
  howpublished = {\url{https://www.manmonthly.com.au/tsmc-turns-on-the-volume-for-its-28nm-process/}},
  key = {TSMC turns on the "volume" for its 28nm process}
}

@inproceedings{das2024Rearchitecting,
  title = {Rearchitecting {{Datacenter Networks}}: {{A New Paradigm}} with {{Optical Core}} and {{Optical Edge}}},
  shorttitle = {Rearchitecting {{Datacenter Networks}}},
  booktitle = {{{IEEE INFOCOM}} 2024 - {{IEEE Conference}} on {{Computer Communications}}},
  author = {Das, Sushovan and Silva, Arlei and Eugene Ng, T. S.},
  year = 2024,
  month = may,
  pages = {1371--1380},
  publisher = {IEEE},
  address = {Vancouver BC Canada},
  issn = {2641-9874},
  doi = {10.1109/INFOCOM52122.2024.10621224},
  urldate = {2026-02-03}
}

@misc{desanti20118021Qbb,
  title = {802.{{1Qbb}} -- {{Priority-based Flow Control}} \textbar},
  author = {DeSanti, Claudio},
  year = 2011,
  urldate = {2025-09-11},
  howpublished = {\url{https://1.ieee802.org/dcb/802-1qbb/}},
  key = {802.1Qbb – Priority-based Flow Control |}
}

@misc{discuss2026Microsoft,
  title = {Microsoft {{Unveils World}}'s {{Biggest AI Datacenter}}, {{Housing Hundreds}} of {{Thousands}} of {{GPUs}}},
  author = {Discuss, AleksandarK},
  year = 2026,
  month = jan,
  journal = {TechPowerUp},
  urldate = {2026-01-26},
  langid = {english},
  howpublished = {\url{https://www.techpowerup.com/341139/microsoft-unveils-worlds-biggest-ai-datacenter-housing-hundreds-of-thousands-of-gpus}},
  key = {Microsoft Unveils World's Biggest AI Datacenter, Housing Hundreds of Thousands of GPUs}
}

@article{esmaeilzadeh2011Dark,
  title = {Dark Silicon and the End of Multicore Scaling},
  author = {Esmaeilzadeh, Hadi and Blem, Emily and St. Amant, Renee and Sankaralingam, Karthikeyan and Burger, Doug},
  year = 2011,
  month = jun,
  journal = {SIGARCH Comput. Archit. News},
  volume = {39},
  number = {3},
  pages = {365--376},
  issn = {0163-5964},
  doi = {10.1145/2024723.2000108},
  urldate = {2026-02-03}
}

@misc{fevrier2023PTX10000,
  title = {{{PTX10000 Power Optimization}}},
  author = {Fevrier, Nicolas},
  year = 2023,
  month = aug,
  urldate = {2026-01-16},
  langid = {english},
  howpublished = {\url{https://community.juniper.net/blogs/nicolas-fevrier/2023/08/29/ptx10000-power-optimization}},
  key = {PTX10000 Power Optimization}
}

@misc{fs2026N960064OD,
  title = {N9600-{{64OD}}, 64-{{Port Ethernet HPC}}/{{AI Data Center Switch}}, 64 x {{800Gb OSFP}}, {{PicOS}}\textregistered, {{Broadcom Tomahawk}} 5 {{Chip}}, {{Front-to-Back Airflow}} - {{FS}}.Com {{Europe}}},
  author = {FS},
  year = 2026,
  month = jan,
  journal = {FS.com},
  urldate = {2026-02-01},
  langid = {english},
  howpublished = {\url{https://www.fs.com/eu-en/products/250955.html}},
  key = {N9600-64OD, 64-Port Ethernet HPC/AI Data Center Switch, 64 x 800Gb OSFP, PicOS®, Broadcom Tomahawk 5 Chip, Front-to-Back Airflow - FS.com Europe}
}

@article{greenberg2009VL2,
  title = {{{VL2}}: A Scalable and Flexible Data Center Network},
  shorttitle = {{{VL2}}},
  author = {Greenberg, Albert and Hamilton, James R. and Jain, Navendu and Kandula, Srikanth and Kim, Changhoon and Lahiri, Parantap and Maltz, David A. and Patel, Parveen and Sengupta, Sudipta},
  year = 2009,
  month = aug,
  journal = {SIGCOMM Comput. Commun. Rev.},
  volume = {39},
  number = {4},
  pages = {51--62},
  issn = {0146-4833},
  doi = {10.1145/1594977.1592576},
  urldate = {2026-02-03}
}

@inproceedings{guo2022Scaling,
  title = {Scaling beyond Packet Switch Limits with Multiple Dataplanes},
  booktitle = {Proceedings of the 18th {{International Conference}} on Emerging {{Networking EXperiments}} and {{Technologies}}},
  author = {Guo, Yibo and Mellette, William M. and Snoeren, Alex C. and Porter, George},
  year = 2022,
  month = nov,
  series = {{{CoNEXT}} '22},
  pages = {214--231},
  publisher = {Association for Computing Machinery},
  address = {New York, NY, USA},
  doi = {10.1145/3555050.3569141},
  urldate = {2025-09-16},
  isbn = {978-1-4503-9508-3}
}

@inproceedings{hamedazimi2014FireFly,
  title = {{{FireFly}}: A Reconfigurable Wireless Data Center Fabric Using Free-Space Optics},
  shorttitle = {{{FireFly}}},
  booktitle = {Proceedings of the 2014 {{ACM}} Conference on {{SIGCOMM}}},
  author = {Hamedazimi, Navid and Qazi, Zafar and Gupta, Himanshu and Sekar, Vyas and Das, Samir R. and Longtin, Jon P. and Shah, Himanshu and Tanwer, Ashish},
  year = 2014,
  month = aug,
  pages = {319--330},
  publisher = {ACM},
  address = {Chicago Illinois USA},
  doi = {10.1145/2619239.2626328},
  urldate = {2024-05-15},
  isbn = {978-1-4503-2836-4},
  langid = {english}
}

@article{han2015Largescale,
  title = {Large-Scale Silicon Photonic Switches with Movable Directional Couplers},
  author = {Han, Sangyoon and Seok, Tae Joon and Quack, Niels and Yoo, Byung-Wook and Wu, Ming C.},
  year = 2015,
  month = apr,
  journal = {Optica},
  volume = {2},
  number = {4},
  pages = {370--375},
  publisher = {Optica Publishing Group},
  issn = {2334-2536},
  doi = {10.1364/OPTICA.2.000370},
  urldate = {2025-01-07},
  copyright = {\copyright{} 2015 Optical Society of America},
  langid = {english}
}

@article{ikeda2020Largescale,
  title = {Large-Scale Silicon Photonics Switch Based on 45-Nm {{CMOS}} Technology},
  author = {Ikeda, Kazuhiro and Suzuki, Keijiro and Konoike, Ryotaro and Namiki, Shu and Kawashima, Hitoshi},
  year = 2020,
  month = jul,
  journal = {Optics Communications},
  volume = {466},
  pages = {125677},
  issn = {0030-4018},
  doi = {10.1016/j.optcom.2020.125677},
  urldate = {2025-01-07}
}

@misc{intel2026Intel,
  title = {{Intel\textregistered{} Tofino™ Reihe -- Programmierbare Ethernet-Switch ASICs}},
  author = {Intel},
  year = 2026,
  month = jan,
  journal = {Intel},
  urldate = {2026-02-01},
  langid = {ngerman},
  howpublished = {\url{https://www.intel.com/content/www/de/de/products/details/ethernet/programmable-ethernet-switch/tofino-series.html}},
  key = {Intel® Tofino™ Reihe – Programmierbare Ethernet-Switch ASICs}
}

@inproceedings{kassing2017fattreesa,
  title = {Beyond Fat-Trees without Antennae, Mirrors, and Disco-Balls},
  booktitle = {Proceedings of the {{Conference}} of the {{ACM Special Interest Group}} on {{Data Communication}}},
  author = {Kassing, Simon and Valadarsky, Asaf and Shahaf, Gal and Schapira, Michael and Singla, Ankit},
  year = 2017,
  month = aug,
  series = {{{SIGCOMM}} '17},
  pages = {281--294},
  publisher = {Association for Computing Machinery},
  address = {New York, NY, USA},
  doi = {10.1145/3098822.3098836},
  urldate = {2025-08-18},
  isbn = {978-1-4503-4653-5}
}

@misc{kassing2025Netbench,
  title = {Netbench},
  author = {Kassing, Simon Arnold},
  year = 2025,
  month = aug,
  urldate = {2025-08-26},
  howpublished = {\url{https://github.com/ndal-eth/netbench}},
  key = {Netbench}
}

@incollection{keslassy2025Petabita,
  title = {Petabit {{Router-in-a-Package}}: {{Rethinking Internet Routers}} in the {{Age}} of {{In-Packaged Optics}} and {{Heterogeneous Integration}}},
  shorttitle = {Petabit {{Router-in-a-Package}}},
  booktitle = {Proceedings of the 24th {{ACM Workshop}} on {{Hot Topics}} in {{Networks}}},
  author = {Keslassy, Isaac and {https://orcid.org/0000-0001-6103-6910} and {View Profile} and Lin, Bill and {https://orcid.org/0000-0003-0965-7247} and {View Profile}},
  year = 2025,
  month = nov,
  series = {{{ACM Conferences}}},
  pages = {245--253},
  publisher = {Association for Computing Machinery},
  address = {College Park, MD, USA},
  doi = {10.1145/3772356.3772398},
  urldate = {2026-01-29},
  isbn = {979-8-4007-2280-6}
}

@incollection{kuhn2010Hungarian,
  title = {The {{Hungarian Method}} for the {{Assignment Problem}}},
  booktitle = {50 {{Years}} of {{Integer Programming}} 1958-2008: {{From}} the {{Early Years}} to the {{State-of-the-Art}}},
  author = {Kuhn, Harold W.},
  editor = {J{\"u}nger, Michael and Liebling, Thomas M. and Naddef, Denis and Nemhauser, George L. and Pulleyblank, William R. and Reinelt, Gerhard and Rinaldi, Giovanni and Wolsey, Laurence A.},
  year = 2010,
  pages = {29--47},
  publisher = {Springer},
  address = {Berlin, Heidelberg},
  doi = {10.1007/978-3-540-68279-0_2},
  urldate = {2026-02-03},
  isbn = {978-3-540-68279-0},
  langid = {english}
}

@misc{lightmatter2026Passage,
  title = {Passage {{M-Series Photonic Superchip}}},
  author = {Lightmatter},
  year = 2026,
  month = jan,
  journal = {Lightmatter\textregistered},
  urldate = {2026-01-22},
  langid = {american},
  howpublished = {\url{https://lightmatter.co/products/m1000/}},
  key = {Passage M-Series Photonic Superchip}
}

@inproceedings{lin2025361,
  title = {36.1 {{A 32Gb}}/s 10.{{5Tb}}/s/Mm 0.{{6pJ}}/b {{UCIe-Compliant Low-Latency Interface}} in 3nm {{Featuring Matched-Delay}} for {{Dynamic Clock Gating}}},
  booktitle = {2025 {{IEEE International Solid-State Circuits Conference}} ({{ISSCC}})},
  author = {Lin, Mu-Shan and Tsai, Chien-Chun and Li, Shenggao and Chen, Wei-Chih and Huang, Wen-Hung and Chen, Yu-Chi and Huang, Yu-Jie and Drake, Alan and Wen, Chin-Hua and Ranucci, Paul and Kuo, Hsin-Hung and Yin, Aidong and Yang, Shu-Chun and Mahmoudi, Farsheed and Ke, Han-Tzung and Li, Chao-Chieh and Cheng, Nai-Chen and Wang, Jimmy and Lin, Kevin and Liao, Harry and Huang, Jie-Ren and Wu, Meng-Hsuan and Hsieh, Kenny Cheng-Hsiang and Amatruda, Nicholas and Polanco, William and King, David and Basso, Todd and Kashem, Anwar},
  year = 2025,
  month = feb,
  volume = {68},
  pages = {586--588},
  publisher = {IEEE},
  address = {San Francisco, CA, USA},
  issn = {2376-8606},
  doi = {10.1109/ISSCC49661.2025.10904767},
  urldate = {2025-08-20}
}

@inproceedings{liu2022HyperParser,
  title = {{{HyperParser}}: {{A High-Performance Parser Architecture}} for {{Next Generation Programmable Switch}} and {{SmartNIC}}},
  shorttitle = {{{HyperParser}}},
  booktitle = {Proceedings of the 5th {{Asia-Pacific Workshop}} on {{Networking}}},
  author = {Liu, Huan and Qiu, Zhiliang and Pan, Weitao and Li, Jiajun and Huang, Jinjian},
  year = 2022,
  month = feb,
  series = {{{APNet}} '21},
  pages = {50--56},
  publisher = {Association for Computing Machinery},
  address = {New York, NY, USA},
  doi = {10.1145/3469393.3469399},
  urldate = {2025-09-16},
  isbn = {978-1-4503-8587-9}
}

@article{markov2014Limits,
  title = {Limits on Fundamental Limits to Computation},
  author = {Markov, Igor L.},
  year = 2014,
  month = aug,
  journal = {Nature},
  volume = {512},
  number = {7513},
  pages = {147--154},
  publisher = {Nature Publishing Group},
  issn = {1476-4687},
  doi = {10.1038/nature13570},
  urldate = {2026-02-03},
  copyright = {2014 Springer Nature Limited},
  langid = {english}
}

@inproceedings{mellette2016PFatTree,
  title = {P-{{FatTree}}: {{A}} Multi-Channel Datacenter Network Topology},
  shorttitle = {P-{{FatTree}}},
  booktitle = {Proceedings of the 15th {{ACM Workshop}} on {{Hot Topics}} in {{Networks}}},
  author = {Mellette, William M. and Snoeren, Alex C. and Porter, George},
  year = 2016,
  month = nov,
  pages = {78--84},
  publisher = {ACM},
  address = {Atlanta GA USA},
  doi = {10.1145/3005745.3005746},
  urldate = {2026-01-19},
  isbn = {978-1-4503-4661-0},
  langid = {english}
}

@inproceedings{mellette2017RotorNet,
  title = {{{RotorNet}}: {{A Scalable}}, {{Low-complexity}}, {{Optical Datacenter Network}}},
  shorttitle = {{{RotorNet}}},
  booktitle = {Proceedings of the {{Conference}} of the {{ACM Special Interest Group}} on {{Data Communication}}},
  author = {Mellette, William M. and McGuinness, Rob and Roy, Arjun and Forencich, Alex and Papen, George and Snoeren, Alex C. and Porter, George},
  year = 2017,
  month = aug,
  series = {{{SIGCOMM}} '17},
  pages = {267--280},
  publisher = {Association for Computing Machinery},
  address = {New York, NY, USA},
  doi = {10.1145/3098822.3098838},
  urldate = {2024-03-14},
  isbn = {978-1-4503-4653-5}
}

@inproceedings{mellette2024Realizinga,
  title = {Realizing {{RotorNet}}: {{Toward Practical Microsecond Scale Optical Networking}}},
  shorttitle = {Realizing {{RotorNet}}},
  booktitle = {Proceedings of the {{ACM SIGCOMM}} 2024 {{Conference}}},
  author = {Mellette, William M. and Forencich, Alex and Athapathu, Rukshani and Snoeren, Alex C. and Papen, George and Porter, George},
  year = 2024,
  month = aug,
  series = {{{ACM SIGCOMM}} '24},
  pages = {392--414},
  publisher = {Association for Computing Machinery},
  address = {New York, NY, USA},
  doi = {10.1145/3651890.3672273},
  urldate = {2024-09-30},
  isbn = {979-8-4007-0614-1}
}

@inproceedings{miao2017SilkRoad,
  title = {{{SilkRoad}}: {{Making Stateful Layer-4 Load Balancing Fast}} and {{Cheap Using Switching ASICs}}},
  shorttitle = {{{SilkRoad}}},
  booktitle = {Proceedings of the {{Conference}} of the {{ACM Special Interest Group}} on {{Data Communication}}},
  author = {Miao, Rui and Zeng, Hongyi and Kim, Changhoon and Lee, Jeongkeun and Yu, Minlan},
  year = 2017,
  month = aug,
  series = {{{SIGCOMM}} '17},
  pages = {15--28},
  publisher = {Association for Computing Machinery},
  address = {New York, NY, USA},
  doi = {10.1145/3098822.3098824},
  urldate = {2026-02-01},
  isbn = {978-1-4503-4653-5}
}

@misc{michael2025Tomahawk,
  title = {Tomahawk 6: {{The}} Industry's First 100-Terabit Switch Chip},
  shorttitle = {Tomahawk 6},
  author = {Michael},
  year = 2025,
  month = jun,
  journal = {Gazettabyte},
  urldate = {2026-01-26},
  langid = {american},
  howpublished = {\url{https://gazettabyte.com/tomahawk-6-the-industrys-first-100-terabit-switch-chip/}},
  key = {Tomahawk 6: The industry’s first 100-terabit switch chip}
}

@misc{moore2019Another,
  title = {Another {{Step Toward}} the {{End}} of {{Moore}}'s {{Law}} - {{IEEE Spectrum}}},
  author = {Moore, Samuel K.},
  year = 2019,
  month = may,
  journal = {IEEE Spectrum},
  publisher = {IEEE Spectrum},
  urldate = {2026-02-03},
  langid = {english},
  howpublished = {\url{https://spectrum.ieee.org/another-step-toward-the-end-of-moores-law}},
  key = {Another Step Toward the End of Moore’s Law - IEEE Spectrum}
}

@inproceedings{narayanan2021Efficient,
  title = {Efficient Large-Scale Language Model Training on {{GPU}} Clusters Using Megatron-{{LM}}},
  booktitle = {Proceedings of the {{International Conference}} for {{High Performance Computing}}, {{Networking}}, {{Storage}} and {{Analysis}}},
  author = {Narayanan, Deepak and Shoeybi, Mohammad and Casper, Jared and LeGresley, Patrick and Patwary, Mostofa and Korthikanti, Vijay and Vainbrand, Dmitri and Kashinkunti, Prethvi and Bernauer, Julie and Catanzaro, Bryan and Phanishayee, Amar and Zaharia, Matei},
  year = 2021,
  month = nov,
  series = {{{SC}} '21},
  pages = {1--15},
  publisher = {Association for Computing Machinery},
  address = {New York, NY, USA},
  doi = {10.1145/3458817.3476209},
  urldate = {2026-02-03},
  isbn = {978-1-4503-8442-1}
}

@incollection{paszke2019PyTorcha,
  title = {{{PyTorch}}: An Imperative Style, High-Performance Deep Learning Library},
  shorttitle = {{{PyTorch}}},
  booktitle = {Proceedings of the 33rd {{International Conference}} on {{Neural Information Processing Systems}}},
  author = {Paszke, Adam and Gross, Sam and Massa, Francisco and Lerer, Adam and Bradbury, James and Chanan, Gregory and Killeen, Trevor and Lin, Zeming and Gimelshein, Natalia and Antiga, Luca and Desmaison, Alban and K{\"o}pf, Andreas and Yang, Edward and DeVito, Zach and Raison, Martin and Tejani, Alykhan and Chilamkurthy, Sasank and Steiner, Benoit and Fang, Lu and Bai, Junjie and Chintala, Soumith},
  year = 2019,
  month = dec,
  number = {721},
  pages = {8026--8037},
  publisher = {Curran Associates Inc.},
  address = {Red Hook, NY, USA},
  urldate = {2026-07-29}
}

@inproceedings{poutievski2022Jupitera,
  title = {Jupiter Evolving: Transforming Google's Datacenter Network via Optical Circuit Switches and Software-Defined Networking},
  shorttitle = {Jupiter Evolving},
  booktitle = {Proceedings of the {{ACM SIGCOMM}} 2022 {{Conference}}},
  author = {Poutievski, Leon and Mashayekhi, Omid and Ong, Joon and Singh, Arjun and Tariq, Mukarram and Wang, Rui and Zhang, Jianan and Beauregard, Virginia and Conner, Patrick and Gribble, Steve and Kapoor, Rishi and Kratzer, Stephen and Li, Nanfang and Liu, Hong and Nagaraj, Karthik and Ornstein, Jason and Sawhney, Samir and Urata, Ryohei and Vicisano, Lorenzo and Yasumura, Kevin and Zhang, Shidong and Zhou, Junlan and Vahdat, Amin},
  year = 2022,
  month = aug,
  series = {{{SIGCOMM}} '22},
  pages = {66--85},
  publisher = {Association for Computing Machinery},
  address = {New York, NY, USA},
  doi = {10.1145/3544216.3544265},
  urldate = {2026-02-02},
  isbn = {978-1-4503-9420-8}
}

@inproceedings{roy2015Social,
  title = {Inside the {{Social Network}}'s ({{Datacenter}}) {{Network}}},
  booktitle = {Proceedings of the 2015 {{ACM Conference}} on {{Special Interest Group}} on {{Data Communication}}},
  author = {Roy, Arjun and Zeng, Hongyi and Bagga, Jasmeet and Porter, George and Snoeren, Alex C.},
  year = 2015,
  month = aug,
  pages = {123--137},
  publisher = {ACM},
  address = {London United Kingdom},
  doi = {10.1145/2785956.2787472},
  urldate = {2024-12-10},
  isbn = {978-1-4503-3542-3},
  langid = {english}
}

@misc{schor2023TSMC,
  title = {{{TSMC N3}}, {{And Challenges Ahead}}},
  author = {Schor, David},
  year = 2023,
  month = may,
  journal = {WikiChip Fuse},
  urldate = {2025-09-06},
  chapter = {Foundries},
  langid = {american},
  howpublished = {\url{https://fuse.wikichip.org/news/7375/tsmc-n3-and-challenges-ahead/}},
  key = {TSMC N3, And Challenges Ahead}
}

@article{seok2016Largescale,
  title = {Large-Scale Broadband Digital Silicon Photonic Switches with Vertical Adiabatic Couplers},
  author = {Seok, Tae Joon and Quack, Niels and Han, Sangyoon and Muller, Richard S. and Wu, Ming C.},
  year = 2016,
  month = jan,
  journal = {Optica},
  volume = {3},
  number = {1},
  pages = {64--70},
  publisher = {Optica Publishing Group},
  issn = {2334-2536},
  doi = {10.1364/OPTICA.3.000064},
  urldate = {2025-01-07},
  copyright = {\copyright{} 2016 Optical Society of America},
  langid = {english}
}

@inproceedings{shrivastav2019Shoal,
  title = {Shoal: {{A Network Architecture}} for {{Disaggregated Racks}}},
  shorttitle = {Shoal},
  booktitle = {16th {{USENIX Symposium}} on {{Networked Systems Design}} and {{Implementation}} ({{NSDI}} 19)},
  author = {Shrivastav, Vishal and Valadarsky, Asaf and Ballani, Hitesh and Costa, Paolo and Lee, Ki Suh and Wang, Han and Agarwal, Rachit and Weatherspoon, Hakim},
  year = 2019,
  pages = {255--270},
  publisher = {USENIX Association},
  address = {Boston, MA, USA},
  urldate = {2026-02-03},
  isbn = {978-1-931971-49-2},
  langid = {english}
}

@misc{si2026Collective,
  title = {Collective {{Communication}} for 100k+ {{GPUs}}},
  author = {Si, Min and Balaji, Pavan and Chen, Yongzhou and Chu, Ching-Hsiang and Gangidi, Adi and Hasan, Saif and Iyengar, Subodh and Johnson, Dan and Liu, Bingzhe and Ren, Regina and Shah, Deep and Shetty, Ashmitha Jeevaraj and Steinbrecher, Greg and Wang, Yulun and Wu, Bruce and Xie, Xinfeng and Yang, Jingyi and Yang, Mingran and Yu, Kenny and Yu, Minlan and Zhao, Cen and Bland, Wes and Boyda, Denis and Gumudavelli, Suman and Kannan, Prashanth and Lumezanu, Cristian and Miao, Rui and Qu, Zhe and Ramesh, Venkat and Samoylov, Maxim and Seidel, Jan and Sundaresan, Srikanth and Tian, Feng and Tan, Qiye and Zhang, Shuqiang and Zhao, Yimeng and Zheng, Shengbao and Zhu, Art and Zeng, Hongyi},
  year = 2026,
  month = jan,
  number = {arXiv:2510.20171},
  eprint = {2510.20171},
  primaryclass = {cs},
  publisher = {arXiv},
  doi = {10.48550/arXiv.2510.20171},
  urldate = {2026-01-26},
  archiveprefix = {arXiv}
}

@article{singh2015Jupiter,
  title = {Jupiter {{Rising}}: {{A Decade}} of {{Clos Topologies}} and {{Centralized Control}} in {{Google}}'s {{Datacenter Network}}},
  shorttitle = {Jupiter {{Rising}}},
  author = {Singh, Arjun and Ong, Joon and Agarwal, Amit and Anderson, Glen and Armistead, Ashby and Bannon, Roy and Boving, Seb and Desai, Gaurav and Felderman, Bob and Germano, Paulie and Kanagala, Anand and Provost, Jeff and Simmons, Jason and Tanda, Eiichi and Wanderer, Jim and H{\"o}lzle, Urs and Stuart, Stephen and Vahdat, Amin},
  year = 2015,
  month = aug,
  journal = {SIGCOMM Comput. Commun. Rev.},
  volume = {45},
  number = {4},
  pages = {183--197},
  issn = {0146-4833},
  doi = {10.1145/2829988.2787508},
  urldate = {2026-02-03}
}

@inproceedings{song2016Parallel,
  title = {Parallel {{Hardware Merge Sorter}}},
  booktitle = {2016 {{IEEE}} 24th {{Annual International Symposium}} on {{Field-Programmable Custom Computing Machines}} ({{FCCM}})},
  author = {Song, Wei and Koch, Dirk and Luj{\'a}n, Mikel and Garside, Jim},
  year = 2016,
  month = may,
  pages = {95--102},
  publisher = {IEEE},
  address = {Washington, D.C, USA},
  doi = {10.1109/FCCM.2016.34},
  urldate = {2025-09-02}
}

@misc{throughput2019Broadcom,
  title = {Broadcom {{Ships First}} 25.{{6Tbps Switch}} on 7nm},
  author = {chips offer superior {throughput}, Arne Verheyde Broadcom's Tomahawk 4},
  year = 2019,
  month = dec,
  journal = {Tom's Hardware},
  urldate = {2026-08-11},
  chapter = {CPUs},
  langid = {english},
  howpublished = {\url{https://www.tomshardware.com/news/broadcom-ships-first-256tbps-switch-on-7nm}},
  key = {Broadcom Ships First 25.6Tbps Switch on 7nm}
}

@inproceedings{tsmots2017FPGA,
  title = {{{FPGA}} Implementation of Vertically Parallel Minimum and Maximum Values Determination in Array of Numbers},
  booktitle = {2017 14th {{International Conference The Experience}} of {{Designing}} and {{Application}} of {{CAD Systems}} in {{Microelectronics}} ({{CADSM}})},
  author = {Tsmots, Ivan and Rabyk, Vasyl and Skorokhoda, Oleksa and Antoniv, Volodymyr},
  year = 2017,
  month = feb,
  pages = {234--236},
  publisher = {IEEE},
  address = {Polyana-Svalyava, Ukraine},
  doi = {10.1109/CADSM.2017.7916123},
  urldate = {2025-09-02}
}

@inproceedings{urata2023Apollo,
  title = {Apollo: {{Large-Scale Deployment}} of {{Optical Circuit Switching}} for {{Datacenter Networking}}},
  shorttitle = {Apollo},
  booktitle = {2023 {{Optical Fiber Communications Conference}} and {{Exhibition}} ({{OFC}})},
  author = {Urata, Ryohei and Liu, Hong and Yasumura, Kevin and Mao, Erji and Berger, Jill and Zhou, Xiang and Lam, Cedric and Bannon, Roy and Hutchinson, Darren and Nelson, Daniel and Poutievski, Leon and Singh, Arjun and Ong, Joon and Vahdat, Amin},
  year = 2023,
  month = mar,
  pages = {1--3},
  publisher = {Optica Publishing Group},
  address = {San Diego, CA, USA},
  doi = {10.1364/OFC.2023.M2G.1},
  urldate = {2025-08-26}
}

@misc{vlsi2021TSMC,
  title = {{{TSMC}} 7nm, 16nm and 28nm {{Technology}} Node Comparisons},
  author = {VLSI, Team},
  year = 2021,
  month = sep,
  journal = {Team VLSI},
  urldate = {2025-09-06},
  langid = {american},
  howpublished = {\url{https://teamvlsi.com/2021/09/tsmc-7nm-16nm-and-28nm-technology-node-comparisons.html}},
  key = {TSMC 7nm, 16nm and 28nm Technology node comparisons}
}

@inproceedings{wang2022RDC,
  title = {{{RDC}}: {{Energy-Efficient Data Center Network Congestion Relief}} with {{Topological Reconfigurability}} at the {{Edge}}},
  shorttitle = {{{RDC}}},
  booktitle = {19th {{USENIX Symposium}} on {{Networked Systems Design}} and {{Implementation}} ({{NSDI}} 22)},
  author = {Wang, Weitao and Wu, Dingming and Das, Sushovan and Rahbar, Afsaneh and Chen, Ang and Ng, T. S. Eugene},
  year = 2022,
  pages = {1267--1288},
  publisher = {USENIX Association},
  address = {Renton, WA, USA},
  urldate = {2025-08-26},
  isbn = {978-1-939133-27-4},
  langid = {english}
}

@misc{wang2022TopoOpta,
  title = {{{TopoOpt}}: {{Co-optimizing Network Topology}} and {{Parallelization Strategy}} for {{Distributed Training Jobs}}},
  shorttitle = {{{TopoOpt}}},
  author = {Wang, Weiyang and Khazraee, Moein and Zhong, Zhizhen and Ghobadi, Manya and Jia, Zhihao and Mudigere, Dheevatsa and Zhang, Ying and Kewitsch, Anthony},
  year = 2022,
  month = sep,
  number = {arXiv:2202.00433},
  eprint = {2202.00433},
  primaryclass = {cs},
  publisher = {arXiv},
  doi = {10.48550/arXiv.2202.00433},
  urldate = {2026-02-06},
  archiveprefix = {arXiv}
}

@inproceedings{wang2023TopoOpt,
  title = {{{TopoOpt}}: {{Co-optimizing Network Topology}} and {{Parallelization Strategy}} for {{Distributed Training Jobs}}},
  shorttitle = {{{TopoOpt}}},
  booktitle = {20th {{USENIX Symposium}} on {{Networked Systems Design}} and {{Implementation}} ({{NSDI}} 23)},
  author = {Wang, Weiyang and Khazraee, Moein and Zhong, Zhizhen and Ghobadi, Manya and Jia, Zhihao and Mudigere, Dheevatsa and Zhang, Ying and Kewitsch, Anthony},
  year = 2023,
  pages = {739--767},
  publisher = {USENIX Association},
  address = {Boston, MA, USA},
  urldate = {2025-08-26},
  isbn = {978-1-939133-33-5},
  langid = {english}
}

@misc{wang2024Railonly,
  title = {Rail-Only: {{A Low-Cost High-Performance Network}} for {{Training LLMs}} with {{Trillion Parameters}}},
  shorttitle = {Rail-Only},
  author = {Wang, Weiyang and Ghobadi, Manya and Shakeri, Kayvon and Zhang, Ying and Hasani, Naader},
  year = 2024,
  month = jul,
  number = {arXiv:2307.12169},
  eprint = {2307.12169},
  primaryclass = {cs},
  publisher = {arXiv},
  urldate = {2024-08-08},
  archiveprefix = {arXiv},
  langid = {english}
}

@inproceedings{wu2016Largescale,
  title = {Large-Scale Silicon Photonic Switches},
  booktitle = {2016 21st {{OptoElectronics}} and {{Communications Conference}} ({{OECC}}) Held Jointly with 2016 {{International Conference}} on {{Photonics}} in {{Switching}} ({{PS}})},
  author = {Wu, Ming C. and Seok, Tae Joon and Han, Sangyoon and Quack, Niels},
  year = 2016,
  month = jul,
  pages = {1--3},
  publisher = {IEEE},
  address = {Niigata, Japan},
  urldate = {2025-01-07}
}

@misc{yeluri2023Chiplets,
  title = {Chiplets - {{The Inevitable Transition}}},
  author = {Yeluri, Sharada},
  year = 2023,
  month = oct,
  urldate = {2025-07-09},
  langid = {english},
  howpublished = {\url{https://community.juniper.net/blogs/sharada-yeluri/2023/10/13/chiplets-the-inevitable-transition}},
  key = {Chiplets - The Inevitable Transition}
}

@inproceedings{zhou2012Mirror,
  title = {Mirror Mirror on the Ceiling: Flexible Wireless Links for Data Centers},
  shorttitle = {Mirror Mirror on the Ceiling},
  booktitle = {Proceedings of the {{ACM SIGCOMM}} 2012 Conference on {{Applications}}, Technologies, Architectures, and Protocols for Computer Communication},
  author = {Zhou, Xia and Zhang, Zengbin and Zhu, Yibo and Li, Yubo and Kumar, Saipriya and Vahdat, Amin and Zhao, Ben Y. and Zheng, Haitao},
  year = 2012,
  month = aug,
  pages = {443--454},
  publisher = {ACM},
  address = {Helsinki Finland},
  doi = {10.1145/2342356.2342440},
  urldate = {2024-05-15},
  isbn = {978-1-4503-1419-0},
  langid = {english}
}

@inproceedings{zilberman2019Stardust,
  title = {Stardust: {{Divide}} and {{Conquer}} in the {{Data Center Network}}},
  shorttitle = {Stardust},
  booktitle = {16th {{USENIX Symposium}} on {{Networked Systems Design}} and {{Implementation}} ({{NSDI}} 19)},
  author = {Zilberman, Noa and Bracha, Gabi and Schzukin, Golan},
  year = 2019,
  pages = {141--160},
  publisher = {USENIX Association},
  address = {Boston, MA, USA},
  urldate = {2025-09-16},
  isbn = {978-1-931971-49-2},
  langid = {english}
}

@inproceedings{zu2024Resiliency,
  title = {Resiliency at {{Scale}}: {{Managing}} \textbraceleft{{Google}}'s\textbraceright{} \textbraceleft{{TPUv4}}\textbraceright{} {{Machine Learning Supercomputer}}},
  shorttitle = {Resiliency at {{Scale}}},
  booktitle = {21st {{USENIX Symposium}} on {{Networked Systems Design}} and {{Implementation}} ({{NSDI}} 24)},
  author = {Zu, Yazhou and Ghaffarkhah, Alireza and Dang, Hoang-Vu and Towles, Brian and Hand, Steven and Huda, Safeen and Bello, Adekunle and Kolbasov, Alexander and Rezaei, Arash and Du, Dayou and Lacy, Steve and Wang, Hang and Wisner, Aaron and Lewis, Chris and Bahini, Henri},
  year = 2024,
  pages = {761--774},
  publisher = {USENIX Association},
  address = {Santa Clara, CA, USA},
  urldate = {2024-04-17},
  isbn = {978-1-939133-39-7},
  langid = {english}
}
\end{small}

\clearpage
\appendix

\newpage

\section{Appendix}

\subsection{Control Plane Efficiency Contd.}
\label{app:control-eff-contd}

The runtime of \Lightpass/ control algorithm can be reduced further through port bundling and hardware acceleration.
Additionally, we compare the runtime of our heuristic with that of the RDC algorithm.

\subsubsection{Bundling Ports}
\label{app:Bundling}
A straightforward way to reduce algorithm runtime is to bundle multiple physical ports into a single ``logical'' port. 
In this case, reconfiguration operates at the group level rather than per port, reducing the effective problem size. 
The optimal matching algorithm scales with $p^3$, so bundling four ports into one lowers the runtime by up to $64\times$.

\Cref{fig:bundle_median} shows the mean throughput and $99^{th}$ percentile FCTs for different inter-ASIC oversubscription ratios. 
At $2{:}1$ oversubscription, bundling four ports results in $<1\,\%$ performance loss compared to the \Ideal/ baseline. 
However, at higher oversubscription ratios, the loss of reconfiguration flexibility becomes more significant, as traffic localization becomes increasingly critical.

\def\addlegend{11}
\begin{figure}
        \centering
        \begin{tikzpicture}
\pgfplotsset{every axis/.append style={font=\large,line width=1pt,tick style={line width=0.8pt}},height=27mm}
\begin{axis}[ 
 title=\textbf{(a)} Mean Throughput (ML), ylabel style={at={(axis description cs:1.15,.5)}, anchor=north, align=center},scaled ticks=false, width=0.8\linewidth,symbolic x coords={2:1,4:1,8:1,16:1,32:1}, xtick=data,ticklabel style={/pgf/number format/fixed,/pgf/number format/.cd, 1000 sep = {}}, xtick pos=left, ytick pos=right,ymin=-0.5, xticklabels={}, name=first plot, legend style={at={(0.3,1.3)}, draw=none, anchor=south}, legend columns=-1] 
\addplot [myblue, mark=*, forget plot] table { 
 
"Inter ASIC Oversub." "Mean Throughput [%]"

2:1 100.0
4:1 100.0
8:1 100.0
16:1 100.0
32:1 100.0
};
\addplot [mygreen, mark=diamond*, forget plot] table { 
 
"Inter ASIC Oversub." "Mean Throughput [%]"

2:1 53.97658870223787
4:1 18.602913224135282
8:1 8.104459755326806
16:1 3.8563921759611555
32:1 1.8898067397233185
};
\addplot [mylightgreen, mark=o, forget plot] table { 
 
"Inter ASIC Oversub." "Mean Throughput [%]"

2:1 99.62443527389807
4:1 95.8501235239097
8:1 72.79912025128644
16:1 53.42880747243201
32:1 43.52255665354894
};
\addplot [mydarkred, mark=triangle, forget plot] table { 
 
"Inter ASIC Oversub." "Mean Throughput [%]"

2:1 99.5996695204183
4:1 92.52781146442472
8:1 64.4738383558932
16:1 44.89824155799773
32:1 37.15432939142078
};
\addplot [mylightblue, mark=x, forget plot] table { 
 
"Inter ASIC Oversub." "Mean Throughput [%]"

2:1 99.57398714616474
4:1 89.58891687819336
8:1 59.09576332995668
16:1 39.012837975876465
32:1 33.38624804869924
};
\addlegendimage{mygreen, mark=diamond*}
\addlegendimage{mylightgreen, mark=o}
\addlegendimage{mydarkred, mark=triangle}
\addlegendimage{mylightblue, mark=x}
\addlegendimage{myblue, mark=*}
\if\addlegend 
 \legend{\StaticPlot/,1-Port,2-Port,4-Port,\IdealPlot/}
 \fi 
\end{axis}
\begin{axis}[ylabel style={at={(axis description cs:-0.75,.5)},anchor=south,align=center}, ylabel={\footnotesize Thrp. [\%]},scaled ticks=false,symbolic x coords={2:1}, xtick=data,xticklabels={},ticklabel style={/pgf/number format/fixed,/pgf/number format/.cd, 1000 sep = {}}, tick pos=left, xmin={[normalized]-0.6}, xmax={[normalized]+0.6},ymin=99,legend style={at={(0.42,1.4)}, draw=none,anchor=south}, legend columns=-1,at=(first plot.west), name=plotfirst, anchor=east, width=0.3\linewidth, xshift=-3mm, xlabel style={yshift=1mm}]\addplot [myblue, mark=*, forget plot] table { 
 
"Inter ASIC Oversub." "Mean Throughput [%]"

2:1 100.0
};
\addplot [mygreen, mark=diamond*, forget plot] table { 
 
"Inter ASIC Oversub." "Mean Throughput [%]"

2:1 53.97658870223787
};
\addplot [mylightgreen, mark=o, forget plot] table { 
 
"Inter ASIC Oversub." "Mean Throughput [%]"

2:1 99.62443527389807
};
\addplot [mydarkred, mark=triangle, forget plot] table { 
 
"Inter ASIC Oversub." "Mean Throughput [%]"

2:1 99.5996695204183
};
\addplot [mylightblue, mark=x, forget plot] table { 
 
"Inter ASIC Oversub." "Mean Throughput [%]"

2:1 99.57398714616474
};
\end{axis}
\begin{axis}[ 
 title=\textbf{(b)} $99^{th}$ Perc. FCT (ML), ylabel style={at={(axis description cs:1.15,.5)}, anchor=north, align=center},xlabel={Inter ASIC Oversub.}, scaled ticks=false, width=0.8\linewidth,symbolic x coords={2:1,4:1,8:1,16:1,32:1}, xtick=data,ticklabel style={/pgf/number format/fixed,/pgf/number format/.cd, 1000 sep = {}}, xtick pos=left, ymin=-10, ymax=65,name=second plot, at=(first plot.south), anchor=north, yshift=-4mm, ytick pos=right, xlabel style={yshift=1mm}]] 
\addplot [myblue, mark=*, forget plot] table { 
 
"Inter ASIC Oversub." "All 99th FCT [ms]"

2:1 1.6351165
4:1 1.6351165
8:1 1.6351165
16:1 1.6351165
32:1 1.6351165
};
\addplot [mygreen, mark=diamond*, forget plot] table { 
 
"Inter ASIC Oversub." "All 99th FCT [ms]"

2:1 2.46669475
4:1 5.8852115
8:1 12.4957555
16:1 25.4290535
32:1 51.38612725
};
\addplot [mylightgreen, mark=o, forget plot] table { 
 
"Inter ASIC Oversub." "All 99th FCT [ms]"

2:1 1.6286915
4:1 1.64103075
8:1 2.0125625
16:1 2.7738705
32:1 3.56660025
};
\addplot [mydarkred, mark=triangle, forget plot] table { 
 
"Inter ASIC Oversub." "All 99th FCT [ms]"

2:1 1.63564375
4:1 1.66724825
8:1 2.26268075
16:1 3.11333025
32:1 3.98711825
};
\addplot [mylightblue, mark=x, forget plot] table { 
 
"Inter ASIC Oversub." "All 99th FCT [ms]"

2:1 1.6400385
4:1 1.69374425
8:1 2.44357
16:1 3.43555475
32:1 4.26145125
};
\end{axis}
\begin{axis}[ylabel style={at={(axis description cs:-0.75,.5)},anchor=south,align=center}, ylabel={\footnotesize FCT [ms]},scaled ticks=false,symbolic x coords={2:1}, xtick=data,ticklabel style={/pgf/number format/fixed,/pgf/number format/.cd, 1000 sep = {}}, tick pos=left, xmin={[normalized]-0.6}, xmax={[normalized]+0.6},ymin=-0.4,legend style={at={(0.42,1.4)}, draw=none,anchor=south}, legend columns=-1,at=(second plot.west), name=plotsecond, anchor=east, width=0.3\linewidth, xshift=-3mm, xlabel style={yshift=1mm}]\addplot [myblue, mark=*, forget plot] table { 
 
"Inter ASIC Oversub." "All 99th FCT [ms]"

2:1 1.6351165
};
\addplot [mygreen, mark=diamond*, forget plot] table { 
 
"Inter ASIC Oversub." "All 99th FCT [ms]"

2:1 2.46669475
};
\addplot [mylightgreen, mark=o, forget plot] table { 
 
"Inter ASIC Oversub." "All 99th FCT [ms]"

2:1 1.6286915
};
\addplot [mydarkred, mark=triangle, forget plot] table { 
 
"Inter ASIC Oversub." "All 99th FCT [ms]"

2:1 1.63564375
};
\addplot [mylightblue, mark=x, forget plot] table { 
 
"Inter ASIC Oversub." "All 99th FCT [ms]"

2:1 1.6400385
};
\end{axis}
\node[draw,rectangle,minimum width=1mm, minimum height=2mm] at (0.43,1) (box) {}; 
\node[draw,rectangle,minimum width=1mm, minimum height=2mm] at (0.43,-1.36) (box2) {}; 
\draw[] (box.south west) -- (plotfirst.south east); 
 \draw[] (box.north west) -- (plotfirst.north east); 
\draw[] (box2.south west) -- (plotsecond.south east); 
 \draw[] (box2.north west) -- (plotsecond.north east); 
\end{tikzpicture}%
        \caption{\Lightpass/ works even with bundled ports, the performance degrades gracefully, and is still better than the \Static/ architecture.}
        \label{fig:bundle_median}
\end{figure}
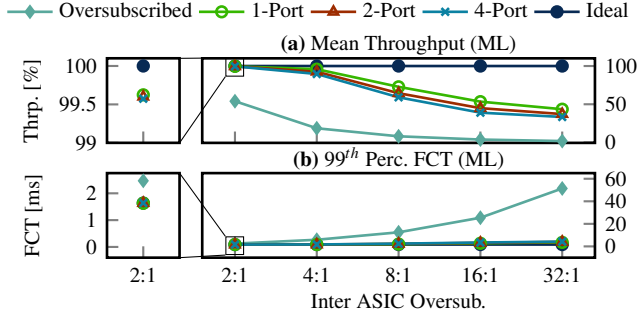

\subsubsection{Specialized Hardware}
Looking forward, specialized hardware can further accelerate the control plane. 
Key operations in both the Hungarian algorithm (e.g., row minima) and the heuristic (e.g., sorting) are inherently parallelizable. 
Implementations on FPGAs or dedicated accelerators~\cite{tsmots2017FPGA,amaru2012High,song2016Parallel,abdelrasoul2021FPGA} can exploit this parallelism to further reduce computation time, making the control plane even more practical at scale.

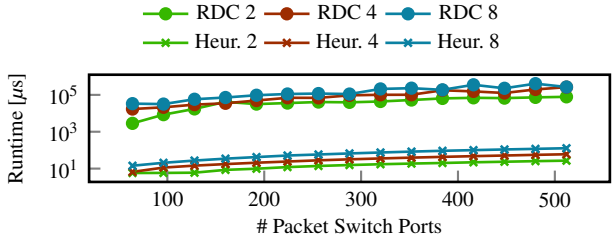
\begin{figure}
    \centering
    \begin{tikzpicture}
\pgfplotsset{every axis/.append style={font=\large,line width=1pt,tick style={line width=0.8pt}},height=30mm}
\begin{axis}[ 
 title=,ylabel style={at={(axis description cs:-0.1,.5)},anchor=south,align=center}, tick pos=left, ylabel={\footnotesize Runtime [$\mu$s]}, xlabel={\# Packet Switch Ports}, scaled ticks=false,ticklabel style={/pgf/number format/fixed,/pgf/number format/.cd, 1000 sep = {}},ymode=log,axis on top, name=first plot, legend style={at={(0.45,1.1)},draw=none,anchor=south},legend columns=-1, xlabel style={yshift=1mm},legend columns=2,transpose legend] 
\addplot [mylightgreen,mark=*] table { 
 
Ports "Time [us]"

512 78279.7
480 73196.0
448 66710.1
416 69148.7
384 64270.8
352 52618.8
320 43816.8
288 39393.6
256 40884.5
224 35944.4
192 31616.7
160 41821.2
128 17314.2
96 8447.9
64 2849.3
};
\addplot [mylightgreen,mark=x] table { 
 
Ports "Time [us]"

512 26.966873
480 25.377641
448 23.558358000000002
416 21.996596
384 20.139525000000003
352 18.741952
320 17.243809000000002
288 15.609706000000001
256 13.883294
224 12.052634
192 9.856334
160 8.52978
128 6.001991
96 5.828282
64 5.766323000000001
};
\addplot [mydarkred,mark=*] table { 
 
Ports "Time [us]"

512 262400.5
480 196647.0
448 126288.1
416 154437.8
384 177604.6
352 101407.8
320 100876.1
288 94278.4
256 68855.6
224 69350.3
192 49336.6
160 34920.3
128 29522.8
96 21131.0
64 17058.5
};
\addplot [mydarkred,mark=x] table { 
 
Ports "Time [us]"

512 59.956807999999995
480 55.716817000000006
448 51.268497
416 47.326616
384 43.249365
352 39.393964
320 35.501112
288 31.822049
256 28.088967
224 24.190226
192 20.840791
160 17.550276999999998
128 14.493191999999999
96 11.336316
64 6.625228
};
\addplot [mylightblue,mark=*] table { 
 
Ports "Time [us]"

512 266643.3
480 405990.0
448 227635.5
416 352155.5
384 187498.6
352 230690.6
320 209669.3
288 110756.2
256 117728.6
224 111438.6
192 95450.7
160 71891.8
128 57834.4
96 31624.0
64 33368.6
};
\addplot [mylightblue,mark=x] table { 
 
Ports "Time [us]"

512 127.323994
480 117.71757000000001
448 109.13928
416 100.41654
384 92.394908
352 83.470832
320 74.724402
288 66.428341
256 58.23838
224 51.228274
192 42.165218
160 34.588193
128 27.259138
96 20.61468
64 14.305301
};
\legend{RDC 2,Heur. 2,RDC 4,Heur. 4,RDC 8,Heur. 8}
\end{axis}
\end{tikzpicture}%
    \caption{Our heuristic is around 3 orders of magnitude faster than the RDC heuristic at the same input size.}
    \label{fig:algo_perf_rdc}
\end{figure}

\subsubsection{RDC Algorithm Is Not Fast Enough}
\label{app:rdcslow}
We compare the runtime of our heuristic to the bipartite graph partitioning heuristic used in the RDC paper~\cite{wang2022RDC}.
\Cref{fig:algo_perf_rdc} shows the runtime for both algorithms for different numbers of ASICs, with the number of ports per device on the x-axis.
We show that the RDC algorithm is around three orders of magnitude slower than our heuristic for the same number of ASICs and switch ports.
This increased algorithm runtime would seriously limit how often a reconfiguration can be performed.

\subsection{Adapting the Number of Hosts}
\label{app:host}
Variation in the radix each architecture achieves would lead to differing topologies even within the same experiment, significantly altering traffic patterns.
This matters especially for structured traffic like Shuffle and Stride, which have a fixed offset at which hosts communicate. 
This create a problem as two communicating hosts might be connected to the same Top-of-Rack (ToR) switch in one architecture, but are separated across different switches in another.
To avoid this problem, we keep the radix fixed for each architecture in our evaluation, as this maintains the same topology and, therefore, the same traffic pattern across architectures. 
We do this by dividing the total usable capacity by the constant radix to obtain the speed of each switch port.
%\Cref{fig:host_web} shows the case for a Web traffic scenario, in which we have adjusted the number of hosts for each architecture.
%We pick the Web traffic scenario since it represents uniform random traffic, which should be largely invariant even across different topologies.
%\Cref{fig:host_web} shows the results for the Web traffic traces, but with varying numbers of hosts depending on the architecture.
%The results are almost identical to those in the main evaluation in \Cref{subsec:performance} with a fixed number of hosts, thereby validating our methodology.
%% Aggressive

In this experiment, we evaluate what would happen if we didn't keep the host count the same over all architectures. 
We use it to compare the performance to the constant host count evaluation setup we use for the other experiments.
The data in \Cref{fig:host_web} shows that, compared to the main evaluation (\Cref{fig:load_web}), the performance of the \Baseline/ is worse when we adapt the host count.
Meaning that our main evaluation, which keeps the number of hosts fixed, is actually favorable to the \Baseline/.
The reason is that the number of hosts in a 2-stage Fat Tree topology scales quadratically with the radix, and the \Baseline/ has a smaller radix compared to the other architectures.
Since we have to keep the host count fixed to maintain consistent traffic patterns, this consequently also neutralizes the radix advantage that \Lightpass/ has over the \Baseline/.

\def\addlegend{11}
\begin{figure}
        \centering
        \begin{tikzpicture}
\pgfplotsset{every axis/.append style={font=\large, line width=1pt, tick style={line width=0.8pt}},width=0.52\linewidth}
\begin{axis}[ 
 title= Mean Throughput (Web), ylabel style={at={(axis description cs:-0.2,.5)}, anchor=south, align=center}, ylabel={\footnotesize Thrp. [\%]}, scaled ticks=false, ticklabel style={/pgf/number format/fixed,/pgf/number format/.cd, 1000 sep = {}}, tick pos=left,ymin=0,axis on top,xlabel={Load [\%]},name=first plot, legend style={at={(1.1,1.26)}, draw=none, anchor=south}, legend columns=-1, xlabel style={yshift=1mm}] 
\addplot [myblue, mark=*, forget plot] table { 
 
"Load [%]" "Mean Throughput [%]"

20 100.0
40 100.0
60 100.0
80 100.0
100 100.0
};
\addplot [myviolet, mark=triangle*, forget plot] table { 
 
"Load [%]" "Mean Throughput [%]"

20 97.47844033122347
40 61.45682330114892
60 21.41586479615323
80 25.962529005781505
100 36.170295019128105
};
\addplot [mygreen, mark=diamond*, forget plot] table { 
 
"Load [%]" "Mean Throughput [%]"

20 99.2652592892513
40 97.81529792134117
60 24.29328765100239
80 27.609032002068446
100 37.88225803839543
};
\addplot [myred, mark=star, forget plot] table { 
 
"Load [%]" "Mean Throughput [%]"

20 99.33064997684372
40 99.45468778021235
60 99.36063445035265
80 98.98655303359675
100 98.94247475978975
};
\addlegendimage{myviolet, mark=triangle*}
\addlegendimage{mygreen, mark=diamond*}
\addlegendimage{myred, mark=star}
\addlegendimage{myblue, mark=*}
\if\addlegend 
 \legend{\BaselinePlot/,\StaticPlot/,\LightpassPlot/,\IdealPlot/}
 \fi 
\end{axis}
\begin{axis}[ 
 title= $99^{th}$ Perc. FCT (Web), ylabel style={at={(axis description cs:-0.15,.5)}, anchor=south, align=center}, ylabel={\footnotesize FCT [ms]}, xlabel={Load [\%]}, scaled ticks=false,ytick distance=1,ticklabel style={/pgf/number format/fixed,/pgf/number format/.cd, 1000 sep = {}}, tick pos=left,ymin=0,axis on top,at=(first plot.east), anchor=west, xshift=15mm, xlabel style={yshift=1mm}] 
\addplot [myblue, mark=*, forget plot] table { 
 
"Load [%]" "All 99th FCT [ms]"

20 0.122524
40 0.128449
60 0.17228700999999977
80 0.426354
100 0.9009540499999988
};
\addplot [myviolet, mark=triangle*, forget plot] table { 
 
"Load [%]" "All 99th FCT [ms]"

20 0.12649700999999977
40 0.289476
60 1.6615870299999993
80 2.081952099999998
100 2.246442049999999
};
\addplot [mygreen, mark=diamond*, forget plot] table { 
 
"Load [%]" "All 99th FCT [ms]"

20 0.123534
40 0.132549
60 1.5475960299999993
80 2.0468070699999985
100 2.2315350199999995
};
\addplot [myred, mark=star, forget plot] table { 
 
"Load [%]" "All 99th FCT [ms]"

20 0.12542805999999865
40 0.1330650099999998
60 0.17417300999999977
80 0.4300781099999975
100 0.9077750199999995
};
\end{axis}
\end{tikzpicture}%
        %\caption{Varying the number of hosts per architecture performs identically to a fixed number of hosts for Web traffic.}
        \caption{Adjusting the host count based on the radix of each architecture hurts the performance of the \Baseline/ architecture.}
        \label{fig:host_web}
\end{figure}
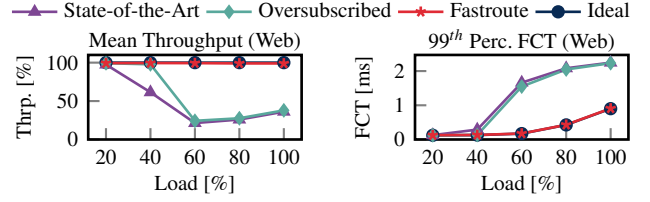

\subsection{Link Failures}
\label{app:failure}

Link failures can alter routing choices and induce traffic patterns that differ from normal operation. 
To evaluate the robustness of \Lightpass/ against failure, we randomly disable $64$ out of $4096$ links. 
\Cref{fig:fail_web_median} shows the resulting FCTs under different traffic loads.
As expected, the reduced network capacity slightly increases FCTs for both \Lightpass/ and the ideal monolithic single-ASIC switch. 
However, \Lightpass/ still closely matches the performance of \Ideal/, since both are equally affected by bandwidth loss. 
The same is also true for the \Static/ and \Baseline/ multi-ASIC designs, which remain largely unaffected.
\lukas{include new baseline}

\def\addlegend{11}
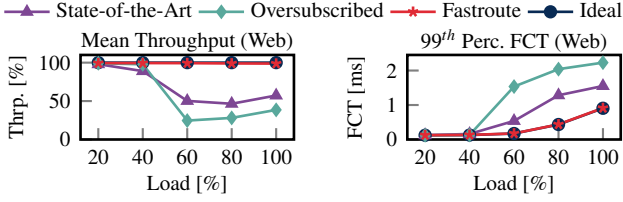
\begin{figure}
        \centering
        \begin{tikzpicture}
\pgfplotsset{every axis/.append style={font=\large, line width=1pt, tick style={line width=0.8pt}},width=0.52\linewidth}
\begin{axis}[ 
 title= Mean Throughput (Web), ylabel style={at={(axis description cs:-0.2,.5)}, anchor=south, align=center}, ylabel={\footnotesize Thrp. [\%]}, scaled ticks=false, ticklabel style={/pgf/number format/fixed,/pgf/number format/.cd, 1000 sep = {}}, tick pos=left,ymin=0,axis on top,xlabel={Load [\%]},name=first plot, legend style={at={(1.1,1.26)}, draw=none, anchor=south}, legend columns=-1, xlabel style={yshift=1mm}] 
\addplot [myblue, mark=*, forget plot] table { 
 
"Load [%]" "Mean Throughput [%]"

20 100.0
40 100.0
60 100.0
80 100.0
100 100.0
};
\addplot [myviolet, mark=triangle*, forget plot] table { 
 
"Load [%]" "Mean Throughput [%]"

20 97.97796624852671
40 88.99057943237291
60 50.21574014926029
80 46.494533213687525
100 57.251624856504826
};
\addplot [mygreen, mark=diamond*, forget plot] table { 
 
"Load [%]" "Mean Throughput [%]"

20 99.2663068208381
40 97.81580028622753
60 24.61809771887765
80 28.10739465883477
100 38.52494757748732
};
\addplot [myred, mark=star, forget plot] table { 
 
"Load [%]" "Mean Throughput [%]"

20 99.33531775678962
40 99.4449855176416
60 99.32348635971154
80 98.94540487037
100 98.73776078751264
};
\addlegendimage{myviolet, mark=triangle*}
\addlegendimage{mygreen, mark=diamond*}
\addlegendimage{myred, mark=star}
\addlegendimage{myblue, mark=*}
\if\addlegend 
 \legend{\BaselinePlot/,\StaticPlot/,\LightpassPlot/,\IdealPlot/}
 \fi 
\end{axis}
\begin{axis}[ 
 title= $99^{th}$ Perc. FCT (Web), ylabel style={at={(axis description cs:-0.15,.5)}, anchor=south, align=center}, ylabel={\footnotesize FCT [ms]}, xlabel={Load [\%]}, scaled ticks=false,ytick distance=1,ticklabel style={/pgf/number format/fixed,/pgf/number format/.cd, 1000 sep = {}}, tick pos=left,ymin=0,axis on top,at=(first plot.east), anchor=west, xshift=15mm, xlabel style={yshift=1mm}] 
\addplot [myblue, mark=*, forget plot] table { 
 
"Load [%]" "All 99th FCT [ms]"

20 0.122535
40 0.12853400999999978
60 0.17298900999999978
80 0.4345210399999991
100 0.905802
};
\addplot [myviolet, mark=triangle*, forget plot] table { 
 
"Load [%]" "All 99th FCT [ms]"

20 0.125545
40 0.16056401999999956
60 0.538562
80 1.2830680699999983
100 1.5514780599999987
};
\addplot [mygreen, mark=diamond*, forget plot] table { 
 
"Load [%]" "All 99th FCT [ms]"

20 0.123544
40 0.132655
60 1.5350452799999939
80 2.0404340599999986
100 2.2290100699999984
};
\addplot [myred, mark=star, forget plot] table { 
 
"Load [%]" "All 99th FCT [ms]"

20 0.12528402999999932
40 0.133133
60 0.1748870099999998
80 0.438397
100 0.9141261199999973
};
\end{axis}
\end{tikzpicture}%
        \caption{\Lightpass/ still works well even with an asymmetric topology due to failed links.}
        \label{fig:fail_web_median}
\end{figure}

\def\addlegend{11}
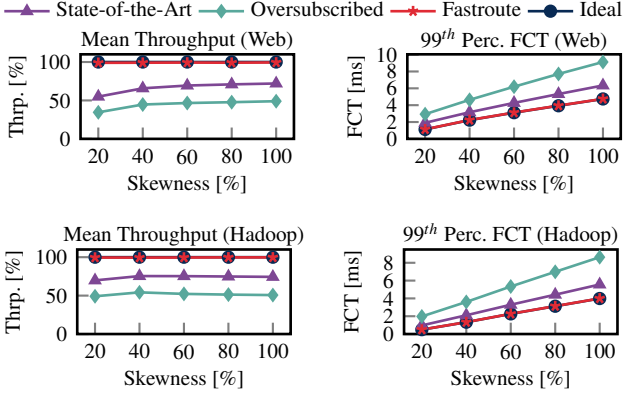
\begin{figure}
    \centering
    \begin{subfigure}{\linewidth}
        \centering
        \begin{tikzpicture}
\pgfplotsset{every axis/.append style={font=\large, line width=1pt, tick style={line width=0.8pt}},width=0.52\linewidth}
\begin{axis}[ 
 title= Mean Throughput (Web), ylabel style={at={(axis description cs:-0.2,.5)}, anchor=south, align=center}, ylabel={\footnotesize Thrp. [\%]}, scaled ticks=false, ticklabel style={/pgf/number format/fixed,/pgf/number format/.cd, 1000 sep = {}}, tick pos=left,ymin=0,axis on top,xlabel={Skewness [\%]},name=first plot, legend style={at={(1.1,1.26)}, draw=none, anchor=south}, legend columns=-1, xlabel style={yshift=1mm}] 
\addplot [myblue, mark=*, forget plot] table { 
 
"Skewness [%]" "Mean Throughput [%]"

20 100.0
40 100.0
60 100.0
80 100.0
100 100.0
};
\addplot [myviolet, mark=triangle*, forget plot] table { 
 
"Skewness [%]" "Mean Throughput [%]"

20 54.94633021763119
40 65.7748517917003
60 69.38737505851495
80 70.90110113644437
100 72.0481928805628
};
\addplot [mygreen, mark=diamond*, forget plot] table { 
 
"Skewness [%]" "Mean Throughput [%]"

20 34.542750803072444
40 44.677547432688776
60 46.6249079541724
80 47.723892994488395
100 49.00748247033789
};
\addplot [myred, mark=star, forget plot] table { 
 
"Skewness [%]" "Mean Throughput [%]"

20 98.93799146850353
40 99.07080732232617
60 99.28229570623232
80 99.06997241044658
100 99.4182291873711
};
\addlegendimage{myviolet, mark=triangle*}
\addlegendimage{mygreen, mark=diamond*}
\addlegendimage{myred, mark=star}
\addlegendimage{myblue, mark=*}
\if\addlegend 
 \legend{\BaselinePlot/,\StaticPlot/,\LightpassPlot/,\IdealPlot/}
 \fi 
\end{axis}
\begin{axis}[ 
 title= $99^{th}$ Perc. FCT (Web), ylabel style={at={(axis description cs:-0.15,.5)}, anchor=south, align=center}, ylabel={\footnotesize FCT [ms]}, xlabel={Skewness [\%]}, scaled ticks=false,ticklabel style={/pgf/number format/fixed,/pgf/number format/.cd, 1000 sep = {}}, tick pos=left,ymin=0,axis on top,at=(first plot.east), anchor=west, xshift=15mm, xlabel style={yshift=1mm}] 
\addplot [myblue, mark=*, forget plot] table { 
 
"Skewness [%]" "All 99th FCT [ms]"

20 1.121703
40 2.224428139999997
60 3.1104071099999975
80 3.93023
100 4.738581069999999
};
\addplot [myviolet, mark=triangle*, forget plot] table { 
 
"Skewness [%]" "All 99th FCT [ms]"

20 1.9056613799999915
40 3.1622130399999993
60 4.269978189999995
80 5.321768029999999
100 6.341718019999999
};
\addplot [mygreen, mark=diamond*, forget plot] table { 
 
"Skewness [%]" "All 99th FCT [ms]"

20 2.9308911899999956
40 4.634020129999997
60 6.192982169999996
80 7.704639149999997
100 9.123970059999998
};
\addplot [myred, mark=star, forget plot] table { 
 
"Skewness [%]" "All 99th FCT [ms]"

20 1.1333810099999997
40 2.2372420699999984
60 3.1223810599999986
80 3.94804501
100 4.7484561199999975
};
\end{axis}
\end{tikzpicture}%
        \label{fig:skew_web}
    \end{subfigure}
    \begin{subfigure}{\linewidth}
        \vspace{-3mm}
        \centering
        \def\addlegend{0}
        \begin{tikzpicture}
\pgfplotsset{every axis/.append style={font=\large, line width=1pt, tick style={line width=0.8pt}},width=0.52\linewidth}
\begin{axis}[ 
 title= Mean Throughput (Hadoop), ylabel style={at={(axis description cs:-0.2,.5)}, anchor=south, align=center}, ylabel={\footnotesize Thrp. [\%]}, scaled ticks=false, ticklabel style={/pgf/number format/fixed,/pgf/number format/.cd, 1000 sep = {}}, tick pos=left,ymin=0,axis on top,xlabel={Skewness [\%]},name=first plot, legend style={at={(1.1,1.26)}, draw=none, anchor=south}, legend columns=-1, xlabel style={yshift=1mm}] 
\addplot [myblue, mark=*, forget plot] table { 
 
"Skewness [%]" "Mean Throughput [%]"

20 100.0
40 100.0
60 100.0
80 100.0
100 100.0
};
\addplot [myviolet, mark=triangle*, forget plot] table { 
 
"Skewness [%]" "Mean Throughput [%]"

20 69.80757436391801
40 75.42658984791824
60 75.32931214829033
80 74.84764457526151
100 74.37457742863145
};
\addplot [mygreen, mark=diamond*, forget plot] table { 
 
"Skewness [%]" "Mean Throughput [%]"

20 49.12999538100087
40 54.234955383478436
60 52.2400570625427
80 51.30408310459359
100 50.68834040448702
};
\addplot [myred, mark=star, forget plot] table { 
 
"Skewness [%]" "Mean Throughput [%]"

20 99.4962803610867
40 99.43105454933175
60 99.53643282997888
80 99.71339914793732
100 99.70840803932134
};
\addlegendimage{myviolet, mark=triangle*}
\addlegendimage{mygreen, mark=diamond*}
\addlegendimage{myred, mark=star}
\addlegendimage{myblue, mark=*}
\if\addlegend 
 \legend{\BaselinePlot/,\StaticPlot/,\LightpassPlot/,\IdealPlot/}
 \fi 
\end{axis}
\begin{axis}[ 
 title= $99^{th}$ Perc. FCT (Hadoop), ylabel style={at={(axis description cs:-0.15,.5)}, anchor=south, align=center}, ylabel={\footnotesize FCT [ms]}, xlabel={Skewness [\%]}, scaled ticks=false,ticklabel style={/pgf/number format/fixed,/pgf/number format/.cd, 1000 sep = {}}, tick pos=left,ymin=0,axis on top,at=(first plot.east), anchor=west, xshift=15mm, xlabel style={yshift=1mm}] 
\addplot [myblue, mark=*, forget plot] table { 
 
"Skewness [%]" "All 99th FCT [ms]"

20 0.49266102999999933
40 1.3292910299999994
60 2.2629101099999978
80 3.123030279999994
100 3.9879660099999996
};
\addplot [myviolet, mark=triangle*, forget plot] table { 
 
"Skewness [%]" "All 99th FCT [ms]"

20 0.9687111399999969
40 2.1067201099999977
60 3.2898567299999835
80 4.418392289999994
100 5.562815119999997
};
\addplot [mygreen, mark=diamond*, forget plot] table { 
 
"Skewness [%]" "All 99th FCT [ms]"

20 1.9752123399999924
40 3.6043036299999858
60 5.3359960399999995
80 6.9856422099999955
100 8.628484239999995
};
\addplot [myred, mark=star, forget plot] table { 
 
"Skewness [%]" "All 99th FCT [ms]"

20 0.4961771699999962
40 1.3328717399999834
60 2.2693420199999994
80 3.134682099999998
100 3.9950650499999987
};
\end{axis}
\end{tikzpicture}%
        \label{fig:skew_hadoop}
    \end{subfigure}
    \caption{The inter-ASIC bottleneck becomes even more pronounced for traffic with higher skewness.}
    \label{fig:skew}
\end{figure}

\subsection{Impact of Traffic Skewness}
\label{app:trafficskewness}

\Cref{fig:skew} shows the mean throughput and $99^{th}$ percentile FCTs of different switch architectures for Web and Hadoop traffic scenarios, respectively, across different levels of skewness at $80\,\%$ network load. 
As observed in~\Cref{subsec:networkload}, the \Static/ architecture already suffers at high network load; while the throughput does not change significantly with higher traffic skewness, the $99^{th}$ percentile FCT slightly increases.
The same is true for the \Baseline/, which performs better than the \Static/ architecture but still struggles over the whole range of traffic skewness.
%Even the alternative cabling has a significant slowdown in FCT, despite reducing the inter-ASIC traffic volume, e.g., the median FCT slowdown is around $60\,\%$ w.r.t. the \Ideal/. 
\Lightpass/, in contrast, can closely match the \Ideal/ single-ASIC switch performance across different traffic skewness by reconfiguring the port-to-ASIC mapping.

\def\addlegend{11}
\begin{figure}
    \begin{subfigure}[]{\linewidth}
        \centering
        \begin{tikzpicture}
\pgfplotsset{every axis/.append style={font=\large, line width=1pt, tick style={line width=0.8pt}},width=0.52\linewidth}
\begin{axis}[ 
 title= Mean Throughput (Hadoop), ylabel style={at={(axis description cs:-0.2,.5)}, anchor=south, align=center}, ylabel={\footnotesize Thrp. [\%]}, scaled ticks=false, symbolic x coords={1:1,3:1,7:1,15:1}, xtick=data,ticklabel style={/pgf/number format/fixed,/pgf/number format/.cd, 1000 sep = {}}, tick pos=left,ymin=0,axis on top,xlabel={Oversubscription},name=first plot, legend style={at={(1.1,1.26)}, draw=none, anchor=south}, legend columns=-1, xlabel style={yshift=1mm}] 
\addplot [myblue, mark=*, forget plot] table { 
 
Oversubscription "Mean Throughput [%]"

1:1 100.0
3:1 100.0
7:1 100.0
15:1 100.0
};
\addplot [myviolet, mark=triangle*, forget plot] table { 
 
Oversubscription "Mean Throughput [%]"

1:1 63.896003225465
3:1 73.30325073177
7:1 75.21453928472413
15:1 72.26252937242617
};
\addplot [mygreen, mark=diamond*, forget plot] table { 
 
Oversubscription "Mean Throughput [%]"

1:1 44.460391725320605
3:1 99.41180576413635
7:1 100.08098562432592
15:1 99.62580750244004
};
\addplot [myred, mark=star, forget plot] table { 
 
Oversubscription "Mean Throughput [%]"

1:1 99.42234846453132
3:1 99.6655621386258
7:1 99.84966994793795
15:1 100.05643318760306
};
\addlegendimage{myviolet, mark=triangle*}
\addlegendimage{mygreen, mark=diamond*}
\addlegendimage{myred, mark=star}
\addlegendimage{myblue, mark=*}
\if\addlegend 
 \legend{\BaselinePlot/,\StaticPlot/,\LightpassPlot/,\IdealPlot/}
 \fi 
\end{axis}
\begin{axis}[ 
 title= $99^{th}$ Perc. FCT (Hadoop), ylabel style={at={(axis description cs:-0.15,.5)}, anchor=south, align=center}, ylabel={\footnotesize FCT [ms]}, xlabel={Oversubscription}, scaled ticks=false,symbolic x coords={1:1,3:1,7:1,15:1}, xtick=data,ticklabel style={/pgf/number format/fixed,/pgf/number format/.cd, 1000 sep = {}}, tick pos=left,ymin=0,axis on top,at=(first plot.east), anchor=west, xshift=15mm, xlabel style={yshift=1mm}] 
\addplot [myblue, mark=*, forget plot] table { 
 
Oversubscription "All 99th FCT [ms]"

1:1 0.3614870299999993
3:1 2.665082129999997
7:1 7.3552500699999985
15:1 15.967732489999989
};
\addplot [myviolet, mark=triangle*, forget plot] table { 
 
Oversubscription "All 99th FCT [ms]"

1:1 0.684348
3:1 3.821551139999997
7:1 9.927569149999997
15:1 21.23108738999997
};
\addplot [mygreen, mark=diamond*, forget plot] table { 
 
Oversubscription "All 99th FCT [ms]"

1:1 1.421185049999999
3:1 2.6617680299999993
7:1 7.342826109999997
15:1 15.98789802
};
\addplot [myred, mark=star, forget plot] table { 
 
Oversubscription "All 99th FCT [ms]"

1:1 0.3640820399999991
3:1 2.6671281099999975
7:1 7.362341039999999
15:1 15.984332129999997
};
\end{axis}
\end{tikzpicture}%
        \caption{2-Stage Network Topology}
        \label{fig:oversub-2}
    \end{subfigure}
    \begin{subfigure}[]{\linewidth}
        \def\addlegend{0}
        \centering
        \begin{tikzpicture}
\pgfplotsset{every axis/.append style={font=\large, line width=1pt, tick style={line width=0.8pt}},width=0.52\linewidth}
\begin{axis}[ 
 title= Mean Throughput (Hadoop), ylabel style={at={(axis description cs:-0.2,.5)}, anchor=south, align=center}, ylabel={\footnotesize Thrp. [\%]}, scaled ticks=false, symbolic x coords={1:1,3:1,7:1,15:1}, xtick=data,ticklabel style={/pgf/number format/fixed,/pgf/number format/.cd, 1000 sep = {}}, tick pos=left,ymin=0,axis on top,xlabel={Oversubscription},name=first plot, legend style={at={(1.1,1.26)}, draw=none, anchor=south}, legend columns=-1, xlabel style={yshift=1mm}] 
\addplot [myblue, mark=*, forget plot] table { 
 
Oversubscription "Mean Throughput [%]"

1:1 100.0
3:1 100.0
7:1 100.0
15:1 100.0
};
\addplot [myviolet, mark=triangle*, forget plot] table { 
 
Oversubscription "Mean Throughput [%]"

1:1 58.850895727638516
3:1 71.99626879576147
7:1 74.7594862207253
15:1 71.29498345447634
};
\addplot [mygreen, mark=diamond*, forget plot] table { 
 
Oversubscription "Mean Throughput [%]"

1:1 36.65983635333739
3:1 52.402641434780875
7:1 55.04346816169273
15:1 46.687818635711245
};
\addplot [myred, mark=star, forget plot] table { 
 
Oversubscription "Mean Throughput [%]"

1:1 99.62586087275854
3:1 100.05855532866575
7:1 99.9679621120692
15:1 99.76156774955277
};
\addlegendimage{myviolet, mark=triangle*}
\addlegendimage{mygreen, mark=diamond*}
\addlegendimage{myred, mark=star}
\addlegendimage{myblue, mark=*}
\if\addlegend 
 \legend{\BaselinePlot/,\StaticPlot/,\LightpassPlot/,\IdealPlot/}
 \fi 
\end{axis}
\begin{axis}[ 
 title= $99^{th}$ Perc. FCT (Hadoop), ylabel style={at={(axis description cs:-0.15,.5)}, anchor=south, align=center}, ylabel={\footnotesize FCT [ms]}, xlabel={Oversubscription}, scaled ticks=false,symbolic x coords={1:1,3:1,7:1,15:1}, xtick=data,ticklabel style={/pgf/number format/fixed,/pgf/number format/.cd, 1000 sep = {}}, tick pos=left,ymin=0,axis on top,at=(first plot.east), anchor=west, xshift=15mm, xlabel style={yshift=1mm}] 
\addplot [myblue, mark=*, forget plot] table { 
 
Oversubscription "All 99th FCT [ms]"

1:1 0.5236820599999986
3:1 5.374625199999995
7:1 14.427258109999997
15:1 31.349894009999996
};
\addplot [myviolet, mark=triangle*, forget plot] table { 
 
Oversubscription "All 99th FCT [ms]"

1:1 1.2101401899999957
3:1 7.708077329999993
7:1 19.396666129999996
15:1 41.44278505999997
};
\addplot [mygreen, mark=diamond*, forget plot] table { 
 
Oversubscription "All 99th FCT [ms]"

1:1 2.89706101
3:1 11.453648099999997
7:1 27.72524839999999
15:1 58.057009849999936
};
\addplot [myred, mark=star, forget plot] table { 
 
Oversubscription "All 99th FCT [ms]"

1:1 0.5244580099999998
3:1 5.363214069999999
7:1 14.426743149999997
15:1 31.40747850999999
};
\end{axis}
\end{tikzpicture}%
        \caption{3-Stage Network Topology}
        \label{fig:oversub-3}
    \end{subfigure}
    \caption{Inter-ASIC bandwidth isn't always the bottleneck, but minimal changes like the number of stages impact the performance if \Lightpass/ is not leveraged.}
    \label{fig:oversub}
\end{figure}
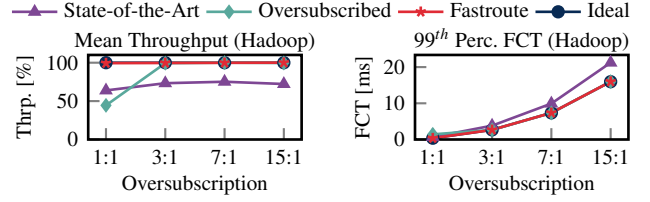
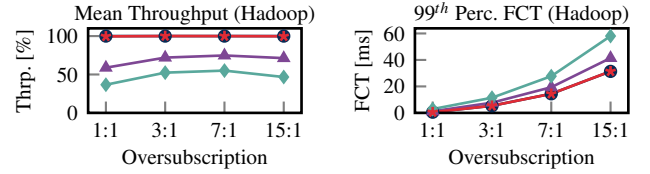

\subsection{Impact of Network Oversubscription}
\label{app:networkoversubscription}
Bandwidth oversubscription at the network core is a common cost-saving practice in modern datacenter networks~\cite{greenberg2009VL2, singh2015Jupiter, roy2015Social, chatzieleftheriou2018Larrya}. 
\Cref{fig:oversub} plots the mean throughput performance of switch architectures at different core oversubscription ratios (OS), considering two-stage and three-stage topologies, respectively. 
As we observe, for traditional DCN traffic, the performance issue with the \Static/ architecture is not prevalent in a two-stage oversubscribed topology (\Cref{fig:oversub-2}), which is mainly because the pattern created by this topology does not stress the inter-ASIC bandwidth. 
Note that this is not the case for other application-oriented traffic scenarios with more regular patterns, as shown in~\Cref{subsec:shuffleandstride}. 
However, for a three-stage oversubscribed topology, the \Static/ switch architecture faces significant performance degradation even at higher oversubscription ratios (\Cref{fig:oversub-3}). 
The primary reason is the high volume of inter-ASIC traffic at the middle stage (aggregation layer).
On the other hand, the performance of \Lightpass/ is unaffected w.r.t. the \Ideal/ across different oversubscribed scenarios for both two-stage and three-stage topologies, as it actively minimizes the inter-ASIC traffic volume.
While the \Baseline/ architecture remains largely unaffected by different oversubscription ratios, aside from slightly higher FCT at high oversubscription, it struggles to keep up across the whole range.
Showcasing again that the \Baseline/ architecture, by being a non-blocking design, is unaffected by traffic patterns, but struggles due to not providing as much outside bandwidth as the oversubscribed designs.

\begin{table}
    \centering
    \caption{More than $4$ ASICs are impractical due to the devices' total power consumption, even with larger savings.}
    \vspace{0.25em}
    \renewcommand{\arraystretch}{1.1}
    \footnotesize
    \tikzmarknode{tab1}{
    \begin{tabularx}{\columnwidth}{Xcccc}
         \toprule
         \textbf{Number of ASIC} & 2 & 4 & 6 & 8 \\
         \textbf{Throughput [Tbps]}              & 136.5   & 234.1     & 335.1  & 436.9 \\
         \midrule
         Total Savings for \Lightpass/ [$W$]              & 172.7   & 444.1  & 420.0    & 406.9 \\
         \bottomrule
         Switch w/o Oversubscription  [$W$]         & 1354     & 3523 & 5044 & 6575 \\
    \end{tabularx}}
    \vspace{0.5em}
    \label{tab:appendixpower}
\end{table}

\subsection{Non-Oversubscribed Switch Power}
\label{app:fabricpowerformula}
The formula shown in~\Cref{eqn:fabric} models the total power consumption of a non-oversubscribed $n$-ASIC switch, and is composed of the components established in~\Cref{subsec:powerandcost}. 
The energy for the packet-switch ASIC is calculated as $750\,W/102.4\,Tbps=7.32\,pJ/bit$ based on the numbers assumed for our baseline chip.
The individual components are in order: the energy per bit from the inter-ASIC links, fabric chips, buffers, and the packet-switch ASIC.
Since there is no oversubscription, all components are multiplied by the switch's total throughput.

\begin{equation}
\label{eqn:fabric}
    \begin{split}
        \text{Power} =& \  \text{Throughput} \cdot (0.6\,pJ/bit + 5.13\,pJ/bit\\  & \ +   2\,pJ/bit + 7.32\,pJ/bit)
    \end{split}
\end{equation}

\end{document}